\documentclass[onecolumn,preprintnumbers,superscriptaddress,pra]{revtex4}
\usepackage{axodraw4j}    
\usepackage{amsmath}      %
\usepackage{amssymb}      
\usepackage{amsfonts}     %
\usepackage{mathtools}    
\usepackage{graphicx}     
\usepackage{shuffle}      
\usepackage{slashbox}     
\usepackage{rotating}     
\usepackage{array}        
\usepackage[arrow, matrix, curve]{xy} 
\usepackage{color}
\usepackage{epsf}
\def\One{\mathbb{I}}
\newcolumntype{C}[1]{>{\centering}p{#1}}
\begin{document}
\title{Filtrations in Dyson-Schwinger equations:
  next-to$^{\{j\}}$-leading log expansions systematically}
\preprint{AEI-2014-061}
\author{Olaf Kr\"uger} \affiliation{Max-Planck-Institut f\"{u}r
  Gravitationsphysik, Albert-Einstein-Institut,\\ Am M\"{u}hlenberg 1,
  D-14476 Potsdam, Germany} \affiliation{Humboldt-Universit\"at zu
  Berlin, Institut f\"ur Physik,\\ Newtonstra\ss e 15, D-12489 Berlin,
  Germany}
\author{Dirk Kreimer
} \affiliation{Humboldt-Universit\"at zu Berlin,
  Institut f\"ur Physik,\\ Newtonstra\ss e 15, D-12489 Berlin,
  Germany}
\date{\today}
\begin{abstract}
  Dyson-Schwinger equations determine the Green functions
  $G^r(\alpha,L)$ in quantum field theory. Their solutions are
  triangular series in a coupling constant $\alpha$ and an external
  scale parameter $L$ for a chosen amplitude $r$, with the order in
  $L$ bounded by the order in the coupling.
  
  Perturbation theory calculates the first few orders in $\alpha$. On
  the other hand, Dyson--Schwinger equations determine
  next-to$^{\{\mathrm{j}\}}$-leading log expansions, $G^r(\alpha,L) =
  1 + \sum_{j=0}^\infty \sum_{\mathcal{M}} p_j^{\mathcal{M}}\alpha^j
  \mathcal{M}(u)$. $\sum_{\mathcal{M}}$ sums for any finite $j$ a finite number of
  functions $\mathcal{M}$ in $u$. Here, $u$ is the one-loop approximation to 
  $G^r$, for example, for the (inverse) propagator in massless Yukawa theory, $u  = \alpha L/2$. 
  
  The leading logs come
  then from the trivial representation $\mathcal{M}(u)
  = \begin{bsmallmatrix}\bullet\end{bsmallmatrix}(u)$ at $j=0$ with
  $p_0^{\begin{bsmallmatrix}\bullet\end{bsmallmatrix}} = 1$. All
  non-leading logs are organized by corresponding  suppressions in powers
  $\alpha^j$.
   
  We describe an algebraic method to derive all
  next-to$^{\{\mathrm{j}\}}$-leading log terms from the knowledge of
  the first $(j+1)$ terms in perturbation theory and their
  filtrations. This implies the calculation of the functions
  $\mathcal{M}(u)$ and periods $p_j^\mathcal{M}$.
  
  In the first part of our paper, we investigate the structure of
  Dyson-Schwinger equations and develop a method to filter their
  solutions. Applying renormalized Feynman rules maps each filtered
  term to a certain power of $\alpha$ and $L$ in the log-expansion.
  
  Based on this, the second part derives the
  next-to$^{\{\mathrm{j}\}}$-leading log expansions. Our method is
  general. Here, we exemplify it using the examples of the propagator
  in Yukawa theory and the photon self-energy in quantum
  electrodynamics. In particular, we give explicit formulas for the
  leading log, next-to-leading log and next-to-next-to-leading log
  orders in terms of at most three-loop Feynman integrals. The reader
  may apply our method to any (set of) Dyson-Schwinger equation(s)
  appearing in renormalizable quantum field theories.
\end{abstract}
\maketitle
%
\section*{Acknowledgments} DK is supported by an Alexander von Humboldt Professorship of the Alexander von Humboldt Foundation and the BMBF. OK thanks the IMPRS for Geometric Analysis, Gravitation \& String Theory (Potsdam Golm) for support.
\section{Introduction and Results}
\label{sec:introduction}
In this section, we give a short introduction as well as a
presentation and discussion on the type of results we obtain in this
paper.
\subsection{Introduction}
\label{subsec:introduction}
The usual way to compute a physical probability amplitude is to
replace each term of the perturbative series in the coupling $\alpha$
by a sum of Feynman graphs. Applying renormalized Feynman rules
$\Phi_R$ to all such graphs translates this sum of graphs to the
physically observable renormalized Feynman amplitude, say a Green
function $G_R(\alpha,L,\theta)$, at least as a formal series.

$\Phi_R$ evaluated on a graph is a polynomial in a suitably chosen
external scale parameter $L=\log S/S_0$, $\Phi_R=\Phi_R(L)$. $L$
includes for example the center of mass energy $S$ given by the
underlying process, with $S_0$ fixing a reference scale for
renormalization.

Further dependences are scattering angles, collected in
$\Phi_R(L,\theta)$ by a set of variables $\theta$. These are
dimensionless parameters incorporating dependences on scalar products
$p_i\cdot p_j/S$ or masses $m_i^2/S$. Throughout, we assume we leave
those scattering angles unchanged for the renormalization point. A
discussion of this point can be found in
\cite{MadridProcKreimer,BrownKreimer}.

Therefore, any renormalized Green function $G_R$ can be written as a
triangular expansion
\begin{equation}
  \label{eq:log-expansion}
  G_R (\alpha, L, \theta) = 1 + \sum_{j=0}^\infty \sum_{i=1}^\infty
  \gamma_{i+j,i}(\theta) \alpha^{i+j} L^i = 1 + \sum_{j=0}^\infty
  \sum_{\mathcal{M}} p_j^{\mathcal{M}} (\theta) \alpha^j
  \mathcal{M}(u), 
\end{equation}
called `log-expansion'. Here, $\alpha$ denotes the perturbative
parameter, a coupling constant say, and $\gamma_{i+j,i}$ are some
functions in $\theta$. We have $p_0^{\begin{bsmallmatrix}
    \bullet \end{bsmallmatrix}} = 1$. For $j\geq1$, the
$p_j^\mathcal{M}(\theta)$ may depend on $\theta$ and are obtained from
the first $(j+1)$ terms of a perturbative expansion in the coupling.
$\mathcal{M}(u) = \sum_{i=1}^\infty q^\mathcal{M}_{i;j} u^i$ is a
series in $u = \alpha L/2$~\footnotemark[101]\footnotetext[101]{Note
  that in general, $u\propto \alpha L$. Since we mainly consider the
  Green function for the Yukawa fermion propagator, the
  proportionality factor is 1/2. For the QED photon self-energy Green
  function that is also considered in this paper, one finds $u = 4/3\,
  \alpha L$.} with $q^\mathcal{M}_{i;j}\in\mathbb{Q}$. These series
are determined from their counterparts in the universal enveloping
algebra of Feynman graphs that we detail below, see also
\cite{MadridProcKreimer}.

Such a form of the Green functions in a renormalizable field theory is
a consequence of the Hopf algebra structure of Feynman graphs
\cite{Connes:1999yr}. One starts from the fact that all superficially
divergent one-particle irreducible (1PI) Feynman graphs of any
physical quantum field theory generate a Hopf algebra $\mathcal{H}$.

This allows to introduce Dyson-Schwinger equations (DSEs) as fix-point
equations for Feynman graphs upon using Hochschild cohomology
\cite{BergbKr}. See \cite{BorCPC} for an effective application of
these mathematical structures to the automatization of perturbative
renormalization, graph generation and graph counting.

Given a quantum field theory, one always finds DSEs whose solution is
simply related to the log-expansion
\textbf{(}Eq.~(\ref{eq:log-expansion})\textbf{)} by applying
renormalized Feynman rules $\Phi_R$. Lo\"{\i}c Foissy classified
exhaustively the structure of possible DSEs \cite{Foissy2011}.

We apply our approach to two exemplary cases: first, to the fermion
propagator
\begin{equation}
  S(q) = \frac{1}{q\!\!\!/\textbf{(}1-\Sigma(\alpha,L) \textbf{)}} =
  \frac{1}{q\!\!\!/ \Phi_R(X_\mathrm{Yuk})}.
\end{equation}
occurring in Yukawa theory. $X_\mathrm{Yuk}$ represents an infinite
sum of graphs that satisfies the DSE
\begin{equation}
  \label{eq:DSE-Yuk}
  X_\mathrm{Yuk} = \mathbb{I} - \sum_{j\geq1} \alpha^j B_+^{\Gamma_j}
  \left( X_\mathrm{Yuk}^{(1-2j)} \right).
\end{equation}
Secondly, to the photon self-energy
\begin{equation}
  \Pi_{\mu\nu}(q) = \frac{g_{\mu\nu} - \hat{q}_\mu \hat{q}_\nu}{q^2
    \textbf{(}1-\Pi(\alpha,L) \textbf{)}} = \frac{g_{\mu\nu}-\hat{q}_\mu
    \hat{q}_\nu}{q^2 \Phi_R(X_\mathrm{QED})} 
\end{equation}
occurring in quantum electrodynamics (QED). The infinite sum of graphs $X_\mathrm{QED}$
satisfies the DSE
\begin{equation}
  \label{eq:DSE-QED}
  X_\mathrm{QED} = \mathbb{I} - \sum_{j\geq1} \alpha^j B_+^{\Gamma_j}
  \left( X_\mathrm{QED}^{(1-j)} \right).
\end{equation}
Note that acting with renormalized Feynman rules $\Phi_R$ on
$X_\mathrm{Yuk}$ and $X_\mathrm{QED}$ yields the log expansions
\textbf{(}Eq.~(\ref{eq:log-expansion})\textbf{)}.

In Yukawa theory, one should consider systems of DSEs because there
are no Ward identities. Here, we restrict to a truncation eliminating
vertex divergences for purposes of presentation.

Our paper consists of two parts. First, in Section
\ref{sec:preliminaries}, we give a brief overview on Hopf algebras and
DSEs. In particular, we relate the Hopf algebra of Feynman graphs to
the Hopf algebra of words by a morphism of Hopf algebras. Once the
solution of a DSE is given in the Hopf algebra of words, we describe
the filtration method in Section \ref{sec:filtration-method}. There,
we rely on  properties of renormalized Feynman rules that we
derive in Section \ref{sec:renormalized-feynman-rules}. Finally, we
present the filtration algorithm in Section
\ref{sec:filtration-algorithm}.

In the second part of our paper, we describe a general method to
derive the next-to$^{\{j\}}$-leading log expansion in Section
\ref{sec:relat-betw-next}. In particular, we exemplify up to $j\leq 2$ for
the Yukawa fermion propagator. In Appendix \ref{sec:qed-results}, we
collect the respective results for the QED photon self-energy. In the
filtrations of the Yukawa fermion propagator $X_\mathrm{Yuk}$ and the
QED photon self-energy $X_\mathrm{QED}$, each term comes with a
multiplicity. We list some resulting series in these multiplicities in
Appendix \ref{sec:summary}.

In the remainder of this section, we summarize and discuss our
results.

\begin{sidewaystable}[!]
  \centering
  \small
  \begin{tabular}{cC{7.5cm}cc}
    \hline\hline\\
    & \textbf{Periods $p_j^\mathcal{M}$} & \textbf{Yukawa generating
      functions $\mathcal{M}(u)$} &
    \textbf{QED generating functions $\mathcal{M}(u)$}\\[13pt] 
    $(i)$ & $ p_0^{\begin{bsmallmatrix} \bullet \end{bsmallmatrix}} =
    1 $&$ \begin{bmatrix}\bullet\end{bmatrix}(u) = 1 - 
    \frac{1}{x} 
    $& $ \begin{bmatrix}\bullet\end{bmatrix}(u) = u $ \\[13pt] 
    $(ii)$ & $p_1^{\begin{bsmallmatrix}
        \bullet & 1
      \end{bsmallmatrix}} = \alpha \Phi_R(\Gamma_2) $&$\begin{bmatrix}
      \bullet & 1
    \end{bmatrix}(u) = x \log x$& $\begin{bmatrix}
      \bullet & 1
    \end{bmatrix}(u) = \log y$ \\[13pt]
    $(iii)$ & $ p_1^{\begin{bsmallmatrix} \bullet \\ 2
      \end{bsmallmatrix}} = \alpha \Phi_R(\Gamma_1)^2 - 2\alpha\Phi_R
    \left( B_+^{\Gamma_1}(\Gamma_1)
    \right)$&$ \begin{bmatrix} \bullet \\ 2
    \end{bmatrix}(u) = -\frac{x}{2} \log x$& $ \begin{bmatrix} \bullet
      \\ 2
    \end{bmatrix}(u) = 0$\\[13pt]
    $(iv)$ & $ p_2^{\begin{bsmallmatrix} \bullet & 0 & 1
      \end{bsmallmatrix}} = \alpha \Phi_R(\Gamma_3)
    $&$ \begin{bmatrix} \bullet & 0 & 1
    \end{bmatrix}(u) = -\frac{x}{2} + \frac{x^3}{2}$&
    $ \begin{bmatrix} \bullet & 0 & 1
    \end{bmatrix}(u) = y - 1$ \\[13pt]
    $(v)$ & $p_2^{\begin{bsmallmatrix} \bullet & 2
      \end{bsmallmatrix}} = \alpha^2
    \Phi_R(\Gamma_2)^2$&$\begin{bmatrix} \bullet & 2
    \end{bmatrix}(u) = \frac{x}{2} - \frac{x^3}{2} + x^3 \log x +
    \frac{x^3}{2} \log^2 x$& $\begin{bmatrix} \bullet & 2
    \end{bmatrix}(u) = 1 - y + y \log y $\\[13pt]
    $(vi)$ & $p_2^{\begin{bsmallmatrix} \bullet \\ 3
      \end{bsmallmatrix}} = \alpha \bigg( 3 \Phi_R\left(
      B_+^{\Gamma_1}\left(B_+^{\Gamma_1}\left(\Gamma_1\right)\right)\right)
    - 3 \Phi_R(\Gamma_1) \Phi_R
    \left(B_+^{\Gamma_1}\left(\Gamma_1\right)\right) +
    \Phi_R(\Gamma_1)^3 \bigg) $&$\begin{bmatrix} \bullet \\ 3
    \end{bmatrix}(u) = \frac{x^3}{2} - \frac{x}{2u} \log x$&
    $\begin{bmatrix} \bullet \\ 3
    \end{bmatrix}(u) = 0$ \\[33pt]
    $(vii)$ & $ p_2^{\begin{bsmallmatrix} \bullet & 0 \\ 1 & 1
      \end{bsmallmatrix}} = \alpha \bigg( -
      \Phi_R\left(B_+^{\Gamma_1}(\Gamma_2)\right) -
      \Phi_R\left(B_+^{\Gamma_2}(\Gamma_1)\right) + \Phi_R(\Gamma_1)
      \Phi_R( \Gamma_2) \bigg)$&$ \begin{bmatrix} \bullet & 0 \\ 1 &
      1
    \end{bmatrix}(u) = \frac{x}{2} - \frac{3x^3}{2} + \frac{x}{u} \log
    x$& $ \begin{bmatrix} \bullet & 0 \\ 1 & 1
    \end{bmatrix}(u) = - y + \frac{1}{u} \log
    y$\\[30pt]
    $(viii)$ & $ p_2^{\begin{bsmallmatrix} \bullet \\ 2 \\ 2
      \end{bsmallmatrix}} = \alpha^2 \left( \Phi_R (\Gamma_1)^2 - 2
      \Phi_R \left(B_+^{\Gamma_1}(\Gamma_1)\right) \right)^2
    $&$ \begin{bmatrix} \bullet \\ 2 \\ 2
    \end{bmatrix}(u) = -\frac{x}{8} - \frac{3x^3}{8} + \frac{x^3}{4} \log
    x + \frac{x}{2u} \log x +
    \frac{x^3}{8} \log^2 x$& $ \begin{bmatrix} \bullet \\ 2 \\ 2
    \end{bmatrix}(u) = 0$\\[20pt]
    $(ix)$ & $ p_2^{\begin{bsmallmatrix} \bullet & 1 \\ 2 & 0
      \end{bsmallmatrix}} = \alpha^2 \Phi_R(\Gamma_2) \left( \Phi_R
      (\Gamma_1)^2 - 2 \Phi_R \left(B_+^{\Gamma_1}(\Gamma_1)\right)
    \right) $&$ \begin{bmatrix} \bullet & 1 \\ 2 & 0
    \end{bmatrix}(u) = x^3 - x^3 \log x - \frac{x}{u} \log x -
    \frac{x^3}{2}\log^2 x$& $ \begin{bmatrix} \bullet & 1 \\ 2 & 0
    \end{bmatrix}(u) = \frac{1}{2} + \frac{y}{2} - \frac{1}{u} \log
    y$\\[13pt]
    $(x)$ & $ p_2^{\begin{bmatrix}\bullet\end{bmatrix}_{[a_1,a_2]}} =
    \alpha \left( \Phi_R\left(B_+^{\Gamma_1}(\Gamma_2)\right) -
      \Phi_R\left(B_+^{\Gamma_2}(\Gamma_1)\right) \right) $&$
    \begin{bmatrix}\bullet\end{bmatrix}_{[a_1,a_2]}(u) = \frac{x^3}{4}
    - \frac{x^5}{4} + \frac{x^3}{4}\log x - \frac{3x^5}{4}\log x$& $
    \begin{bmatrix}\bullet\end{bmatrix}_{[a_1,a_2]}(u) = \frac{y}{2} -
    \frac{y^2}{2}$\\[13pt]
    $(xi)$ &
    $p_2^{\begin{bmatrix}\bullet\end{bmatrix}_{[a_1,\Theta(a_1,a_1)]}}
    = \alpha \bigg( 2\Phi_R 
    \left( B_+^{\Gamma_1} \left( \Gamma_1 \cup \Gamma_1 \right)
    \right) - \Phi_R \left(
      B_+^{\Gamma_1}\left(B_+^{\Gamma_1}\left(\Gamma_1\right)\right)
    \right) -
    \Phi_R(\Gamma_1)\Phi_R\left(B_+^{\Gamma_1}\left(\Gamma_1\right)\right)
    \bigg)
    $&$\begin{bmatrix}\bullet\end{bmatrix}_{[a_1,\Theta(a_1,a_1)]}(u)
    = -\frac{x^3}{8} + 
    \frac{x^5}{8} - \frac{x^3}{8}\log x + \frac{3x^5}{8}\log x $&
    $\begin{bmatrix}\bullet\end{bmatrix}_{[a_1,\Theta(a_1,a_1)]}(u) = 0 $\\[35pt]
    \hline\hline
  \end{tabular}
  \caption{The results obtained in this paper: a renormalized Green
    function $G_R$ is given by the log-expansion in
    Eq.~(\ref{eq:log-expansion}). We calculate $G_R$ up to
    next-to-next-to-leading log order ($j\leq2$). This table gives the
    periods $p_j^\mathcal{M}(\theta)$ in the first column. These are
    general for any Green function in any quantum field theory. The
    second column shows the generating functions $\mathcal{M}(u)$ for
    the Yukawa fermion propagator. Here, $u = \alpha L/2$ and we
    abbreviate $x = 1/\sqrt{1-2u}$. The third column collects the
    generating functions for the QED photon self-energy, where $u =
    4/3 \, \alpha L$ and $y = 1/(1-u)$. The periods
    are calculated implicitly in
    Eqs.~(\ref{eq:GR-ll1},\ref{eq:GR-nll1},\ref{eq:GR-nnll1}). They
    can also be obtained independently in three ways: first, from the
    Feynman rules acting on primitive elements $\Gamma_j\in \mathcal{H}$, which define an accompanying
    period. Secondly, they are obtained from primitives generated by
    the Dynkin operator $S\star Y$ applied to shuffles of primitives.
    Finally, they are obtained from concatenation multi-commutators of
    primitives of either sort. The generating functions in Yukawa
    theory are obtained from
    Eqs.~(\ref{eq:bullet1},\ref{eq:bullet2},\ref{eq:bullet5},\ref{eq:bullet3},\ref{eq:bullet4},\ref{eq:bullet6},\ref{eq:bullet7},\ref{eq:bullet8},\ref{eq:bullet9},\ref{eq:bullet10},\ref{eq:bullet11}),
    The generating functions for the QED photon self-energy Green
    function are given in
    Eqs.~(\ref{eq:bulletQED1},\ref{eq:bulletQED2},\ref{eq:bulletQED3},\ref{eq:bulletQED4},\ref{eq:bulletQED5},\ref{eq:bulletQED6},\ref{eq:bulletQED7},\ref{eq:bulletQED8},\ref{eq:bulletQED9}).}
  \label{tab:results}  
\end{sidewaystable}
\clearpage

\subsection{Results and Discussion}
\label{subsec:resdisc}

Let us first concentrate on Yukawa theory, Eq.~(\ref{eq:DSE-Yuk}). To
find the leading log expansion we can simplify to
\begin{equation}
  X_\mathrm{Yuk} = \mathbb{I} - \alpha B_+^{\Gamma_1}
  \left( X_\mathrm{Yuk}^{-1}
  \right).
\end{equation}
Feynman rules $\Phi_R$ with massless internal propagators and a
momentum scheme for subtraction turn Eq.~(\ref{eq:DSE-Yuk}) into a
differential equation for the corresponding anomalous dimension, which
can be solved implicitly \cite{BroadhurstKreimer}. If we are only
interested in the leading log expansion the situation is even simpler.
We only have to solve
\begin{equation}
  \label{eq:ll-solution}
  \mathcal{M}(u) = 1 - \int^u\frac{\mathrm{d}x}{1 -
    \mathcal{M}(x)},\qquad \mathcal{M}(0) = 1 \qquad \Rightarrow
  \qquad \mathcal{M}(u) = 1 - \sqrt{1 - 2u}. 
\end{equation}
Indeed, only inserting the one-loop correction for the Yukawa fermion
propagator,
\begin{equation}
  \Phi_R(\Gamma_1)=\frac{1}{2}L  
\end{equation}
into itself in all possible ways give graphs that contain leading log
contributions. 

In our paper, we introduce a convenient matrix notation.
$[]$-bracketed matrices with a dot in the upper left entry denote the
functions $\mathcal{M}$ in one variable, say $z\in\mathbb{R}$. These
functions occur in the log-expansions
\textbf{(}Eq.~(\ref{eq:log-expansion})\textbf{)} setting
\begin{equation}
  z \to \alpha \Phi_R(\Gamma_1) = \frac{\alpha L}{2} = u.
\end{equation}
In Section \ref{sec:relat-betw-next}, we explain the notation and
derive ordinary first order differential equations for these
objects. The respective differential equations depend
on the corresponding DSEs \textbf{(}in our case,
Eq.~(\ref{eq:DSE-Yuk})\textbf{)} and are solved for Yukawa theory in
Section \ref{sec:relat-betw-next}. For the leading log order, we
define
\begin{equation}
  \begin{bmatrix}\bullet\end{bmatrix}(u) = 1 - \sqrt{1-2u}
\end{equation}
with corresponding period
$p_0^{\begin{bsmallmatrix}\bullet\end{bsmallmatrix}} = 1$. The leading
log expansion \textbf{(}$j=0$ in
Eq.~(\ref{eq:log-expansion})\textbf{)} finally yields
\begin{equation}
  \label{eq:GR-ll}
  G_R(X_\mathrm{Yuk})\big|_{\mathrm{l.l.}} =
  p_0^{\begin{bsmallmatrix}\bullet\end{bsmallmatrix}} \begin{bmatrix}
    \bullet \end{bmatrix}(u). 
\end{equation}
$p_0^{\begin{bsmallmatrix}\bullet\end{bsmallmatrix}}$ and $
\begin{bsmallmatrix}
  \bullet
\end{bsmallmatrix}
(u)$ are also given in the first line
of Table \ref{tab:results}.

Eq.~(\ref{eq:ll-solution}) is obvious as Feynman rules in a momentum
scheme map the Hochschild one-co-cycle $B_+^{\Gamma_1}$ to the
Hochschild one-co-cycle $\int^u \mathrm{d}x$ on polynomials in the
variable $u = \alpha L/2$ \cite{BlKrlimMHS}.

To get the next-to-leading log expansion, we consider two
contributions: the insertion of the one-loop propagator graph into
itself gives a contribution $ u^2/2 + p_1^{
  \begin{bsmallmatrix}
    \bullet \\ 2
  \end{bsmallmatrix}
} u $. The first term correspond to the leading log and the second
term to the next-to-leading log expansion. There is also a
contribution $p_1^{ \begin{bsmallmatrix} \bullet & 1
  \end{bsmallmatrix}
} u$ from the next Hochschild one-co-cycle provided by $\Gamma_2$. The
latter contribution occurs in Eq.~(\ref{eq:DSE-Yuk}) that have a single
appearance of $\Gamma_2$ (beside several $\Gamma_1$). We can therefore
simplify to
\begin{equation}
  X_\mathrm{Yuk} = \mathbb{I} - \alpha B_+^{\Gamma_1}
  \left( X_\mathrm{Yuk}^{-1} 
  \right)-\alpha^2 B_+^{\Gamma_2}\left( X_\mathrm{Yuk}^{-3} 
  \right).
\end{equation}

This gives the next-to-leading log expansion \textbf{(}$j=1$ in
Eq.~(\ref{eq:log-expansion})\textbf{)},
\begin{equation}
  \label{eq:GR-nll}
  G_R(X_\mathrm{Yuk})\big|_{\mathrm{n.l.l.}} = \alpha p_1^{
    \begin{bsmallmatrix}
      \bullet & 1
    \end{bsmallmatrix}
  }
  \begin{bmatrix}
    \bullet & 1
  \end{bmatrix}(u) + \alpha p_1^{
    \begin{bsmallmatrix}
      \bullet \\ 2
    \end{bsmallmatrix}
  }
  \begin{bmatrix}
    \bullet \\ 2
  \end{bmatrix}(u).
\end{equation}
The functions $
\begin{bsmallmatrix}
  \bullet & 1
\end{bsmallmatrix}(u) $ and $
\begin{bsmallmatrix}
  \bullet \\ 2
\end{bsmallmatrix}(u)$ as well as the corresponding periods are listed
in lines $(ii)$ and $(iii)$ of Table \ref{tab:results}.

For the next-to-next-to-leading log expansion, there are several
contributions, which we do not summarize here. In Section
\ref{sec:next-tonext-leading}, we derive
\begin{align}
  \label{eq:GR-nnll}
  G_R(X_\mathrm{Yuk})\big|_\mathrm{n.n.l.l.} =& \alpha^2 p_2^{
    \begin{bsmallmatrix}
      \bullet & 0 & 1
    \end{bsmallmatrix}
  }
  \begin{bmatrix}
    \bullet & 0 & 1
  \end{bmatrix}(u) + \alpha^2 p_2^{ \begin{bsmallmatrix} \bullet & 2
    \end{bsmallmatrix}}
  \begin{bmatrix}
    \bullet & 2
  \end{bmatrix}(u) + \alpha^2 p_2^{ \begin{bsmallmatrix} \bullet \\ 3
    \end{bsmallmatrix}}
  \begin{bmatrix}
    \bullet \\ 3
  \end{bmatrix}(u) + \alpha^2 p_2^{
    \begin{bsmallmatrix}
      \bullet & 0 \\ 1 & 1 
    \end{bsmallmatrix}}
  \begin{bmatrix}
    \bullet & 0 \\ 1 & 1
  \end{bmatrix}(u) + \alpha^2 p_2^{
    \begin{bsmallmatrix}
      \bullet \\ 2 \\ 2
    \end{bsmallmatrix}}
  \begin{bmatrix}
    \bullet \\ 2 \\ 2
  \end{bmatrix}(u) \nonumber\\
  &+ \alpha^2 p_2^{
    \begin{bsmallmatrix}
      \bullet & 1 \\ 2 & 0
    \end{bsmallmatrix}
  }
  \begin{bmatrix}
    \bullet & 1 \\ 2 & 0
  \end{bmatrix}(u) + \alpha^2
  p_2^{\begin{bsmallmatrix}\bullet\end{bsmallmatrix}_{[a_1,a_2]}} 
  \begin{bmatrix}\bullet\end{bmatrix}_{[a_1,a_2]}(u) + \alpha^2
  p_2^{\begin{bsmallmatrix}\bullet\end{bsmallmatrix}_{[a_1,\Theta(a_1,a_1)]}}
  \begin{bmatrix}\bullet\end{bmatrix}_{[a_1,\Theta(a_1,a_1)]}(u)
\end{align}
with generating functions and periods given in lines $(iv)-(xi)$ of
Table \ref{tab:results}.

{\it Eqs.~(\ref{eq:GR-ll},\ref{eq:GR-nll},\ref{eq:GR-nnll}) and the
  respective periods (see the first column of Table \ref{tab:results})
  are in general, valid for any DSE. However, the generating
  functions (2nd and 3rd column of Table \ref{tab:results}) depend on the DSE.}

For the QED photon self-energy, the corresponding DSE is given in
Eq.~(\ref{eq:DSE-QED}). The respective generating functions are listed
in the third column of Table \ref{tab:results} and are much simpler.
This is because there is no insertion point in the Hochschild
one-co-cycle $B_+^{\Gamma_1}$ in QED. It follows that the mentioned
differential equations for generating functions turn out to be
ordinary equations. Note that in QED, we have $u = 4/3\,\alpha L$
instead of $\alpha L/2$.

The usual perturbative approach to quantum field theory does not
suffice in high-energy regimes, where $\alpha L \sim 1$. There, it
becomes significant to consider the log-expansion instead of
perturbation theory.

Our results simplify the calculation of the log-expansion in
Eq.~(\ref{eq:log-expansion}) drastically. One usually needs to compute
an infinite number of Feynman integrals of any loop-number. Using our
results, the complete next-to$^{\{j\}}$-leading log expansion only
depends on at most $(j+1)$-loop graphs. For example, the leading log
order \textbf{(}Eq.~(\ref{eq:GR-ll})\textbf{)} only depends on the
Feynman graph $\Gamma_1$. The next-to-leading log Green function in
Eq.~(\ref{eq:GR-nll}) only depends on the graphs $\Gamma_1$,
$\Gamma_2$ and $B_+^{\Gamma_1}(\Gamma_1)$. The next-to-next-to-leading
log expansion only depends on the Feynman graphs
\begin{equation}
  \label{eq:feynman-graphs}
  \Gamma_1, \qquad \Gamma_2, \qquad
  B_+^{\Gamma_1}\left(\Gamma_1\right), \qquad
  \Gamma_3,\qquad B_+^{\Gamma_1}\left(\Gamma_2\right),\qquad
  B_+^{\Gamma_2}\left(\Gamma_1\right),\qquad B_+^{\Gamma_1} \left(
    B_+^{\Gamma_1}\left(\Gamma_1\right) \right), \qquad B_+^{\Gamma_1}
  \left( \Gamma_1 \cup \Gamma_1 \right),
\end{equation}
see Eq.~(\ref{eq:GR-nnll}) and the periods in the first column of
Table \ref{tab:results}.

In summary, we filter the images of Feynman graphs \textbf{(}as in
Eq.~(\ref{eq:feynman-graphs})\textbf{)} in a suitable universal
enveloping algebra that we construct below. This decomposition then
yields ordinary first order differential equations for generating
functions $\mathcal{M}$. We solve for the functions $\mathcal{M}$,
which are the coefficients for the periods $p_j^\mathcal{M}$ obtained
in these filtrations.

The methods presented here are valid in general. One can apply the
described techniques to any DSE in any quantum field theory. One could
also compute the log-expansion for systems of DSEs. We expect that the
generating functions then become solutions in systems of ordinary
first order differential equations. Here, we are content in exhibiting
our approach. A structural analysis of its mathematical underpinnings
is left to future work.

We start with the introduction of necessary preliminaries.

\section{Preliminaries}
\label{sec:preliminaries}

\subsection{Hopf algebra of Feynman graphs}
\label{sec:hopf-algebra-feynman}

Let $\mathcal{H}$ be the vector space of 1PI Feynman graphs and their
disjoint unions in a given quantum field theory.
$(\mathcal{H},m,\mathbb{I})$ forms an associative and unital algebra,
where $\mathbb{I}$ denotes the empty graph and $m$ is the disjoint
union of graphs, serving as a product.

This algebra is graded by the loop number (the first Betti number) as
an infinite sum of finite dimensional vector spaces
\begin{equation}
  \mathcal{H}=\oplus_{j=0}^\infty \mathcal{H}^{(j)},  
\end{equation}
with $\mathcal{H}^{(0)}=\mathbb{Q}\One$ and augmentation ideal 
\begin{equation}
  \mathcal{A}_H=\oplus_{j=1}^\infty \mathcal{H}^{(j)}.
\end{equation}
Note that Cartesian products $\mathcal{A}^j_H:=\mathcal{A}_H^{\times
  j}\subset \mathcal{H}$ deliver a decreasing filtration
$\mathcal{A}^{j+1}_H\subsetneq\mathcal{A}^j_H$.

The associated graded spaces $\mathfrak{gr}_j(\mathcal{A}_H
)=\mathcal{A}^j_H/\mathcal{A}^{j+1}_H$ contain
$<\Gamma>:=\mathcal{A}_H/\mathcal{A}_H^{\times 2}\equiv
\mathfrak{gr}_1(\mathcal{A}_H)$ as the linear span of graphs in first
degree. We set $\mathfrak{gr}_\bullet(\mathcal{A}_H) =
\oplus_{j}\mathfrak{gr}_j(\mathcal{A}_H)$.

$\mathcal{H}$ acquires a co-algebraic structure by introducing a
co-product $\Delta:\mathcal{H} \to \mathcal{H} \otimes \mathcal{H}$
that acts on 1PI graphs $\Gamma$ as
\begin{equation}
  \label{eq:Delta-FG}
  \Delta (\Gamma) = \mathbb{I} \otimes \Gamma + \Gamma \otimes
  \mathbb{I} + \sum_{\gamma \in \mathcal{P}(\Gamma)} \gamma \otimes
  \Gamma / \gamma.
\end{equation}
$\mathcal{P}(\Gamma)$ is the set of all proper sub-graphs $\gamma
\subsetneq \Gamma$ such that $\gamma$ is the disjoint union of
divergent 1PI sub-graphs in $\Gamma$. The action of $\Delta$ on
$\mathbb{I}$ and products of graphs is given by
\begin{equation}
  \Delta \mathbb{I} = \mathbb{I} \otimes \mathbb{I},\quad
  \Delta \circ m = (m \otimes m) \circ \tau_{(2,3)} \circ (\Delta
  \otimes \Delta),
\end{equation}
where $\tau_{(2,3)}$ flips the second and third element of the
fourfold tensor product. The co-product is co-associative
\cite{Connes:1999yr}. Together with a co-unit $\hat{\mathbb{I}} :
\mathcal{H} \to \mathbb{K}$ that assigns a non-zero value only for
$\mathbb{I}$ and $\hat{\mathbb{I}} (\mathbb{I}) = 1$, $\left(
  \mathcal{H}, \Delta, \hat{ \mathbb{I}} \right)$ forms a
co-associative and co-unital co-algebra.

The given construction implies that $(\mathcal{H}, m, \Delta,
\mathbb{I}, \hat{\mathbb{I}})$ forms a bi-algebra as well.

Finally, $(\mathcal{H}, m, \Delta, \mathbb{I}, \hat{\mathbb{I}}, S)$
forms a Hopf algebra, abbreviated by $\mathcal{H}$. $S:\mathcal{H} \to
\mathcal{H}$ is the antipode that fulfills
\begin{equation}
  m \circ (S \otimes \mathrm{id}) \circ \Delta = m \circ
  (\mathrm{id} \otimes S) \circ \Delta = \mathbb{I} \circ
  \hat{\mathbb{I}}.
\end{equation}
$\mathcal{H}$ is called the Hopf algebra of Feynman graphs. There have
been various introductions to Hopf algebras and in particular, to the
Hopf algebra of Feynman graphs \cite{BorCPC}.

Graphs without sub-divergences form primitive elements. By
Eq.~(\ref{eq:Delta-FG}), their reduced co-product,
\begin{equation}
  \tilde{\Delta}(\Gamma) := \Delta(\Gamma) - \mathbb{I} \otimes \Gamma
  - \Gamma \otimes \mathbb{I}
\end{equation}
vanishes. 

\begin{figure}[!]
  \centering $ \gamma_1 = \begin{picture}(27.6,20.7)(0,14) 
    \Photon(0,15)(20,15){2.25}{2.5}
    \Vertex(20,15){1.3} 
    \Line(20,15)(40,30) 
    \Vertex(34,25.5){1.3}
    \ArrowLine(40,30)(46,34.5)
    \Line(40,0)(20,15) 
    \ArrowLine(46,-4.5)(40,0)
    \Vertex(34,4.5){1.3}
    \Photon(34,25.5)(34,4.5){2.25}{2.5}
  \end{picture}
  \qquad\quad\, \gamma_2 = \begin{picture}(27.6,20.7)(0,14)
    \Photon(0,15)(20,15){2.25}{2.5} \Vertex(20,15){1.3}
    \Line(20,15)(40,30) \Vertex(34,25.5){1.3}
    \ArrowLine(40,30)(46,34.5) \Line(40,0)(20,15)
    \ArrowLine(46,-4.5)(40,0) \Vertex(34,4.5){1.3}
    \Vertex(26,10.5){1.3} \Vertex(26,19.5){1.3}
    \Photon(34,25.5)(26,10.5){1}{3.5}
    \CCirc(28.5,15){1.5}{White}{White}
    \Photon(26,19.5)(34,4.5){1}{3.5}
  \end{picture} \qquad\quad\, \gamma_{1,a}
  = \begin{picture}(27.6,20.7)(0,14)
    \Photon(0,15)(20,15){2.25}{2.5}
    \Vertex(20,15){1.3} 
    \Line(20,15)(40,30) 
    \Vertex(34,25.5){1.3}
    \ArrowLine(40,30)(46,34.5)
    \Line(40,0)(20,15) 
    \ArrowLine(46,-4.5)(40,0)
    \Vertex(34,4.5){1.3}
    \Photon(34,25.5)(34,4.5){2.25}{2.5}
    \PhotonArc(34,25.5)(6.25,37,217){1.5}{3.5}
  \end{picture} \qquad\quad\, \gamma_{1,b}
  = \begin{picture}(27.6,20.7)(0,14) 
    \Photon(0,15)(20,15){2.25}{2.5}
    \Vertex(20,15){1.3} 
    \Line(20,15)(40,30) 
    \Vertex(34,25.5){1.3}
    \ArrowLine(40,30)(46,34.5)
    \Line(40,0)(20,15) 
    \ArrowLine(46,-4.5)(40,0)
    \Vertex(34,4.5){1.3}
    \Photon(34,25.5)(34,4.5){2.25}{2.5}
    \PhotonArc(34,4.5)(6.25,143,323){1.5}{3.5}
  \end{picture} \qquad\quad\, \gamma_{1,c}
  = \begin{picture}(27.6,20.7)(0,14)
    \Photon(0,15)(20,15){2.25}{2.5}
    \Vertex(20,15){1.3} 
    \Line(20,15)(40,30) 
    \Vertex(34,25.5){1.3}
    \ArrowLine(40,30)(46,34.5)
    \Line(40,0)(20,15) 
    \ArrowLine(46,-4.5)(40,0)
    \Vertex(34,4.5){1.3}
    \Vertex(26,10.5){1.3}
    \Vertex(26,19.5){1.3}
    \Photon(34,25.5)(34,4.5){2.25}{2.5}
    \Photon(26,19.5)(26,10.5){2.25}{1.5}
  \end{picture} \qquad\quad\, \gamma_{1,D}
  = \begin{picture}(27.6,20.7)(0,14) 
    \Photon(0,15)(20,15){2.25}{2.5}
    \Vertex(20,15){1.3} 
    \Line(20,15)(40,30) 
    \Vertex(34,25.5){1.3}
    \ArrowLine(40,30)(46,34.5)
    \Line(40,0)(20,15) 
    \ArrowLine(46,-4.5)(40,0)
    \Vertex(34,4.5){1.3}
    \Vertex(26,10.5){1.3}
    \Vertex(26,19.5){1.3}
    \Photon(34,25.5)(34,4.5){2.25}{2.5}
    \Photon(26,19.5)(26,10.5){2.25}{1.5}
    \PhotonArc(34,4.5)(6.25,143,323){1.5}{3.5}
    \PhotonArc(34,25.5)(6.25,37,217){1.5}{3.5}
  \end{picture}$ \vspace{15pt}
  \caption{Some QED Feynman graphs that are used within this paper}
  \label{fig:graphs}
\end{figure}
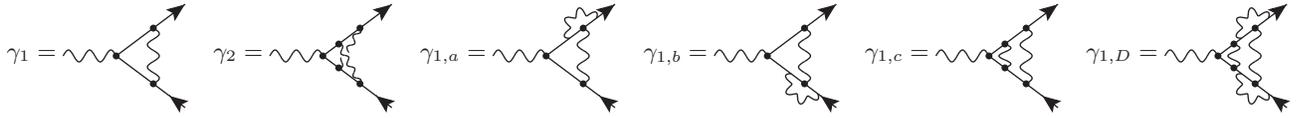

For each such primitive element $\Gamma \in \mathcal{H}$, $\tilde
\Delta (\Gamma) = 0$, we define a grafting operator $B_+^\Gamma :
\mathcal{H} \to \mathcal{H}$ that linearly inserts its argument
graph(s) into $\Gamma$. For example,
\begin{equation}
  B_+^{\gamma_1}(\gamma_1 \cup \gamma_1 \cup \gamma_1) = \gamma_{1,D},
\end{equation}
see Figure \ref{fig:graphs} for the QED Feynman graphs $\gamma_1$ and
$\gamma_{1,D}$. 

Furthermore, we have $B_+^\Gamma(\mathbb{I}) := \Gamma$. If the
insertion is not unique, the result is a sum over all possibilities,
for example
\begin{equation}
  B_+^{\gamma_1}(\gamma_1) = \frac{1}{3}
  \left(
    \gamma_{1,a} + \gamma_{1,b} + \gamma_{1,c}
  \right).
\end{equation}
The graphs $\gamma_1$, $\gamma_{1,a}$, $\gamma_{1,b}$ and
$\gamma_{1,c}$ are defined in Figure \ref{fig:graphs}.

The $B_+^\Gamma$ are Hochschild one-co-cycles by definition
\cite{anatomy,KvS}
\begin{equation}
  \label{eq:Hochschild-1-co-cycle}
  \Delta \circ B_+^{\Gamma} 
  \left(
    \cdot
  \right) =
  B_+^{\Gamma}
  \left(
    \cdot
  \right) \otimes \mathbb{I} +     
  \left(
    \mathrm{id} \otimes B_+^{\Gamma}
  \right) \circ \Delta
  \left(
    \cdot
  \right).
\end{equation}

They generate the co-radical filtration and the associated grading by
sub-divergences. To define the co-radical filtration, set
$G_{-1}^H=\emptyset$ and for $j\geq 0$ set
\begin{equation}
  G_j^H=\Delta^{-1}\left(\mathcal{H}
    \otimes\mathcal{H}^{(0)}+G_{j-1}^H\otimes \mathcal{H}\right).   
\end{equation}
This is an increasing filtration, $G_j^H\subsetneq G_{j+1}^H$, and we set
$\mathfrak{gr}^\bullet(\mathcal{H})=\oplus_{j=0}^\infty G_j^H/G_{j-1}^H$, 
so the first degree elements are given as
\begin{equation}
  \mathfrak{gr}^1(\mathcal{H})=G_1^H/G_0^H=\{\Gamma\in
  \mathcal{A}_H|\tilde{\Delta}(\Gamma)=0\}. 
\end{equation}
In this filtration,
\begin{equation}
  2B_+^\Gamma\circ B_+^\Gamma(\One)\sim m \left( B_+^\Gamma(\One)
    \otimes B_+^\Gamma(\One)\right),
\end{equation}
so they are the same element in
$\mathfrak{gr}^2(\mathcal{H})=G_2^H/G_1^H$, an algebraic fact that by
$\Phi_R$ is the starting point for the existence of the
renormalization group \cite{BlKrlimMHS,BrownKreimer}.

Hochschild closedness
\textbf{(}Eq.(\ref{eq:Hochschild-1-co-cycle})\textbf{)} will also be
essential when we relate a general DSE and its solution to the Hopf
algebra of words in Section \ref{sec:isom-betw-mathc}. This relation
is reminiscent of the flag-decomposition \cite{BlKrlimMHS} which
appears when analyzing renormalized amplitudes as a limiting mixed
Hodge structure.

We now turn to the Hopf algebra of words, for which the decreasing and
increasing filtrations $\mathfrak{gr}_\bullet$ and
$\mathfrak{gr}^\bullet$ above exist analogously.

\subsection{Hopf algebra of words}
\label{sec:hopf-algebra-words}

Let $\mathcal{H}_W$ be the vector space of words and $\mathcal{H}_L
\subset \mathcal{H}_W$ be the sub-space of letters. We need to collect
some properties of $\mathcal{H}_W$. In the remainder of our paper, we
abbreviate letters by $a,b,c,a_1,a_2,\ldots$ and words by
$u,v,w,w_0,w_1,w_2,\ldots$~\footnotemark[102]\footnotetext[102]{Note
  that the symbol $u$ is used twice in this paper: It denotes the
  word $u \in \mathcal{H}_W$ and it abbreviates $u = \alpha L/2 $.
  In the following sections, we only use $u \in \mathcal{H}_W$.}.
Concatenation of letters creates words, e.g., $ab$ and $au$ form new
words.

Let furthermore $\Theta:\mathcal{H}_L \times \mathcal{H}_L \to
\mathcal{H}_L$ be a commutative and associative map that assigns a new
letter to any two given letters. It is always assumed that
$\mathcal{H}_L$ is completed if necessary so that it contains all
images of $\Theta$.

Using this map, we can recursively define the shuffle product
$m_W:\mathcal{H}_W \otimes \mathcal{H}_W \to \mathcal{H}_W$ (also
denoted by $\shuffle_\Theta$) as
\begin{gather}
  \label{eq:shuffle1}
  m_W(au \otimes bv) := au \shuffle_\Theta bv = a(u\shuffle_\Theta bv)
  + b(au \shuffle_\Theta v) + \Theta(a,b)(u\shuffle_\Theta v),
  \\ \label{eq:shuffle2} m_W(u\otimes\mathbb{I}_W) := u =:
  m_W(\mathbb{I}_W \otimes u),
\end{gather}
where on the rhs of Eq.~(\ref{eq:shuffle1}) any word in brackets is
concatenated from right to the respective letter. The so defined
shuffle product is commutative and
associative~\footnotemark[103]\footnotetext[103]{In some references,
  this product is called quasi-shuffle product.} \cite{reutenauer}.
Thus, $(\mathcal{H}_W,m_W = \shuffle_\Theta, \mathbb{I}_W)$ forms an
associative and unital algebra.

$\mathcal{H}_W$ acquires a co-algebraic structure by introducing the
co-product $\Delta_W: \mathcal{H}_W \to \mathcal{H}_W \otimes
\mathcal{H}_W$ and the co-unit $\hat{\mathbb{I}}_W : \mathcal{H}_W \to
\mathbb{K}$ acting on words as
\begin{equation}
  \label{eq:DeltaW}
  \Delta_W (w) = \sum_{v u = w} u \otimes v, \quad
  \hat{\mathbb{I}}_W(w) = 
  \begin{cases}
    1,& w = \mathbb{I}_W\\
    0,& \mathrm{else}
  \end{cases}.
\end{equation}
For example, the co-product of the word $abc$ is
\begin{equation}
  \Delta_W
  \left(
    a b c
  \right) =  \mathbb{I}_W \otimes a b c + c \otimes a b +
  b c \otimes a + a b c \otimes \mathbb{I}_W.
\end{equation}

$\big( \mathcal{H}_W, m_W, \mathbb{I}_W, \Delta_W, \hat{\mathbb{I}}_W
\big)$ forms a bi-algebra as well.

Finally, $(\mathcal{H}_W, m_W, \Delta_W, \mathbb{I}_W,
\hat{\mathbb{I}}_W, S_W)$ forms a Hopf algebra, called the Hopf
algebra of words. The antipode $S_W:\mathcal{H}_W \to \mathcal{H}_W$
fulfills
\begin{equation}
  m_W \circ (S_W \otimes \mathrm{id}) \circ \Delta_W = m_W \circ
  (\mathrm{id} \otimes S_W) \circ \Delta_W = \mathbb{I}_W \circ
  \hat{\mathbb{I}}_W.
\end{equation}
For more details on the Hopf algebra of words, the reader may consult
the textbook of Reutenauer \cite{reutenauer}.

We finally introduce the grafting operators. The primitive elements of
$\mathcal{H}_W$ are all letters, since $\Delta_W(a) = \mathbb{I}_W
\otimes a + a \otimes \mathbb{I}$. We then define
\begin{equation}
  \label{eq:B+}
  B_+^a(u) := au,
\end{equation}
which means that $B_+^a$ concatenates the letter $a$ from left to its
argument word (or sum of words since $B_+^a$ acts linearly). The
reader may check using Eqs.~(\ref{eq:DeltaW},\ref{eq:B+}), that the
$B_+^a$ are indeed, Hochschild one-co-cycles,
\begin{equation}
  \Delta_W \circ B_+^{a} 
  \left(
    u
  \right) =
  B_+^{a}
  \left(
    u
  \right) \otimes \mathbb{I}_W +     
  \left(
    \mathrm{id} \otimes B_+^{a}
  \right) \circ \Delta_W
  \left(
    u
  \right).
\end{equation}

Note that $\mathfrak{gr}_1(\mathcal{A}_{\mathcal{H}_W})$ is the linear
span of words that can not be written as a shuffle product, in analogy
to the definitions for $\mathcal{H}$. Similarly,
$\mathfrak{gr}^1({\mathcal{H}_W})$ can be identified with the
completed set of letters.

\subsection{Isomorphism between $\mathcal{H}$ and $\mathcal{H}_W$}
\label{sec:isom-betw-mathc}

We can now relate the Hopf algebra of Feynman graphs to the Hopf
algebra of words. Indeed, there exists a unique Hopf algebra morphism
$\Upsilon:\mathcal{H} \to \mathcal{H}_W$ that fulfills
\begin{align}
  \label{eq:morphism-begin}
  \Upsilon \left( \mathbb{I}
  \right) =&\, \mathbb{I}_W,\\
  m_W \circ ( \Upsilon \otimes \Upsilon ) =&\, \Upsilon \circ m,\\
  \hat{\mathbb{I}}_W \circ \Upsilon =&\, \Upsilon \circ \hat{\mathbb{I}},\\
  \Delta_W \circ \Upsilon =&\, (\Upsilon \otimes \Upsilon) \circ \Delta,\\
  S_W \circ \Upsilon =&\, \Upsilon \circ S,\\
  \label{eq:morphism-end}
  B_+^{a_n} \circ \Upsilon =&\, \Upsilon \circ B_+^{\Gamma_n}.
\end{align}
It respects the Hopf algebraic structures. The existence of such a
morphism is guaranteed by the fact that the grafting operators are
Hochschild one-co-cycles \cite{Connes:1998qv}. For example,
Eqs.~(\ref{eq:morphism-begin}, \ref{eq:morphism-end}) give
\begin{equation}
  \Upsilon \left( \Gamma_n \right) = \Upsilon \circ B_+^{\Gamma_n}
  (\mathbb{I}) = B_+^{a_n} \circ \Upsilon(\mathbb{I}) = B_+^{a_n}
  \left(
    \mathbb{I}_W
  \right) = a_n
\end{equation}
($\Upsilon$ assigns the letter $a_n$ to the primitive Feynman graph
$\Gamma_n$).

We give another example. Consider the last graph in Figure
\ref{fig:graphs}. With $\Theta \left(a_1, a_1, a_1 \right):= \Theta
\left( a_1, \Theta \left( a_1, a_1 \right) \right)$, we find
\begin{align}
  \Upsilon(\gamma_{1,D}) =&\, \Upsilon \circ B_+^{\Gamma_1} \left(
    \Gamma_1 \cup \Gamma_1 \cup \Gamma_1
  \right) \nonumber\\
  =&\, B_+^{a_1} \circ \Upsilon \circ m \textbf{(} \Gamma_1 \otimes m \left(
      \Gamma_1 \otimes \Gamma_1 \right) \textbf{)}
  \nonumber\\
  =&\, B_+^{a_1} \circ m_W \circ \textbf{(} \Upsilon \left( \Gamma_1 \right)
    \otimes \Upsilon \circ m \left( \Gamma_1 \otimes \Gamma_1 \right)
  \textbf{)} \nonumber\\
  =&\, B_+^{a_1} \circ m_W \circ \textbf{(} a_1 \otimes m_W \left( a_1
      \otimes a_1 \right)
  \textbf{)}\nonumber\\
  =&\, B_+^{a_1} \textbf{(} a_1 \shuffle_\Theta \left( a_1 \shuffle_\Theta
      a_1 \right) \textbf{)} \nonumber\\
  =&\, B_+^{a_1} \textbf{(} a_1 \shuffle_\Theta \left( 2 a_1 a_1 + \Theta
      \left( a_1,a_1 \right) \right) \textbf{)} \nonumber\\
  =&\, B_+^{a_1} \textbf{(} 6a_1 a_1 a_1 + 3 a_1 \Theta \left( a_1, a_1
  \right) + 3 \Theta \left( a_1, a_1 \right) a_1 + \Theta \left( a_1,
    a_1, a_1 \right)
  \textbf{)} \nonumber\\
  =&\, 6 a_1 a_1 a_1 a_1 + 3 a_1 a_1 \Theta \left( a_1, a_1 \right) + 3
  a_1 \Theta \left( a_1, a_1 \right) a_1 + a_1 \Theta \left( a_1, a_1,
    a_1 \right).
\end{align}

The Hopf algebra morphism $\Upsilon$ allows us to translate any DSE to
the Hopf algebra of words. In particular, applying $\Upsilon$ to
Eqs.~(\ref{eq:DSE-Yuk},\ref{eq:DSE-QED}) and using
Eqs.~(\ref{eq:morphism-begin}-\ref{eq:morphism-end}) yields
\begin{align}
  \label{eq:DSE-Yuk-words}
  W_\mathrm{Yuk} :=& \Upsilon
  \left(
    X_\mathrm{Yuk}
  \right) = \mathbb{I}_W - \sum_{j\geq1} \alpha^j B_+^{a_j}
  \left( W_\mathrm{Yuk}^{\shuffle_\Theta (1-2j)}
  \right),\\
  \label{eq:DSE-QED-words}
  W_\mathrm{QED} :=&\Upsilon
  (
  X_\mathrm{QED}
  ) = \mathbb{I}_W - \sum_{j\geq 1} \alpha^j B_+^{a_j}
  \left( 
    W_\mathrm{QED}^{\shuffle_\Theta(1-j)}
  \right).
\end{align}
Here, we drop any further subscript on letters, hence
$a^\mathrm{QED}_j = a_j$ and $a^\mathrm{Yuk}_j = a_j$. 

We finally solve Eqs.~(\ref{eq:DSE-Yuk-words}, \ref{eq:DSE-QED-words})
via the Ans\"atze
\begin{equation}
  \label{eq:DSE-solution}
  W_\mathrm{Yuk} = w_0^\mathrm{Yuk} - \sum_{n\geq 1} \alpha^n w_n^\mathrm{Yuk}, \qquad\qquad
  W_\mathrm{QED} = w_0^\mathrm{QED} - \sum_{n\geq 1} \alpha^n w_n^\mathrm{QED}
\end{equation}
and obtain
\begin{align}
  \label{eq:DSE-solution-Yuk}
  w_0^\mathrm{Yuk} =& \mathbb{I}_W, \qquad w_n^\mathrm{Yuk} = a_n +
  \sum_{j=1}^{n-1}\sum_{k=1}^{n-j}
  \begin{pmatrix}
    2j - 2 +k \\ k
  \end{pmatrix} B_+^{a_j} \left( \sum_{t_1 + \ldots + t_k = n -
    j}^{t_i
    \geq 1} w_{t_1}^\mathrm{Yuk} \shuffle_\Theta \ldots
  \shuffle_\Theta w_{t_k}^\mathrm{Yuk} \right),\\
  \label{eq:DSE-solution-QED}
  w_0^\mathrm{QED} =& \mathbb{I}_W, \qquad w_n^\mathrm{QED} = a_n +
  \sum_{j=2}^{n-1}\sum_{k=1}^{n-j}
  \begin{pmatrix}
    j - 2 +k \\ k
  \end{pmatrix} B_+^{a_j} \left( \sum_{t_1 + \ldots + t_k = n -
    j}^{t_i
    \geq 1} w_{t_1}^\mathrm{QED} \shuffle_\Theta \ldots
  \shuffle_\Theta w_{t_k}^\mathrm{QED} \right).
\end{align}
The first orders are $w_1^\mathrm{QED} = a_1$, $w_2^\mathrm{QED} =
a_2$ and $w_1^\mathrm{Yuk} = a_1$, all others are recursively given.
{\it This  Ansatz can also be used for any other DSE
  \cite{BergbKr}.}

Renormalized Feynman rules linearly act on words as
\begin{equation}
  \Psi_R = \Phi_R \circ \Upsilon^{-1}  
\end{equation}
and $\Psi_R \left( \alpha^n w_n \right) =
\alpha^n \Psi_R(w_n) \propto \alpha^n$ in the log-expansion
\textbf{(}Eq.~(\ref{eq:log-expansion})\textbf{)}. The remaining
question is now: which part of $W$ maps to which power of the external
scale parameter $L$ in the log-expansion
\textbf{(}Eq.~(\ref{eq:log-expansion})\textbf{)}? We will answer this
question in the next section, in full accordance with the blow-ups
needed for Feynman integrands from the viewpoint of algebraic geometry
\cite{BlKrlimMHS,BrownKreimer}.

\section{Filtrations in Dyson-Schwinger equations}

In the following, we filter the coefficients $w_n$ in the solution of
any DSE \textbf{(}as occurring in
Eq.~(\ref{eq:DSE-solution})\textbf{)}. Each filtered term then maps to
a certain power of $L$ in the log-expansion
\textbf{(}Eq.~(\ref{eq:log-expansion})\textbf{)}.

We derive filtration rules for words (in general, $w_n$ is a sum of
words) by considering their dual elements in the universal enveloping
algebra $\mathcal{U}_{\mathcal{L}}$. We introduce
$\mathcal{U}_\mathcal{L}$ as the dual Hopf algebra to $\mathcal{H}_W$
in the next section. Section \ref{sec:renormalized-feynman-rules}
states and proves the two most important properties of renormalized
Feynman rules. This also explains how the filtration of words works.
In Section \ref{sec:filtration-algorithm}, we finally present a
canonical filtration algorithm for arbitrary words and prove that it
is free of redundancies.

\subsection{The dual Hopf algebra to $\mathcal{H}_W$}
\label{sec:filtration-method}

Let $\mathcal{L}$ be a vector space over a field $\mathbb{K}$. Let
furthermore $[\cdot,\cdot]: \mathcal{L} \otimes \mathcal{L} \to
\mathcal{L}$ be a bi-linear map that fulfills $ \left[x,x\right] = 0 $
as well as the Jacobi-identity
\begin{equation}
  [x,[y,z]] + [y,[z,x]] + [z,[x,y]] = 0
\end{equation}
$\forall x,y,z \in \mathcal{L}$. $(\mathcal{L}, [\cdot,\cdot])$ (or
shortly $\mathcal{L}$) is called a Lie algebra and the bi-linear map
$[\cdot,\cdot]$ (called Lie bracket) is antisymmetric.

$\mathcal{L}$ acquires a descending series of sub-algebras $
\mathcal{L} = \mathcal{L}_1 \unrhd \mathcal{L}_2 \unrhd \mathcal{L}_3
\unrhd \ldots $, where $\mathcal{L}_{n+1}$ is generated by all $[x,y]$
with $x \in \mathcal{L}$ and $y \in \mathcal{L}_n$. In the following,
we denote all basis elements of some Lie algebra $\mathcal{L}$ that
are not in $\mathcal{L}_2$ by $x_1,x_2,x_3,\ldots$ and ask for a
complete basis in terms of these $x_i$. For example, is $[x_1,x_2]$ or
$[x_2,x_1] = -[x_1,x_2]$ a basis element of $\mathcal{L}$? Both are
linearly dependent.

We therefore consider the Hall basis \cite{Hall:1950}, which exists
for any Lie algebra. It requires a lexicographical ordering of all
elements in $\mathcal{L}$, for example, let $x_1 < x_2 < \ldots <
[x_1,x_2]< \ldots$ (it does not matter which ordering we take as long
as we choose one). We then define $[x,x']$ to be a (Hall) basis
element of $\mathcal{L}$ iff both,
\begin{enumerate}
\item $x, x' \in \mathcal{L}$ are (Hall) basis elements with $x < x'$,
\item if $x' = [x'',x''']$, then $x\geq x''$
\end{enumerate}
are fulfilled. For example, $[x_1,x_2]$, $[x_2,[x_1,x_3]]$ and
$[x_3,[x_1,x_2]]$ are (Hall) basis elements while $[x_2,x_1]$,
$[x_1,[x_2,x_3]]$ and $[[x_1,x_2],x_3]$ etc. are not.

The bracket in $(\mathcal{L}, [\cdot,\cdot])$ does not comprise an
associative product. However, one can construct enveloping algebras,
i.e. an \textit{algebra} $ \left( \mathcal{A}_\mathcal{L},
  m_\mathcal{A}, \mathbb{I}_\mathcal{A} \right)$ such that there
exists a homomorphism $ \rho_\mathcal{A}: \mathcal{L} \to
\mathcal{A}_\mathcal{L}$ fulfilling
\begin{equation}
  \label{eq:enveloping-algebra}
  m_\mathcal{A}
  \textbf{(}
  \rho_\mathcal{A}(x) \otimes \rho_\mathcal{A}(y) 
  \textbf{)} - m_\mathcal{A}
  \textbf{(}  
  \rho_\mathcal{A}(y) \otimes \rho_\mathcal{A}(x) 
  \textbf{)} =
  \rho_\mathcal{A}\textbf{(}\left[x,y\right]\textbf{)} 
\end{equation}
$\forall x,y \in \mathcal{L}$. 

There may be several enveloping algebras but we can always find a
unique \textit{universal enveloping algebra} $ \textbf{(}
\mathcal{U}(\mathcal{L}), m_\mathcal{U}, \mathbb{I}_\mathcal{U}
\textbf{)}$ up to isomorphism: for each
enveloping algebra $ \left( \mathcal{A}_\mathcal{L}, m_\mathcal{A},
  \mathbb{I}_\mathcal{A} \right) $ there exists a unique algebra
homomorphism $\rho_{\mathcal{U} \to \mathcal{A}}$ such that the
following diagram
\begin{equation}
  \label{eq:commuting-diagram}
  \begin{split}
    \begin{xy} \xymatrix{ \mathcal{L} \ar[r]^{\rho_\mathcal{U}}
        \ar[dr]_{\rho_\mathcal{A}} & \mathcal{U}(\mathcal{L})
        \ar[d]^{\rho_{\mathcal{U} \to
            \mathcal{A}}}  \\
        & \mathcal{A}_\mathcal{L} }
    \end{xy}  
  \end{split}
\end{equation}
commutes. $\mathcal{U}(\mathcal{L})$ is unique \textbf{(}assume that
there are two universal enveloping algebras
$\mathcal{U}_1(\mathcal{L})$ and $\mathcal{U}_2(\mathcal{L})$, then
the homomorphism $\rho_{\mathcal{U}_1 \to \mathcal{U}_2}$ turns out to
be an isomorphism\textbf{)}.

Let us now construct the universal enveloping algebra for any Lie
algebra $\mathcal{L}$, which will prove its existence as well.
Consider the tensor algebra $\textbf{(}T(\mathcal{L}), \otimes ,
1\textbf{)}$, where
\begin{equation}
  T(\mathcal{L}) = \bigoplus_{n\geq0} \mathcal{L}^{\otimes n} =
  \mathbb{K} \oplus \mathcal{L} \oplus (\mathcal{L} \otimes
  \mathcal{L}) \oplus \ldots 
\end{equation}
We define the sub-space $I \subset T(\mathcal{L})$ as
\begin{equation}
  I := \{ s \otimes \left( x \otimes y - y
    \otimes x - \left[ x,y \right] \right) \otimes t| x, y \in
  \mathcal{L} ;\, s,t \in T(\mathcal{L})\}
\end{equation}
($I$ is a 2-sided ideal) and introduce equivalent classes of
$T(\mathcal{L})$,
\begin{equation}
  [t] = \left\{ s \in
    T(\mathcal{L}) | s-t \in I
  \right\}.
\end{equation}
For example, $\textbf{[} [ x_1,x_2 ] \textbf{]} = \{ [ x_1,x_2 ], x_1
\otimes x_2 - x_2 \otimes x_1, \ldots \}$ and $ [ x_1 \otimes x_2
\otimes x_3 ] = \{ x_1 \otimes x_2 \otimes x_3, [ x_1,x_2 ] \otimes
x_3 + x_2 \otimes x_1 \otimes x_3, \ldots \}$ etc.

All such equivalent classes together form a vector space, denoted by
$T(\mathcal{L})/I$. The sum of two elements $[s],[t] \in
T(\mathcal{L})/I$ is well defined as $[s+t]$. We carefully abbreviate
$T(\mathcal{L})/I$ by $\mathcal{U}(\mathcal{L})$ and define an
associative product
\begin{equation}
  m_\mathcal{U}: \mathcal{U}(\mathcal{L}) \otimes
  \mathcal{U}(\mathcal{L}) \to \mathcal{U}(\mathcal{L})  
\end{equation}
acting on equivalent classes as \begin{equation}m_\mathcal{U} \textbf{(}[s] \otimes
[t]\textbf{)} := [s \otimes t].\end{equation}

Thus, $\textbf{(} \mathcal{U}(\mathcal{L}), m_\mathcal{U}, [1]
\textbf{)}$ forms an algebra and together with the homomorphism
$\rho_\mathcal{U}: \mathcal{L} \to \mathcal{U}(\mathcal{L})$ defined
as $\rho_\mathcal{U}(x) = [x]$ $\forall x \in \mathcal{L}$, it is an
enveloping algebra \textbf{(}Eq.~(\ref{eq:enveloping-algebra})
holds\textbf{)}. Finally, $\mathcal{U}(\mathcal{L})$ is even the
universal enveloping algebra of $\mathcal{L}$, since $T(\mathcal{L})$
fulfills the universality property in Eq.~(\ref{eq:commuting-diagram})
as well \cite{Milnor-Moore:1965}.

It turns out that the universal enveloping algebra of a Lie algebra
acquires a Hopf algebra structure. Upon setting
\begin{equation}
  \Delta_{\mathcal{U}}([x])=[x]\otimes\One+\One\otimes [x] \qquad
  \forall x \in \mathcal{L},
\end{equation}
it is determined from compatibility with the product $m_\mathcal{U}$.

For the Hopf algebra of words $\mathcal{H}_W$, there exists a Lie
algebra $\mathcal{L}$ such that $\mathcal{U}(\mathcal{L})$ is dual to
$\mathcal{H}_W$ (Milnor-Moore theorem \cite{Milnor-Moore:1965}).

The indicated Lie algebra is easily constructed. For each letter
$a_1,a_2,\ldots,\Theta(a_1.a_2),\ldots \in \mathcal{H}_L \subset
\mathcal{H}_W$, we name one respective element
$x_1,x_2,\ldots,\Theta(x_1,x_2),\ldots \in \mathcal{L}/\mathcal{L}_2$.
Note that $\mathcal{L}$ contains more elements than $\mathcal{H}_L$
(For example, there is no element $[l_1,l_2]$ in $\mathcal{H}_L$, but
$[x_1,x_2] \in \mathcal{L}$). This will be crucial in the following.

Duality between $\mathcal{H}_W$ and $\mathcal{U}(\mathcal{L})$ allows
us to uniquely define a linear and  invertable map $\eta: \mathcal{H}_W \to
\mathcal{U}(\mathcal{L})$ (see Section \ref{sec:hall-words}), such
that
\begin{equation}
  \label{eq:dual-map}
  \eta(a_i) = [x_i], \qquad \eta\textbf{(}\Theta(a_i, a_j)\textbf{)} =
  [\Theta(x_i,x_j)], \qquad \eta( a_i a_j) = [x_i \otimes x_j], \ldots
\end{equation}
$\forall i,j \in \mathbb{N}$. In general, concatenation of words is
the dual operation of multiplication in $\mathcal{U}(\mathcal{L})$.

\subsection{Renormalized Feynman rules: how the filtration of words
  works}
\label{sec:renormalized-feynman-rules}

We now give two crucial properties of renormalized Feynman rules,
which we need for our filtration method. The proofs are collected
below. \vspace{6 pt}
\begin{enumerate}
\item Let $u\in \mathcal{H}_W$ and $[x]\in \mathcal{U}(\mathcal{L})$
  be its dual element $\textbf{(}\eta(u) = [x]\textbf{)}$. If $x \in
  T(\mathcal{L})$ is also an element of $\mathcal{L} \subset
  T(\mathcal{L})$, then renormalized Feynman rules map $u$ to the
  $L$-linear part of the log-expansion in
  Eq.~(\ref{eq:log-expansion}),
  \begin{equation}
    \label{eq:state1}
    \Psi_R (u) \propto L.
  \end{equation}
\item Renormalized Feynman rules are character-like,
  \begin{equation}
    \label{eq:state2}
    \Psi_R (u \shuffle_\Theta v) = \Psi_R (u) \cdot \Psi_R (v)
  \end{equation}
  $\forall u,v \in \mathcal{H}_W$, where the dot on the rhs of
  Eq.~(\ref{eq:state2}) represents usual multiplication. Furthermore,
  $\Psi_R(a)/L$ is a period $\forall a\in\mathcal{H}_L$.
\end{enumerate}

Let us consider some examples before we turn to the proofs (in the
following, $i,j \in \mathbb{N}$). Each letter $a \in \mathcal{H}_L$
has a dual element $\eta(a) = [x]\in \mathcal{U}(\mathcal{L})$ such
that $x \in \mathcal{L}$ \textbf{(}by construction of the dual
elements, see Eq.~(\ref{eq:dual-map})\textbf{)}. Thus, $\Psi_R (a_i)
\propto L$ and $\Psi_R \textbf{(}\Theta(\cdot,\cdot)\textbf{)} \propto
L$. More interesting is the dual element of $a_i a_j - a_j a_i \in
\mathcal{H}_W$. It is
\begin{equation}
  \eta(a_i a_j - a_j a_i) = \eta(a_i a_j) - \eta(a_j a_i) = [x_i
  \otimes x_j] - [x_j \otimes 
  x_i] = \textbf{[} [x_i, x_j] \textbf{]}
\end{equation}
and $[x_i,x_j] \in \mathcal{L}$. Thus, $\Psi_R (a_i a_j - a_j a_i))
\propto L^1$ although $\Psi_R (a_i a_j)$ also contains terms $\propto
L^2$ \cite{MadridProcKreimer}.

Finally, we \textit{filter} the word $a_i a_j$ to
\begin{equation}
  \label{eq:filtration-example}
  a_i a_j = \frac{1}{2} a_i \shuffle_\Theta a_j + \frac{1}{2} [a_i,
  a_j] - \frac{1}{2} \Theta(a_i,a_j).
\end{equation}
Here, we abbreviated $a_i a_j - a_j a_i$ by the concatenation
commutator $[a_i, a_j]$ discussed in Section (\ref{sec:hall-words}).
We treat concatenation (multi-)commutators of letters as a letter
itself, since the respective dual Hopf algebra element is primitive.

Note the information content of our filtration. The first term of
Eq.~(\ref{eq:filtration-example}) maps to $L^2$ under renormalized
Feynman rules and Eq.~(\ref{eq:state2}) tells us how to determine this
$L^2$-term simply by calculating $\Psi_R (a_i)$ and $\Psi_R (a_j)$.
The other two terms of Eq.~(\ref{eq:filtration-example}) map to $L$.

The possibility to calculate the $L^2$-term in
Eq.~(\ref{eq:filtration-example}) out of the renormalized Feynman
amplitudes for $a_i$ and $a_j$
\textbf{(}Eq.(\ref{eq:state2})\textbf{)} finally leads to the desired
relations between next-to$^{\{j\}}$-leading log orders and terms up to
$\mathcal{O}(\alpha^{j+1})$ in the log-expansion
\textbf{(}Eq.~(\ref{eq:log-expansion})\textbf{)}. We explore this in
great detail in Section \ref{sec:relat-betw-next}.

However, we first give the proofs for the necessary properties of
Feynman rules stated above in light in particular of the duality of
$\mathcal{H}_W$ and $\mathcal{U}(\mathcal{L})$. Terms linear in $L$
can be interpreted in $\rho_{\mathcal{U}}(\mathcal{L})\subset
\mathcal{U}(\mathcal{L})$, and higher powers in $L$ reflect terms in
the quotient algebra $\mathcal{U}(\mathcal{L})$.

\subsubsection{Proof of Claim 1}
\label{sec:proof-state-1}

As we stated before, renormalized Feynman rules map an element $u$
that is dual to a Lie algebra element as above to the $L$-linear part
of the log-expansion in Eq.~(\ref{eq:log-expansion}),
\begin{equation}
  \Psi_R (u) \propto L,
\end{equation}
see Eq.~(\ref{eq:state1}).

This is a direct consequence of the renormalization group action on a
single graph. In fact, let $\Gamma$ be a Hopf algebra element of fixed
co-radical degree $r_\Gamma$, $\Gamma\in \mathfrak{gr}_{r_\Gamma}(G)$.
Then, it allows for an expansion
\begin{equation}
  \Phi_R(\Gamma)=\sum_{j=1}^{r_\Gamma} c_j^\Gamma(\theta)L^j.
\end{equation}
By the renormalization group
\begin{equation}\label{RGE}
  c_j^\Gamma={\bf c}_1^{\otimes j}
  \tilde{\Delta}^{j-1}(\Gamma),
\end{equation}
where ${\bf c}_1$ is the function ${\bf c}_1: \Gamma\to c^\Gamma_1$.
Here, we identified the tensor-product of values with their product
($\mathbb{C}\otimes_{\mathbb{C}}\mathbb{C}\simeq\mathbb{C}$):
\begin{equation}
  {\bf c}_1^{\otimes j}: \underbrace{\mathcal{H} \otimes \cdots
    \otimes\mathcal{H}}_{j\;\mathrm{times}}\to\mathbb{C}.
\end{equation}
Note that ${\bf c}_1^{\otimes j}$ is a symmetric function by
construction.

This leads to a strict inequality on the co-radical degrees
\begin{equation}
  r_{[\Gamma_1,\Gamma_2]} < r_{\Gamma_1}+r_{\Gamma_2},
\end{equation}
which implies the result, Eq.~(\ref{eq:state1}).

\subsubsection{Proof of Claim 2}
\label{sec:proof-state-2}

We also have that renormalized Feynman rules are character-like,
\textbf{(}Eq.~(\ref{eq:state2})\textbf{)},
\begin{equation}
  \Psi_R (u \shuffle_\Theta v) = \Psi_R (u) \cdot \Psi_R (v)
\end{equation}
$\forall u,v \in \mathcal{H}_W$. This is a direct consequence of
Chen's Lemma \citep{BrownKreimer,Chen} in this context.

To show that $\Psi_R(a)/L$ is a period $\forall a\in\mathcal{H}_L$ is
non-trivial only for letters in the image of $\Theta$. Therefore, it
suffices to consider Feynman graphs (with fixed labels on their edges)
that are nested insertions of primitive graphs into each other: a Hopf
algebra element $\Gamma$ is a flag if there exists a sequence of
primitive graphs $\gamma_i$, $1\leq i\leq r_\Gamma$, with
\begin{equation}
  \tilde{\Delta}^{r_\Gamma-1}(\Gamma)=\gamma_1\otimes\cdots
  \otimes\gamma_{r_\Gamma}. 
\end{equation}

Similarly, we say that a sum $G$ of $r_\Gamma$ flags $G_i$,
\begin{equation}
  \label{symmflag}
  G=\sum_i q_i G_i,\,q_i\in\mathbb{Q},
\end{equation} 
is a symmetrized flag if there exists a sequence of primitive graphs
$\gamma_i$, $1\leq i\leq r_\Gamma$, with
\begin{equation}
  \tilde{\Delta}^{r_\Gamma-1}(G) = \sum_\sigma\gamma_{\sigma(1)} \otimes
  \cdots \otimes \gamma_{\sigma(r_\Gamma)}.
\end{equation}
The $\sum_\sigma$-sum is over $r_\Gamma !$ unsigned permutations.
Instead of having the full permutation group acting, one could also
make do with permutations so as to make the rhs co-commutative, if so
desired.

For a given flag $\Gamma$, and hence given sequence of primitive
graphs $\gamma_i$, $1\leq i\leq r_\Gamma$, let $n_\Gamma$ be the
cardinality of the set
\begin{equation}
  X_\Gamma:=\{\Gamma\;\mathrm{a} \,
  \mathrm{flag}|\tilde{\Delta}^{r_\Gamma-1}(\Gamma) = 
  \gamma_1 \otimes \cdots\otimes\gamma_{r_\Gamma}\},  
\end{equation}
so $n_\Gamma=|X_\Gamma|$.

A symmetrized flag is complete if $q_i=1/n_{G_i}$ in
Eq.(\ref{symmflag}).

Finally, renormalized Feynman rules are a forest sum
\cite{BrownKreimer} in graph polynomials $\psi,\phi$ (see
\cite{BrownKreimer} for notation):
\begin{equation}
  \Phi_R(\Gamma) = \int \sum_{f\in\mathcal{F}_\Gamma} (-1)^{|f|}
  \frac{\log \frac{\phi_{\Gamma/f}\psi_f + \phi^0_f \psi_{\Gamma/f}}{
      \phi^0_{\Gamma/f} \psi_f + \phi^0_f
      \psi_{\Gamma/f}}}{\psi^2_{\Gamma/f} \psi^2_f}.
\end{equation}
Here, $\phi_\emptyset=0,\psi_\emptyset=1$.

The coefficient $\Phi_R^1$ of the $L$-linear term  is
\begin{equation}
  \label{LinL}
  \Phi_R^1(\Gamma) = \int \sum_{f\in\mathcal{F}_\Gamma} (-1)^{|f|}
  \frac{1}{\psi^2_{\Gamma/f}\psi^2_f} \frac{\phi_{\Gamma/f}\psi_f}{
    \phi_{\Gamma/f} \psi_f+\phi_f\psi_{\Gamma/f}},
\end{equation}
if the renormalization point preserves scattering angles.

We then have the following result on the angle-independence of
symmetrized flags: for any symmetrized flag $G$,
\begin{equation}
  \label{eq:result-sym-flag}
  \Phi_R^1(G):=\sum_i q_i\Phi_R^1(\Gamma_i)=\sum_i q_i 
  \int
  \sum_{f\in\mathcal{F}_{\Gamma_i}} (-1)^{|f|}
  \frac{1}{\psi^2_{\Gamma_i/f}\psi^2_f}. 
\end{equation}

This justifies that $\Upsilon^{-1}(\Theta(\cdot,\cdot))$ is primitive
in the Hopf algebra of Feynman graphs: to any set $S$ of letters, we
can assign a unique complete symmetrized flag $G_S$ of Feynman graphs
corresponding to the letters in $S$. We set $\Theta(S)$ such that
\begin{equation}
  \Psi_R\textbf{(}\Theta(S)\textbf{)} = -\Phi_R^1(G_S).  
\end{equation}
>From Eq.~(\ref{eq:result-sym-flag}) follows that $\Theta(S)$ can
indeed be regarded as a new letter in $\mathcal{H}_W$ because it is
independent of scattering angles by construction.

We now prove Eq.~(\ref{eq:result-sym-flag}): it follows immediately
from writing Eq.(\ref{LinL}) as elementary symmetric polynomials in
the variables $\phi_x,\psi_x$, with $x$ ranging over all necessary
forests and co-forests, which is possible precisely for symmetrized
flags. Indeed, the denominator in Eq.(\ref{LinL})
\begin{equation}
  \phi_{\Gamma/f}\psi_f+\phi_f\psi_{\Gamma/f}  
\end{equation}
is symmetric under exchange of $\phi\leftrightarrow\psi$, while in
symmetrized flags, we also have co-commutativity which ensures
symmetry under $\Gamma/f\leftrightarrow f$. Hence, in the sum for
symmetrized flags, the factor
\begin{equation}
  \frac{\phi_{\Gamma/f}\psi_f}{\phi_{\Gamma/f}\psi_f +
    \phi_f\psi_{\Gamma/f}}
\end{equation}
in Eq.(\ref{LinL}) turns to unity. This proves Eq.(\ref{eq:result-sym-flag}).

Let us consider an example: Have a look at the first two graphs in
Figure \ref{fig:graphs}. The graph $\gamma_1$ has three vertices, the
graph $\gamma_2$ has five.

Accordingly, there are three graphs $\gamma_{1,i}$, $1\leq i\leq 3$,
obtained by replacing a vertex of $\gamma_1$ by $\gamma_2$, and five
graphs $\gamma_{2,i}$, $1\leq i\leq 5$, obtained by replacing a vertex
of $\gamma_2$ by $\gamma_1$.

We have the reduced co-products
\begin{equation}
  \tilde\Delta \gamma_{1,i}=\gamma_2\otimes\gamma_1,\,\forall 1\leq
  i\leq 3, \qquad \tilde\Delta \gamma_{2,i}=\gamma_1\otimes\gamma_2,
  \, \forall 1\leq i\leq 5.
\end{equation}
Set 
\begin{equation}
  X = \frac{1}{3}\left(\sum_{i=1}^3 \gamma_{1,i}\right)
  + \frac{1}{5}\left(\sum_{i=1}^5 \gamma_{2,i}\right).  
\end{equation}
We have $\tilde
\Delta(X)=\gamma_1\otimes\gamma_2+\gamma_2\otimes\gamma_1$, so $X$ is
a symmetrized flag, and it is complete. Then,
\begin{equation}
  \Phi_R(X) = \frac{1}{3}\left(\sum_{i=1}^3 \Phi_R(\gamma_{1,i})\right)
  +\frac{1}{5}\left(\sum_{i=1}^5 \Phi_R(\gamma_{2,i})\right).
\end{equation}
Using Eq.(\ref{LinL}) and the reduced co-products above, we indeed find
that the second Symanzik polynomial appearing in Eq.(\ref{LinL}) drops
out in this co-commutative sum of symmetrized graph insertions
\begin{equation}
  \Phi_R^1(X)=\int\left( \frac{1}{3}\left(\sum_{i=1}^3
      \frac{1}{\psi^2(\gamma_{1,i})}\right)
    +\frac{1}{5}\left(\sum_{i=1}^5 \frac{1}{\psi^2(\gamma_{2,i})}\right)
    - \frac{1}{\psi^2(\gamma_1)\psi^2(\gamma_2)}\right),
\end{equation}
which is an example of the above result (see also
\cite{MadridProcKreimer}).

\subsection{Filtration algorithm}
\label{sec:filtration-algorithm}

\subsubsection{Presentation of the filtration algorithm}
\label{sec:pres-filtr-algor}

It is not difficult to filter $w_1$ and $w_2$ in the solution of any
DSE \textbf{(}as occurring in Eq.~(\ref{eq:DSE-solution})\textbf{)}.
However, to filter higher order coefficients ($w_n$ for $n>2$)
requires a canonical algorithm that we give here.

Consider for example $w_n^\mathrm{Yuk}$ for $n=1,2,3$ \textbf{(}see
Eq.~(\ref{eq:DSE-solution-Yuk})\textbf{)},
\begin{align}
  w_1^\mathrm{Yuk} =&\, a_1,\\ \label{eq:w2-Yuk} 
  w_2^\mathrm{Yuk} =&\, a_2 + a_1 a_1 = 
  \frac{1}{2} a_1 \shuffle_\Theta a_1 - \frac{1}{2}\Theta(a_1, a_1) +
  a_2,\\  \label{eq:w3-Yuk}
  w_3^\mathrm{Yuk} =&\, a_3 + 3 a_2 a_1 + a_1 a_2 + 3 a_1 a_1 a_1 + a_1
  \Theta(a_1,a_1).
\end{align}
We already filtered $w_1^\mathrm{Yuk}$ and $w_2^\mathrm{Yuk}$ without
much effort but it is not obvious to see the filtration for
$w_3^\mathrm{Yuk}$. The required filtration algorithm is the following
loop over the length $k$ of occurring words: \vspace{6 pt}
\begin{enumerate}
\item Bring all words with length $k$ into lexicographical order using
  the concatenation commutator (respect the Hall basis). This
  introduces words with length $(k-1)$, as we treat concatenation
  (multi-)commutators as own letters.
\item Repeat step 1. for the full shuffle products of the $k$
  corresponding letters and insert them into the expression. All words
  with length $k$ drop out, the introduced full shuffle products
  remain untouched in the remainder.
\end{enumerate}
We start with the maximal length of occurring words down to $k=2$.
Hence, in the case of $w_n$, we perform the above loop for $k =
n,\ldots,2$.

Let us illustrate this for $w_3^\mathrm{Yuk}$. The only word with
length $k = 3$ in Eq.~(\ref{eq:w3-Yuk}) is $a_1a_1a_1$ and it is already
given in lexicographical order. We calculate the corresponding full
shuffle product
\begin{equation}
  a_1 \shuffle_\Theta a_1 \shuffle_\Theta a_1 = 6 a_1 a_1 a_1 + 3 a_1
  \Theta(a_1, a_1) + 3 \Theta(a_1, a_1) a_1 + \Theta(a_1, a_1, a_1)
\end{equation}
and insert it into $w_3^\mathrm{Yuk}$ such that the word $a_1 a_1 a_1$
drops out. Hence,
\begin{equation}
  w_3^\mathrm{Yuk} = a_3 - \frac{3}{2} \Theta(a_1, a_1) a_1 -
  \frac{1}{2} a_1 \Theta(a_1,a_1) + 3 a_2 a_1 + a_1 a_2 -
  \frac{1}{2} \Theta(a_1, a_1, a_1) + \frac{1}{2} a_1 \shuffle_\Theta
  a_1 \shuffle_\Theta a_1\,\, 
\end{equation}
and we proceed with $k=2$. The term $a_1\shuffle_\Theta a_1
\shuffle_\Theta a_1$ remains untouched during the rest of the
filtration.

The words $a_2 a_1$ and $\Theta(a_1, a_1) a_1$ are not in
lexicographical order, we write
\begin{equation}
  \label{eq:w3-Yuk2}
  w_3^\mathrm{Yuk} = a_3 + \frac{3}{2} [ a_1,\Theta(a_1, a_1)] - 2 a_1
  \Theta(a_1,a_1) - 3 [a_1, a_2] + 4 a_1a_2 - \frac{1}{2}
  \Theta(a_1, a_1, a_1) + \frac{1}{2} a_1 \shuffle_\Theta a_1
  \shuffle_\Theta a_1,
\end{equation}
where we only introduced Hall basis elements ($[a_1, a_2]$ instead of
$[a_2, a_1]$ etc.). The respective shuffle products are
\begin{align}
  \label{eq:sh-example1}
  a_1 \shuffle_\Theta a_2 =& 2a_1 a_2 - [a_1, a_2] + \Theta(a_1,a_2),
  \\ \label{eq:sh-example2}
  a_1 \shuffle_\Theta \Theta(a_1,a_1) =& 2 a_1 \Theta(a_1,a_1) -
  [a_1, \Theta(a_1,a_1)] + \Theta(a_1,a_1,a_1)
\end{align}
(they are already brought into lexicographical order using the
concatenation commutator). Inserting Eqs.~(\ref{eq:sh-example1},
\ref{eq:sh-example2}) into Eq.~(\ref{eq:w3-Yuk2}) finally results in
\begin{align}
  w_3^\mathrm{Yuk} =& \frac{1}{2} a_1 \shuffle_\Theta a_1
  \shuffle_\Theta a_1 + 2 a_1 \shuffle_\Theta a_2 - a_1
  \shuffle_\Theta \Theta(a_1,a_1) + a_3 + \frac{1}{2} [
  a_1,\Theta(a_1, a_1)] - [a_1, a_2] + \frac{1}{2} \Theta(a_1, a_1,
  a_1) \nonumber\\
  &- 2 \Theta(a_1,a_2).
\end{align}

The explicit filtration algorithm is the basis for Section
\ref{sec:relat-betw-next}. There, we derive the relations for
next-to$^{\{j\}}$-leading log terms in the log expansion
\textbf{(}Eq.~(\ref{eq:log-expansion})\textbf{)}. However, we first
give some basics to Hall words and concatenation (multi-) commutator
letters. This explains, why the filtration algorithm described above
is redundancy-free.

\subsubsection{Hall words and concatenation (multi-)commutators}
\label{sec:hall-words}
In the Hopf algebra of Feynman graphs $\mathcal{H}$, any $\Gamma\in
\mathcal{H}$ evaluates to $\Phi_R(\Gamma) = \sum_{j=1}^{r_\Gamma}
c_j^\Gamma L^j$. Here, $\Gamma \in \mathfrak{gr}^{r_\Gamma}(
\mathcal{H} )$ and the $c_j^\Gamma$ are given through Eq.(\ref{RGE}).
In particular, this amounts for $j=r_\Gamma$ to an evaluation of
products of primitive elements.

Through $\Upsilon$, we inherit the same properties for words
\cite{BlKrlimMHS}: any $u\in \mathcal{H}_W$ evaluates to $\Psi_R(u) =
\sum_{j=1}^{r_u} d_j^u L^j$, where $u\in\mathfrak{gr}^{r_u}(
\mathcal{H}_W )$ and
\begin{equation}
  d_j^u =
    c_j^{\Upsilon^{-1}(u)}.
\end{equation}
In particular, this amounts for $j=r_u$ to an evaluation of products
of letters. Note that $\Upsilon$ preserves the co-radical degree.

The above filtration algorithm answers the question how to obtain
non-leading logs, $j < r_u$, from the letters that constitute $u$. For
this, we first have to consider the lower central series filtration
$\mathfrak{gr}_\bullet(\mathcal{L})$ of the Lie algebra $\mathcal{L}$,
$\mathfrak{gr}_k(\mathcal{L})=\mathcal{L}_k/\mathcal{L}_{k+1}$ and its
associated grading. Secondly, we have to consider the filtrations and
gradings of the universal enveloping algebra: $\mathfrak{gr}_\bullet(
\mathcal{U}_\mathcal{L})$ by its augmentation and
$\mathfrak{gr}^\bullet(\mathcal{U}_\mathcal{L})$ by its co-radical.

We will use that $\mathfrak{gr}_k( \mathcal{U}_\mathcal{L})$ is
isomorphic to the $k$-fold symmetric tensor-power of $\mathcal{L}$ by
the Poincar\'e--Birkoff--Witt theorem:
\begin{equation}
  \mathfrak{gr}_k( \mathcal{U}_\mathcal{L})\sim
  \mathfrak{Sym}\left(\rho_{\mathcal{U}}(\mathcal{L})^{\otimes k}\right).
\end{equation}

Let $\eta:\mathcal{H}_W\to \mathcal{U}_\mathcal{L}$ as before. We have
the commutative diagrams
\begin{equation}
  \label{eq:eta-commuting-diagram1}
  \begin{split}
    \begin{xy} \xymatrix{ \mathcal{H}_W\otimes
        \mathcal{H}_W\ar[r]^{\shuffle_\Theta} \ar[d]_{\eta\otimes\eta}
        & \mathcal{H}_W
        \ar[d]^{\eta}  \\
        \mathcal{U}_{\mathcal{L}}\otimes \mathcal{U}_{\mathcal{L}}
        \ar[r]^{ m_{\mathcal{U}}} & \mathcal{U}_{\mathcal{L}} }
    \end{xy}
    \qquad\qquad\qquad
    \begin{xy} \xymatrix{ \mathcal{H}_W\otimes \mathcal{H}_W
        \ar[d]_{\eta\otimes\eta} & \mathcal{H}_W\ar[l]^{\Delta}
        \ar[d]^{\eta}  \\
        \mathcal{U}_{\mathcal{L}}\otimes \mathcal{U}_{\mathcal{L}} &
        \mathcal{U}_{\mathcal{L}}\ar[l]^{ \Delta_{\mathcal{U}}} }
    \end{xy}  
  \end{split}
\end{equation}
and two more, by replacing $\eta\to\eta^{-1}$ and downward pointing
arrows by upward pointing ones. These determine the action of $\eta$
once we have defined it on letters $a\in \mathcal{H}_L$
\textbf{(}$\eta(a_i) = [x_i]$, $\eta(\Theta(a_i,a_j)) =
[\Theta(x_i,x_j)]\ldots$\textbf{)}. For example, the degree two image
$\eta(aa)$ of the word $aa$ with respect to $\mathfrak{gr}_\bullet(
\mathcal{U}_\mathcal{L})$ is $\frac{1}{2} [x\otimes x]$, with
$\eta(a)=[x]$.

We define Feynman rules for elements $[s] \in
\mathcal{U}_{\mathcal{L}}$ by
\begin{equation}
  \Psi_R^\eta: [s] \to \Psi_R\textbf{(}\eta^{-1}([s])\textbf{)}.
\end{equation}
In particular for homogeneous elements $[s] \in \mathfrak{gr}_j(
\mathcal{U}_\mathcal{L})$ we have $\Psi_R^\eta([s])\sim L^j$, by
construction.

{\it The structure of renormalized Feynman rules then allows us to
  regain the above filtration algorithm on words $w$ as}
\begin{equation}\label{strucfil}
  w\to \Theta_{\mathcal{U}}\left(\sum_{j=1}^{|w|} P_j\left(\eta^{\otimes
        |w|}\tilde{\Delta}^{|w|-1}(w)\right)\right).
\end{equation}
Here, $P_j$ is the projection into the grade $j$ piece,
$P_j\textbf{(}\eta(w)\textbf{)}\in \mathfrak{gr}_j( \mathcal{U}_\mathcal{L}),\forall
\eta(w)\in \mathcal{U}_{\mathcal{L}}$. $\Theta_{\mathcal{U}}$ is the
map $[x_1\otimes\ldots\otimes x_j]\to
\eta\circ\Theta\textbf{(}\eta^{-1}([x_1]),\ldots,\eta^{-1}([x_j])\textbf{)}$,
where the $\eta^{-1}([x_i])\in \mathcal{H}_L$ are letters by
construction ($x_i \in \mathcal{L}$).

The lower central series filtration
$\mathfrak{gr}_k(\mathcal{L})=\mathcal{L}_k/\mathcal{L}_{k+1}$ filters
in particular $\mathfrak{gr}_1(
\mathcal{U}_\mathcal{L})\sim\mathcal{L}$. Thus, using the Hall basis
of $\mathcal{U}_{\mathcal{L}}$ and the invertibility of $\eta$ 
  finally allows us to write the filtration algorithm in the word
  algebra $\mathcal{H}_W$ with concatenation multi-commutators.
  
Indeed, the rhs of Eq.(\ref{strucfil}) is of degree one by construction as it is in the image of $\Theta_{\mathcal{U}}$.
This suffices, as the degree-$j$ piece is a product of the corresponding $j$ degree-one pieces obtained in $\tilde{\Delta}^{j-1}(w)$.

Let us consider an example. Words on three letters $a_1,a_2,a_3$ have
a Hall basis, which {\it for their degree one part} can be written in
\begin{gather}
  \label{eq:Hall-basis-deg1}
  \{y_1:=\Theta(a_1,a_2,a_3),\quad y_2:= \Theta(a_1,[a_2,a_3]),\quad
  y_3:=\Theta(a_2,[a_1,a_3]),\quad y_4:=\Theta(a_3,[a_1,a_2]),
  \nonumber\\
  y_5:=[a_2,[a_1,a_3]],\quad y_6:=[a_3,[a_1,a_2]]\}.
\end{gather}
In degree one this is the inverse image $\eta^{-1}$ of the elements
\begin{equation}
  \{x_1x_2x_3, x_1[x_2,x_3],
  x_2[x_1,x_3], x_3[x_1,x_2],[x_2,[x_1,x_3]], [x_3,[x_1,x_2]]\}
\end{equation}
in $\mathcal{U}_\mathcal{L}$ written in Hall basis notation (ordered and omitting the symmetric tensor product). 
These form a standard Hall basis on three `letters'
$x_1,x_2,x_3$ dual to $a_1,a_2,a_3$ in $\mathcal{U}_\mathcal{L}$.

There are six words on three distinct letters $a_1,a_2,a_3$. For their
degree one part, these can be written in the basis above:
\begin{align}
  a_1a_2a_3 = & (y_1 + 3y_2 + 3y_3 + 3y_4 + 2y_5 + 4y_6)/6,\\
  a_1a_3a_2 = & (y_1 - 3y_2 + 3y_3 + 3y_4 + 2y_5 + 4y_6)/6,\\
  a_2a_1a_3 = & (y_1 + 3y_2 + 3y_3 - 3y_4 - 4y_5 - 2y_6)/6,\\
  a_2a_3a_1 = & (y_1 + 3y_2 - 3y_3 - 3y_4 - 4y_5 - 2y_6)/6,\\
  a_3a_1a_2 = & (y_1 - 3y_2 - 3y_3 + 3y_4 + 2y_5 - 2y_6)/6,\\
  a_3a_2a_1 = & (y_1 - 3y_2 - 3y_3 - 3y_4 + 2y_5 - 2y_6)/6.
\end{align}
Inverting these equations expresses the degree one elements $y_i$
through the six words on the left.
These correspond to linear combinations of Feynman
graphs $B_+^{\Gamma_i}\circ B_+^{\Gamma_j}\circ B_+^{\Gamma_k}(\One)$,
$i,j,k\in\{1,2,3\}$ which map under $\Upsilon$ to the corresponding
words. For example,
\begin{equation}
  y_2\equiv\Theta(a_1,[a_2,a_3])=\eta^{-1}P_1\eta(a_1a_2a_3-a_1a_3a_2+a_2a_3a_1-a_3a_2a_1), 
\end{equation}
with
\begin{equation}
  \Psi_R(y_2)=\Psi_R^1(a_1a_2a_3-a_1a_3a_2+a_2a_3a_1-a_3a_2a_1).
\end{equation}
Furthermore $P_3(a_1a_2a_3-a_1a_3a_2+a_2a_3a_1-a_3a_2a_1)=0$ so there is no term
$\sim L^3$, whilst the term in $L^2$ is
\begin{equation}
  \Psi_R^1(a_1)\Psi_R^1([a_2,a_3])-\frac{1}{2}\left(\Psi_R^1(a_2)\Psi_R^1([a_1,a_3])
    +\Psi_R^1(a_3)\Psi_R^1([a_2,a_1])\right).
\end{equation}
This gives us a definition in terms of Feynman diagrams for
\begin{align}
  \Upsilon^{-1}(y_2) = & \left(
    B_+^{\Gamma_1}B_+^{\Gamma_2}B_+^{\Gamma_3}(\One)-
    B_+^{\Gamma_1}B_+^{\Gamma_3}B_+^{\Gamma_2}(\One)
    +B_+^{\Gamma_2}B_+^{\Gamma_3}B_+^{\Gamma_1}(\One)-
    B_+^{\Gamma_3}B_+^{\Gamma_2}B_+^{\Gamma_1}(\One)\right)\nonumber\\ & - 
  B_+^{\Gamma_1}(\One)\left(B_+^{\Gamma_2}B_+^{\Gamma_3}(\One) -
    B_+^{\Gamma_3}B_+^{\Gamma_2}(\One)\right)\nonumber\\
  & +\frac{1}{2}\left(
    B_+^{\Gamma_2}(\One)\left(B_+^{\Gamma_1}B_+^{\Gamma_3}(\One) -
      B_+^{\Gamma_3}B_+^{\Gamma_1}(\One)\right) +
    B_+^{\Gamma_3}(\One)\left(B_+^{\Gamma_2}B_+^{\Gamma_1}(\One) -
      B_+^{\Gamma_1}B_+^{\Gamma_2}(\One)\right) \right),
\end{align}
which defines a $L$-linear term.

Let us now describe the standard Hall basis for a set of words on $n$
distinct letters in general. The case of repeated letters follows
easily. There are $n!$ words we can form. First, we count with the
help of the M\"obius function $\mu$ the number of available
concatenation multi-commutators $C_n$.
\begin{equation}
  \label{mcs}
  C_n=\sum_{j=2}^n \left({n\atop j}\right)(-1)^{n-j}C_n^j=(n-1)!,
\end{equation}
with $C_n^j=\frac{1}{n}\sum_{d|n}\mu(d) j^{n/d}$ the well-known number
of multi-commutators of degree $n$ on an alphabet of size $j$.

Let $\mathfrak{P}(n)$ be the set of partitions $p$ of the integer $n$
with the following properties:
\begin{equation}
  n = p_1+\cdots+p_k,\, p_i\leq p_{i+1},\, p_i\geq 2, i\geq 2.
\end{equation}
We allow for at most one such $p_i$ to be marked, which we indicate as
$ \dot{p_i}$. Furthermore, if $p_1=1$, $p_1$ must be marked. If $p$
contains such a distinguished marked $\dot{p_i}$, we say $p$ is marked
at $i$, else it is unmarked.

The marking reflects the fact that $\mathfrak{gr}_1(\mathcal{L})=\mathcal{L}/[\mathcal{L},\mathcal{L}]$
is distinguished amongst all $\mathfrak{gr}_j(\mathcal{L})$.

We say that a partition $q$ of a set $A_n$ of $n$ letters is
compatible with the partition $p$ of the integer $n$, if it is a
disjoint union of sets $A_{p_i}$ of $p_i$ letters according to $p$.
Letters in $A_{\dot{p}_i}$ are completely symmetrized, while all other
sit in multi-commutators of degree $p_j$.

Assume $p\in \mathfrak{P}(n)$ is unmarked. Then we assign a set of
letters
\begin{equation}
  X_q=\{\Theta_{\mathcal{U}}(l_1,\cdots,l_k)\}    
\end{equation}
on $k$ multi-commutators $l_i\in C_{p_i}$ on letters $p_i\in A_{p_i}$
to it.

If $p$ is marked at $i$, we assign a set of letters
\begin{equation}
  X_q = \{\Theta_{\mathcal{U}}(l_1,\cdots,l_k,a_1,\cdots,a_{p_i}),\,
  a_j\in A_{\dot{p}_i}\}.
\end{equation}
Then, summing over all partitions $p$ and all partitions of letters
$q$ compatible with it, we get $n!$ different words which form a base
for the degree one elements of $\mathcal{U}_{\mathcal{L}}$.

If a partition $p$ contains an unmarked integer $p_i$ say $r_i$ times,
the symmetry factor $S(p)$ of $p$ is $S(p):=\prod_i r_i!$. Then, we
indeed count
\begin{equation}
  n! = \sum_p \left(n\atop p_1\cdots p_k
  \right)\frac{1}{S(p)}\prod_{i=1}^k N_i, 
\end{equation}
where $N_i=C_{p_i}$ if $p$ is not marked at $i$, and $N_i=1$ if it is
marked at $i$.

We complete this section by giving a final example.
For four distinct letters, we can have the partition 
$p=4+0$ with $p$ unmarked. So it will provide  six elements in $C_4$. For the partition $p=\dot{1}+3$ we
have $4=\frac{4!}{3!}$ possibilities to choose three letters for $C_3$
which itself has two elements, whilst the fourth letter belongs to
$\dot{1}$. For the partition $p=\dot{2}+2$ we have $6=\frac{4!}{2!2!}$ possibilities
to choose two letters for a one-element $C_2$ while the other two
letters constitute $\dot{2}$. For $p=2+2$ we get a non-trivial
symmetry factor and have $3=\frac{4!}{2!2!}\frac{1}{2!}$ possibilities
to form the product $C_2\times C_2$. Finally, we have the partition
$\dot{4}$. This gives a single element - the symmetric sum over all permutations of four letters.

Counting, we get
\begin{equation}
  24=1\times 6+4\times 2+6\times 1+3\times 1\times 1+1.
\end{equation}
All such words are independent by construction and using
Eq.(\ref{mcs}) repeatedly there are $n!$ of them. They hence form a
base.

\section{Relations for the log expansion}
\label{sec:relat-betw-next}

We now present the main result of our work: how to write the
next-to$^{\{j\}}$-leading log order as a function of terms up to
$\mathcal{O}( \alpha^{j+1})$ in the log-expansion
\textbf{(}Eq.~(\ref{eq:log-expansion})\textbf{)}? We first introduce a
convenient notation for multiplicities of full shuffle products in a
filtered word. Secondly, we derive generating functions for these
multiplicities using the example of the Yukawa propagator. We derive
the generating functions for the QED photon self-energy in Appendix
\ref{sec:qed-results}.

\subsection{Notation}
\label{sec:notation}

We represent the multiplicity of shuffle products in a filtered word
by $[]$-bracketed matrices $m$, for example
\begin{equation}
  w_2^\mathrm{Yuk} = m_1 a_1\shuffle_\Theta a_1 + m_2 \Theta(a_1,a_1)
  + m_3 a_2.
\end{equation}
Each matrix denotes a number, in our case $m_1 = -m_2 = 1/2$, $m_3 =
1$ \textbf{(}see Eq.~(\ref{eq:w2-Yuk})\textbf{)}. In the following we
say that a matrix $m$ {\it belongs to} a shuffle product $S$, when it
gives the multiplicity of $S$ in a filtered word. We also say
that $S$ is the {\it respective shuffle product} to $m$.

Each matrix $m$ with corresponding shuffle product $S$ is built as
follows: the first row contains the numbers of letters $a_1$,
$a_2$,$\ldots$ in $S$. The other rows represent one letter
$\Theta(\ldots)$ in $S$ each, s.t. $\ldots$ contains $m_{ij}$ letters
$a_j$ for the $i$-th row. For example, the filtered word $w_{14}$
contains the term,
\begin{equation}
  w_{14} = 
  \begin{bmatrix}
    4 & 1 & 0 \\ 2 & 0 & 1 \\ 1 & 1 & 0
  \end{bmatrix} a_1^{\shuffle_\Theta 4}\shuffle_\Theta a_2
  \shuffle_\Theta \Theta(a_1,a_1,a_3) \shuffle_\Theta \Theta(a_1,a_2)
  + \ldots
\end{equation}

We index matrices $m$ if the respective shuffle $S$ contains (multi-)commutator letters $[\cdot,\cdot]$ by the
commutator letters themselves. For example, the filtered word $w_6$
contains the term
\begin{equation}
  w_6 = \begin{bmatrix}
    1 \\ 2
  \end{bmatrix}_{[a_1,a_2]} a_1 \shuffle_\Theta
  \Theta(a_1,a_1)\shuffle_\Theta [a_1,a_2] + \ldots
\end{equation}
We treat indexed matrices separately in Section \ref{sec:commutators}.
First though, we only treat index-free matrices and hence, full shuffle
products {\it without (multi-)commutator letters}.

Now, consider an {\it unfiltered word} $w_n$ that is recursively
defined via a DSE. In our case of the Yukawa propagator, this is
Eq.~(\ref{eq:DSE-solution-Yuk}). A matrix with $\{\}$ brackets
represents the number of words in $w_n$ that consists of a given set
of letters. The matrix entries encode the particular set of letters in
full analogy to the case of $[]$ brackets above. For example,
\begin{equation}
  \begin{Bmatrix}
    4 & 1 & 0 \\ 2 & 0 & 1 \\ 1 & 1 & 0
  \end{Bmatrix}
\end{equation}
represents the number of words with four letters $a_1$, one $a_2$, one
$\Theta(a_1,a_1,a_3)$ and one $\Theta(a_1,a_2)$ in the unfiltered word
$w_n$. As another example, $w_3^\mathrm{Yuk}$ contains the terms
$3a_2a_1$ and $a_1a_2$ \textbf{(}see Eq.~(\ref{eq:w3-Yuk})\textbf{)},
hence $
\begin{Bsmallmatrix}
  1 & 1
\end{Bsmallmatrix} = 3 + 1 = 4$.

We always represent a $[]$-bracketed matrix by a lower case letter and
the same matrix with $\{\}$ brackets by the corresponding capital
letter ($m \to M$).

The defined matrices can have any sizes. Filling zeros does not change
the multiplicity of the corresponding shuffle product.

We call a row `$\Theta$-row' when it is not the first row of a matrix.
We consider two matrices with two $\Theta$-rows interchanged to be the
same object, since they represent the same number ($\shuffle_\Theta$
is symmetric). For example
\begin{equation}
  \begin{bmatrix}
    1 & 1 \\ 2 & 0 \\ 1 & 1
  \end{bmatrix} \equiv \begin{bmatrix} 1 & 1 \\ 1 & 1 \\ 2 & 0
  \end{bmatrix}.
\end{equation}

Let two matrices $m_1$, $m_2$ with corresponding shuffle products
$S_1$ and $S_2$ be given. We define the matrix $m_1 \oplus m_2$ such
that it belongs to the shuffle product $S_1\shuffle_\Theta S_2$. This
defines a special summation of matrices, denoted by $\oplus$. This is
realized as follows: $\oplus$ adds up each first row as in ordinary
matrix summation and writes the $\Theta$-rows one below the other. We
can even add matrices with different sizes by filling zeros. For
example,
\begin{equation}
  \begin{bmatrix}
    4 & 1 & 0 \\ 2 & 0 & 1 \\ 1 & 1 & 0
  \end{bmatrix} = 
  \begin{bmatrix}
    2 & 1 \\ 1 & 1
  \end{bmatrix} \oplus
  \begin{bmatrix}
    0 & 0 & 0 \\ 2 & 0 & 1
  \end{bmatrix} \oplus
  \begin{bmatrix}
    2
  \end{bmatrix}.
\end{equation}

We define matrices
\begin{equation}
  p_j =
  \begin{bmatrix}
    \delta_{1j} & \delta_{2j} & \delta_{3j} & \ldots 
  \end{bmatrix},
\end{equation}
as well as the useful vectors
\begin{equation}
  \mathbf{u} =
  \begin{pmatrix}
    1 & 1 & 1 & \ldots
  \end{pmatrix}, \qquad \mathbf{v} =
  \begin{pmatrix}
    1 & 2 & 3 & \ldots
  \end{pmatrix}, \qquad
  \mathbf{w} =
  \begin{pmatrix}
    1 & 0 & 0 & \ldots
  \end{pmatrix}
\end{equation}
with appropriate sizes. We use these vectors to notate some properties
of matrices. We calculate products of matrices and vectors by ordinary
matrix multiplication and not by replacing the matrix by the
corresponding number. For example, a matrix $m$ that occurs in the
filtered word $w_n$ of a DSE fulfills $\mathbf{u} m \mathbf{v}^T = n$.
Let $n_\Theta(m)$ be the number of nonzero $\Theta$-rows. Then, we
define
\begin{equation}
  |m| := \mathbf{w} m \mathbf{u}^T + n_\Theta(m),
\end{equation}
which is the number of letters in the respective shuffle product.

Note that once, a matrix representing a number is contracted with a
vector, the result is regarded as a real vector,
\begin{equation}
  \mathbf{u} m_1 + \mathbf{u} m_2 = \mathbf{u} (m_1 \oplus m_2). 
\end{equation}

We introduce a function $\mathcal{S}$ acting on two matrices, say
$m_1$ and $m_2$ with respective shuffle products $S_1$ and $S_2$.
$\mathcal{S}(m_1,m_2)$ counts the number of combinatorial
possibilities to work out some shuffle products in $S_2$ such that the
resulting expression contains one term $S_1$. For example,
\begin{equation}
  \mathcal{S}
  \left(
    \begin{bmatrix}
      0 & 0 \\ 1 & 1
    \end{bmatrix},
    \begin{bmatrix}
      1 & 1
    \end{bmatrix}
  \right) = 1, \qquad
  \mathcal{S}
  \left(
    \begin{bmatrix}
      3 & 0 \\ 1 & 1
    \end{bmatrix},
    \begin{bmatrix}
      4 & 1
    \end{bmatrix}
  \right) =
  \begin{pmatrix}
    4 \\ 1
  \end{pmatrix} = 4 , \qquad \mathcal{S} \left(
    \begin{bmatrix}
      2 & 0 & 1 \\ 2 & 0 & 0 \\ 1 & 1 & 0 \\ 2 & 0 & 1
    \end{bmatrix},
    \begin{bmatrix}
      5 & 1 & 1 \\ 2 & 0 & 1
    \end{bmatrix}
  \right) =
  \begin{pmatrix}
    5 \\ 1
  \end{pmatrix}
  \cdot
  \begin{pmatrix}
    4 \\ 2
  \end{pmatrix}
  = 30.
\end{equation}
We comment on the second example: the corresponding shuffle product of $
\begin{bsmallmatrix}
  4 & 1
\end{bsmallmatrix}
$ is $a_1^{\shuffle_\Theta\, 4} \shuffle_\Theta a_2$ and there are 4
possibilities to work out one of the shuffle products to derive
\begin{equation}
  a_1^{\shuffle_\Theta\, 4} \shuffle_\Theta a_2 =
  a_1^{\shuffle_\Theta\, 3} \shuffle_\Theta a_1 a_2 +
  a_1^{\shuffle_\Theta\, 3} \shuffle_\Theta a_2 a_1 +
  a_1^{\shuffle_\Theta\, 3} \shuffle_\Theta \Theta(a_1,a_2).
\end{equation}
The last term on the rhs is the respective shuffle product of $
\begin{bsmallmatrix}
  3 & 0 \\ 1 & 1
\end{bsmallmatrix}
$.

We relate $\{\}$-bracketed matrices to $[]$-bracketed matrices.
Consider for example $m =
\begin{bsmallmatrix}
  1 & 1 & 1
\end{bsmallmatrix}
$ and the respective shuffle product $S = a_1\shuffle_\Theta
a_2\shuffle_\Theta a_3$. Computing $S$ gives $3!$ words that consists
of the letters $a_1$, $a_2$ and $a_3$. Ordering these words using the
filtration algorithm,
\begin{equation}
  a_2a_1 \to a_1a_2 - [a_1,a_2] 
\end{equation}
does not change the overall number of words with letters $a_1$, $a_2$
and $a_3$. It only introduces new words with commutator letters. We
have thus,
\begin{equation}
  \label{eq:curly-square-1}
  \begin{Bmatrix}
    1 & 1 & 1
  \end{Bmatrix} = 3! 
  \begin{bmatrix}
    1 & 1 & 1
  \end{bmatrix}.
\end{equation}
In general,
\begin{equation}
  \label{eq:curly-square-2}
  M = \sum_{m'} |m|! \, \mathcal{S}(m,m')\, m',
\end{equation}
where the sum is over all possible matrices $m'$. If $m$ has only one
row, $\mathcal{S}(m,m') = 1$ if $m' = m$ and $\mathcal{S}(m,m') = 0$
otherwise. Eq.~(\ref{eq:curly-square-2}) then reads $M = |m|!\, m$, as
in Eq.~(\ref{eq:curly-square-1}). If $m$ consists of more than one
row, the reader may verify that indeed, Eq.~(\ref{eq:curly-square-2})
is the correct generalization. Note that Eq.~(\ref{eq:curly-square-2})
only holds for index-free matrices. Indeed, the number of words that
consist of a certain (commutator-free) set of letters does not change
when introducing concatenation commutators during the filtration
algorithm.

We introduce generating functions for the multiplicities (matrices)
$m$. Let $\mathcal{M}$ be a matrix with integer entries except for the
upper left entry, which is just a dot. A matrix $m$ is said to be
equivalent to $\mathcal{M}$, $m \sim \mathcal{M}$, if replacing the
upper left entry to a dot yields $\mathcal{M}$. We define
\begin{equation}
  \label{eq:def-generating-function}
  \mathcal{M} = \mathcal{M}(z) = \sum_{m \sim \mathcal{M}}^{m \neq
    \begin{bsmallmatrix}
      0
    \end{bsmallmatrix}
    } m \, z^{|m|}.
\end{equation}
$\mathcal{M}(z)$ is a function in $z$, represented by a matrix. It
generates all $m \sim \mathcal{M}$,
\begin{equation}
  \label{eq:matrices-of-generating-functions}
  m = \frac{1}{|m|!} \left(
    \frac{\mathrm{d}}{\mathrm{d}z}\right)^{|m|}
  \mathcal{M}(z)\Big|_{z=0}.
\end{equation}
Examples of generating functions are
\begin{equation}
  \begin{bmatrix}\bullet\end{bmatrix} = \sum_{N = 1}^\infty
  \begin{bmatrix}
    N
  \end{bmatrix} \, z^N, \qquad
  \begin{bmatrix}
    \bullet & 1 \\ 1 & 1
  \end{bmatrix} = \sum_{N = 0}^\infty
  \begin{bmatrix}
    N & 1 \\ 1 & 1
  \end{bmatrix} z^{N+2}, \qquad   
  \begin{bmatrix}
    \bullet & 2 & 0 \\ 2 & 0 & 0 \\ 1 & 0 & 1
  \end{bmatrix} = \sum_{N = 0}^\infty
  \begin{bmatrix}
    N & 2 & 0 \\ 2 & 0 & 0 \\ 1 & 0 & 1
  \end{bmatrix} z^{N+4}.
\end{equation}

We translate two properties of matrices $m$ to generating functions
$\mathcal{M}$. First, we sum generating functions $\mathcal{M}$ in the
same special way as matrices $m$. The dot remains untouched, for
example
\begin{equation}
  \begin{bmatrix}
    \bullet & 1 \\ 1 & 1
  \end{bmatrix} \oplus
  \begin{bmatrix}
    \bullet & 0 & 0 \\ 2 & 0 & 1
  \end{bmatrix} \oplus \begin{bmatrix}\bullet\end{bmatrix} =
  \begin{bmatrix}
    \bullet & 1 & 0 \\ 2 & 0 & 1 \\ 1 & 1 & 0
  \end{bmatrix}.
\end{equation}
Secondly, let $\tilde m \sim \mathcal{M}$ with $\tilde m_{11} = 0$. We
define $|\mathcal{M}| := |\tilde m|$. Summation and absolute value of
generating functions will be useful in Section
\ref{sec:index-free-n0}.

\subsection{Derivation of the master differential equation}
\label{sec:deriv-mast-equat}

We find another relation between a $\{\}$-bracketed matrix $M$ and
$[]$-bracketed matrices by use of the recursive equation
Eq.~(\ref{eq:DSE-solution-Yuk}). Let the lhs of
Eq.~(\ref{eq:DSE-solution-Yuk}) be an unfiltered word $w_n$ and let
the words $w_{t_i}$ on the rhs be filtered words. Let $M$ be the
number of words consisting of a certain set of letters on the lhs, a
$\{\}$-bracketed matrix hence. On the rhs, only terms that constitute
these words are taken into account. Note that we have $\mathbf{u} M
\mathbf{v}^T = \mathbf{u} m \mathbf{v}^T = n$.

The recursive relation for $M$ is
\begin{equation}
  \label{eq:curly-square-3}
  M = \delta_{|m|1}\delta_{n_\Theta(m)0} + \sum_{j = 1}^{\mathbf{u}
    m \mathbf{v}^T - 1} \left( 1 - \delta_{m_{1j} \, 0} \right)
  \sum_{k = 1}^{\mathbf{u} m \mathbf{v}^T - j}
  \begin{pmatrix}
    2j - 2 + k \\ k
  \end{pmatrix} 
  \sum^{(*)} (|m|-1)! \, \mathcal{S}\left(m \ominus p_j, \bigoplus_i
    m_i \right) m_1 m_2 \ldots m_k,
\end{equation}
where $m$ is still, the same matrix as $M$ but with $[]$ brackets.
$(*)$ sums integers $t_i$ as in Eq.~(\ref{eq:DSE-solution-Yuk}) and
matrices $m_i$ such that
\begin{equation}
  \label{eq:star}
  (*):\quad t_i \geq 1,\quad i= 1 \ldots k, \quad \sum_{i=1}^{k} t_i =
  \mathbf{u} m \mathbf{v}^T - j,\quad \mathbf{u} m_i \mathbf{v}^T =
  t_i,\quad \mathbf{u} p_j + \sum_i \mathbf{u} m_i = \mathbf{u} m.
\end{equation}
We explain the different terms in Eq.~(\ref{eq:curly-square-3})
individually. First, $\delta_{|m|1}\delta_{n_\Theta(m)0}$ corresponds
to the $a_n$ term in Eq.~(\ref{eq:DSE-solution-Yuk}). The integers $j$
and $k$ range over the same numbers as in
Eq.~(\ref{eq:DSE-solution-Yuk}). The term $(1-\delta_{m_{1j}0})$ gives
1(0) if the respective words to $M$ do (not) contain the letter $a_j$.
Only if they do contain $a_j$, they may arise from the term
$B_+^{a_j}(\ldots)$ in Eq.~(\ref{eq:DSE-solution-Yuk}). Since we
introduce filtered words $w_{t_i}$ into the rhs in
Eq.~(\ref{eq:DSE-solution-Yuk}), we obtain expressions $B_+^{a_j}(S)$,
where $S$ is a full shuffle product. $S$ is built out of $k$ shuffles,
namely terms in $w_{t_1}$, $w_{t_2}$, $\ldots$, $w_{t_k}$. We
therefore claim that $S$ is the respective shuffle product to
$\bigoplus_{i=1\ldots k} m_i$. Each matrix $m_i$ belongs to a full
shuffle product in $w_{t_i}$. Condition $(*)$ in Eq.~(\ref{eq:star})
consists of two parts. The first three relations correspond to the
third sum in Eq.~(\ref{eq:DSE-solution-Yuk}). The fourth equation
together with the factor $\mathcal{S}(\cdot,\cdot)$ in
Eq.~(\ref{eq:curly-square-3}) ensure that the letters in $S$ together
with the letter $a_j$ (matrix $p_j$) constitute the set of letters in
$M$. $S$ consists of $(|m| - 1)$ letters, which gives rise to the
factor of $(|m|-1)!$ in Eq.~(\ref{eq:curly-square-3}).

We now derive an inhomogeneous linear differential equation for the
corresponding generating function to $m$, i.e. $\mathcal{M}(z)$
\textbf{(}see Eq.~(\ref{eq:def-generating-function})\textbf{)}.
Therefore, inserting Eq.~(\ref{eq:curly-square-2}) into
Eq.~(\ref{eq:curly-square-3}) yields
\begin{align}
  |m| \, m =& -\sum_{m' \neq m} |m| \mathcal{S}(m,m') m' +
  \delta_{|m|1}\delta_{n_\Theta(m)0} \nonumber\\
  &+ \sum_{j =
    1}^{\mathbf{u} m \mathbf{v}^T - 1} \left( 1 -
    \delta_{m_{1j} \, 0} \right) \sum_{k = 1}^{\mathbf{u}
    m \mathbf{v}^T - j}
  \begin{pmatrix}
    2j - 2 + k \\ k
  \end{pmatrix} 
  \sum^{(*)} \mathcal{S}\left(m \ominus p_j, \bigoplus_i m_i \right)
  m_1 m_2 \ldots m_k.
\end{align}
The final step is to multiply with $z^{|m|-1}$ and to sum over all
matrices that are equivalent to $m$. {\it This gives the master
  differential equation.} Indeed, on the lhs we obtain
$\mathcal{M}(z)'$,
\begin{align}
  \label{eq:master}
  \mathcal{M}(z)' =& \sum_{m \sim \mathcal{M}} \left( -\sum_{m' \neq
      m} \frac{\mathrm{d}}{\mathrm{d}z} z^{|m| - |m'|}
    \mathcal{S}(m,m') m' z^{|m'|} + \delta_{|m|1}\delta_{n_\Theta(m)0}
    + \sum_{j = 1}^{\mathbf{u} m \mathbf{v}^T - 1} \,\, \sum_{k =
      1}^{\mathbf{u} m \mathbf{v}^T - j}
    \begin{pmatrix}
      2j - 2 + k \\ k
    \end{pmatrix} \right. \nonumber\\
  &\left. \times\sum^{(*)} z^{|m| - 1 - \sum_i |m_i|}
    \mathcal{S}\left(m \ominus p_j, \bigoplus_i m_i \right) m_1
    z^{|m_1|} m_2 z^{|m_2|} \ldots m_k z^{|m_k|} \times
    \begin{cases}
      1,&  j = 1 \\ ( 1 - \delta_{m_{1j} \, 0} ),& \mathrm{else}
    \end{cases}
  \right).
\end{align}
See Eq.~(\ref{eq:star}) for the summation $(*)$ of integers $t_i$ and
matrices $m_i$. From Eq.~(\ref{eq:def-generating-function}), we read
off the initial condition
\begin{equation}
  \label{eq:initial-condition}
  \mathcal{M}(0) = 0.
\end{equation}

Let us consider a first example: for $\mathcal{M}(z)
= \begin{bsmallmatrix}\bullet\end{bsmallmatrix}$, the first term in
Eq.~(\ref{eq:master}) vanishes. The last term only gives a non-zero
value for $j=1$. $\forall i \leq k$, $m_i
=\begin{bsmallmatrix}t_i\end{bsmallmatrix}$. The $\mathcal{S}$
function gives 1 for $\bigoplus_i m_i \oplus p_j = m$ and 0 otherwise.
Together with the initial condition in
Eq.~(\ref{eq:initial-condition}), we find
\begin{equation}
  \label{eq:bullet1}
  \begin{bmatrix}\bullet\end{bmatrix}' = 1 +
  \sum_{k\geq1} \begin{bmatrix}\bullet\end{bmatrix}^k = \frac{1}{1
    - \begin{bmatrix}\bullet\end{bmatrix}} 
  \qquad \Rightarrow \qquad \begin{bmatrix}\bullet\end{bmatrix} = 1 -
  \sqrt{1-2z}. 
\end{equation}

We can now derive the homogeneous part of the differential master
equation Eq.~(\ref{eq:master}). On the rhs, the only terms including
the function $\mathcal{M}(z)$ itself occur in the sum for $j=1$, when
$(k-1)$ of the matrices $m_i$ are equivalent to
$\begin{bsmallmatrix}\bullet\end{bsmallmatrix}$ and the $k$-th matrix
is equivalent to $m$. Using Eq.~(\ref{eq:bullet1}), we obtain
\begin{equation}
  \mathcal{M}(z)'|_\mathrm{hom.} = \sum_{k\geq1}
  k \begin{bmatrix}\bullet\end{bmatrix}^{k-1} 
  \mathcal{M}(z) = \frac{1}{1 - 2z} \mathcal{M}(z).
\end{equation}
Hence, the differential equation Eq.~(\ref{eq:master}) reduces to an
integration using the Ansatz
\begin{equation}
  \mathcal{M}(z) = \frac{\mathcal{C}(z)}{\sqrt{1-2z}}, \qquad
  \mathcal{C}(0) = 0.
\end{equation}
We read off the initial condition for $\mathcal{C}$ from
Eq.~(\ref{eq:initial-condition}). In particular, the integration is
\begin{equation}
  \label{eq:generating-function-final}
  \mathcal{M}(z) = \frac{1}{\sqrt{1-2z}} \left( \int
    \mathcal{M}(z)'|_\mathrm{inhom.} \sqrt{1-2z} \, \mathrm{d}z + c
  \right),
\end{equation}
where we obtain $\mathcal{M}(z)'|_\mathrm{inhom.}$ from
Eq.~(\ref{eq:master}). $c$ is an appropriate integration constant such
that $\mathcal{M}(0) = 0$.

Further general simplifications of the differential master equation
are not obvious. The problem is that the functions $\mathcal{S}$ in
Eq.~(\ref{eq:master}) give individual numbers that do not generalize
and so have to be worked out case-by-case. They result in an overall
differential operator acting on whatever follows. We demonstrate this
in several examples, which will give us next-to and
next-to-next-to-leading log generating
functions.\footnotemark[104]\footnotetext[104]{We call a function
  $\mathcal{M}(z)$ next-to$^{\{j\}}$-leading log generating function
  when it occurs in the log-expansion
  \textbf{(}Eq.~(\ref{eq:log-expansion})\textbf{)} for a certain value
  of $j$.}

\subsection{Generating functions for index-free matrices with
  $n_\Theta(m) = 0$}
\label{sec:index-free-n0}

One exception is the case that $\mathcal{M}$ only contains one row,
$n_\Theta(m) = 0$. These generate the matrices $m$ that belong to full
shuffle products $S$ without $\Theta(\cdot,\cdot)$ letters.
`Index-free' means that $S$ also does not contain $[\cdot,\cdot]$
letters.

Here, $\mathcal{S}(m,m')$ reduces to 1 if $m = m'$ and to 0
otherwise. Thus, the first term in Eq.~(\ref{eq:master}) is zero. The
other $\mathcal{S}$ term constrains
\begin{equation}
  m_1 \oplus m_2 \oplus \ldots m_k \oplus p_j = m
\end{equation}
hence, $\sum_i |m_i| + 1 = |m|$. We denote the generating functions of
$m_i$ and $p_j$ by $\mathcal{M}_i$ and $\mathcal{P}_j$ to be
consistent. We obtain
\begin{equation}
  \label{eq:master-special}
  \mathcal{M}(z)' = \delta_{|\mathcal{M}|0} + \delta_{|\mathcal{M}|1} +
  \sum_{j,k = 1}^\infty
  \begin{pmatrix}
    2j - 2 + k \\ k
  \end{pmatrix}
  \sum^{(**)}\mathcal{M}_1(z) \mathcal{M}_2(z) \ldots \mathcal{M}_k(z)
  \times
  \begin{cases}
    1,& j = 1 \\ ( 1 - \delta_{\mathcal{M}_{1j} \, 0} ),&
    \mathrm{else}
  \end{cases},
\end{equation}
where $(**)$ sums the generating functions $\mathcal{M}_1$, $\ldots$,
$\mathcal{M}_k$ such that
\begin{equation}
  \label{eq:starstar}
  (**): \quad \mathcal{M}_1 \oplus \mathcal{M}_2 \oplus \ldots \oplus
  \mathcal{M}_k \oplus \mathcal{P}_j = \mathcal{M}.
\end{equation}
In the following, we give some examples that constitute the
next-to$^{\{j\}}$-leading log expansions.

\subsubsection{The generating function $\protect\begin{bsmallmatrix}
    \bullet & 1 \protect\end{bsmallmatrix}$}
\label{sec:bullet2}

The first example is $\mathcal{M}(z) =
\begin{bsmallmatrix}
  \bullet & 1
\end{bsmallmatrix}
$. The respective shuffle products to the matrices $m \sim
\mathcal{M}$ are $a_1^{\shuffle_\Theta N}\shuffle_\Theta a_2$ for $N
\in \mathbb{N}$. The sum in Eq.~(\ref{eq:master-special}) only gives
non-zero values if $j = 1$ or $j = 2$. For $j = 1$, $\mathcal{P}_j =
\mathcal{P}_1 = \begin{bsmallmatrix}\bullet\end{bsmallmatrix}$ and
Eq.~(\ref{eq:starstar}) is only fulfilled if one $\mathcal{M}_i$
matches $\mathcal{M}$. This part belongs to the homogeneous
differential equation.

For $j = 2$, $\mathcal{P}_j = \mathcal{P}_2 = \mathcal{M}$ and we find
$\mathcal{M}_i(z) = \begin{bsmallmatrix}\bullet\end{bsmallmatrix}$
$\forall i \leq k$ from Eq.~(\ref{eq:starstar}).

Thus, the inhomogeneous part of Eq.~(\ref{eq:master-special}) reads
\begin{equation}
  \begin{bmatrix}
    \bullet & 1
  \end{bmatrix}'\big|_\mathrm{inhom.} = 1 + \sum_{k = 1}^\infty
  \begin{pmatrix}
    2 + k \\ k
  \end{pmatrix} \begin{bmatrix}\bullet\end{bmatrix}^k = \frac{1}{\left(1
    - \begin{bmatrix}\bullet\end{bmatrix}\right)^3} = 
  \frac{1}{\sqrt{1-2z}^3}.
\end{equation}
We used Eq.~(\ref{eq:bullet1}) in the last line. We insert this result
into Eq.~(\ref{eq:generating-function-final}) and obtain
\begin{equation}
  \label{eq:bullet2}
  \begin{bmatrix}
    \bullet & 1
  \end{bmatrix} = \frac{1}{\sqrt{1-2z}} \log \left(
    \frac{1}{\sqrt{1-2z}} \right).
\end{equation}

\subsubsection{The generating function $\protect\begin{bsmallmatrix}
    \bullet & 0 & 1 \protect\end{bsmallmatrix}$}

Matrices $m \sim \mathcal{M}(z) =
\begin{bsmallmatrix}
  \bullet & 0 & 1
\end{bsmallmatrix}
$ belong to the shuffle products $a_1^{\shuffle_\Theta
  N}\shuffle_\Theta a_3$ for $N \in \mathbb{N}$. The sum in
Eq.~(\ref{eq:master-special}) gives non-zero values only if $j = 1$ or
$j = 3$. As in the previous example, the $j = 1$ part belongs to the
homogeneous differential equation.

For $j = 3$, $\mathcal{P}_j = \mathcal{P}_3 = \mathcal{M}$ and we find
$\mathcal{M}_i(z) = \begin{bsmallmatrix}\bullet\end{bsmallmatrix}$
$\forall i \leq k$ from Eq.~(\ref{eq:starstar}).

Thus, the inhomogeneous part of Eq.~(\ref{eq:master-special}) reads
\begin{equation}
  \begin{bmatrix}
    \bullet & 0 & 1
  \end{bmatrix}'\big|_\mathrm{inhom.} = 1 + \sum_{k = 1}^\infty
  \begin{pmatrix}
    4 + k \\ k
  \end{pmatrix} \begin{bmatrix}\bullet\end{bmatrix}^k =
  \frac{1}{\left(1 - \begin{bmatrix}\bullet\end{bmatrix}\right)^5} =
  \frac{1}{\sqrt{1-2z}^5}.
\end{equation}
We insert this result into Eq.~(\ref{eq:generating-function-final})
and obtain
\begin{equation}
  \label{eq:bullet3}
  \begin{bmatrix}
    \bullet & 0 & 1
  \end{bmatrix} = -\frac{1}{2\sqrt{1-2z}} + \frac{1}{2\sqrt{1-2z}^3}.
\end{equation}

\subsubsection{The generating function $\protect\begin{bsmallmatrix}
    \bullet & 2 \protect\end{bsmallmatrix}$}

$ \mathcal{M}(z) =
\begin{bsmallmatrix} 
  \bullet & 2
\end{bsmallmatrix}
$ is the final next-to-next-to-leading log generating function for
matrices $m\sim\mathcal{M}$ with $n_\Theta(m) = 0$. The matrices $m$
belong to the shuffle products $a_1^{\shuffle_\Theta N}\shuffle_\Theta
a_2\shuffle_\Theta a_2$ for $N \in \mathbb{N}$. Thus, the sum in
Eq.~(\ref{eq:master-special}) gives non-zero values only if $j = 1$ or
$j = 2$.

If $j = 1$, $\mathcal{P}_j = \mathcal{P}_1
= \begin{bsmallmatrix}\bullet\end{bsmallmatrix}$. There are two
possibilities to choose the matrices $\mathcal{M}_i$ such that
Eq.~(\ref{eq:starstar}) is fulfilled. However, only one of these
belongs to the inhomogeneous part of the differential equation: for
two integers $ i \leq k$, $\mathcal{M}_i=
\begin{bsmallmatrix}
  \bullet & 1
\end{bsmallmatrix}
$ and for all other $i$, $\mathcal{M}_i
= \begin{bsmallmatrix}\bullet\end{bsmallmatrix}$.

If $j = 2$, Eq.~(\ref{eq:starstar}) is only satisfied if one of the
matrices $\mathcal{M}_i$ is $
\begin{bsmallmatrix}
  \bullet & 1
\end{bsmallmatrix}
$ and the others are equal to
$\begin{bsmallmatrix}\bullet\end{bsmallmatrix}$.

We finally find the inhomogeneous part of
Eq.~(\ref{eq:master-special}),
\begin{align}
  \begin{bmatrix} 
    \bullet & 2
  \end{bmatrix}'\big|_\mathrm{inhom.} =&\, \sum_{k = 1}^\infty
  \begin{pmatrix}
    k \\ 2
  \end{pmatrix} \begin{bmatrix}\bullet\end{bmatrix}^{k-2}
  \begin{bmatrix}
    \bullet & 1
  \end{bmatrix}^2 + \sum_{k = 1}^\infty   
  \begin{pmatrix}
    2 + k \\ k
  \end{pmatrix} k
  \begin{bmatrix}\bullet\end{bmatrix}^{k-1}
  \begin{bmatrix}
    \bullet & 1
  \end{bmatrix} \nonumber\\
  =&\, \frac{1}{\left(1- \begin{bmatrix}\bullet\end{bmatrix}\right)^3}
  \begin{bmatrix}
    \bullet & 1
  \end{bmatrix}^2 +
  \frac{3}{\left(1- \begin{bmatrix}\bullet\end{bmatrix}\right)^4}   
  \begin{bmatrix}
    \bullet & 1
  \end{bmatrix} \nonumber\\
  =&\, \frac{1}{\sqrt{1-2z}^5} \left( \log^2 \left(
      \frac{1}{\sqrt{1-2z}} \right) + 3 \log \left(
      \frac{1}{\sqrt{1-2z}} \right) \right).
\end{align}
We used the previous results in
Eqs.~(\ref{eq:bullet1},\ref{eq:bullet2}) in the last line. Inserting
this into Eq.~(\ref{eq:generating-function-final}) and performing the
integration finally results in
\begin{equation}
  \label{eq:bullet4}
  \begin{bmatrix}
    \bullet & 2
  \end{bmatrix}
  = \frac{1}{2\sqrt{1-2z}} - \frac{1}{2\sqrt{1-2z}^3} +
  \frac{1}{\sqrt{1-2z}^3} \log
  \left(
    \frac{1}{\sqrt{1-2z}}
  \right) +
  \frac{1}{2\sqrt{1-2z}^3} \log^2
  \left(
    \frac{1}{\sqrt{1-2z}}
  \right).
\end{equation}

\subsection{Generating functions for index-free matrices with
  $n_\Theta(m) \neq 0$}
\label{sec:index-free-n1}

The respective full shuffle products of index-free matrices $m$ with
$n_\Theta(m) \neq 0$ contain at least one letter $\Theta(\cdot,\cdot)$
but no $[\cdot,\cdot]$ letters. Here, we have to proceed from the
master differential equation Eq.~(\ref{eq:master}) to obtain
$\mathcal{M}(z)'|_\mathrm{inhom.}$ in
Eq.~(\ref{eq:generating-function-final}). We treat the different
next-to$^{\{j\}}$-leading log generating functions in separate
subsections.

\subsubsection{The generating function $\protect\begin{bsmallmatrix}
    \bullet \protect\\ 2 \protect\end{bsmallmatrix}$}

The first example is the next-to-leading log generating function $
\mathcal{M}(z) =
\begin{bsmallmatrix}
  \bullet \\ 2 
\end{bsmallmatrix}
$. Matrices $m\sim\mathcal{M}$ belong to the shuffle products
$a_1^{\shuffle_\Theta\,N}\shuffle_\Theta \Theta(a_1,a_1)$ for
$N\in\mathbb{N}$.

In Eq.~(\ref{eq:master}), we replace the sum over $m\sim\mathcal{M}$
by a sum over $N\in\mathbb{N}$ such that $m = \begin{bsmallmatrix} N
  \\ 2
\end{bsmallmatrix}
$. In the first term on the rhs, only $m' =
\begin{bsmallmatrix}
  N+2
\end{bsmallmatrix}
$ yields a
non-vanishing function $\mathcal{S}$. In particular,
\begin{equation}
  \mathcal{S}\left(
    \begin{bmatrix}
      N \\ 2
    \end{bmatrix},
    \begin{bmatrix}
      N+2
    \end{bmatrix}
  \right) =
  \begin{pmatrix}
    N+2 \\ 2
  \end{pmatrix} = \frac{(N+2)(N+1)}{2}.
\end{equation}
The other sum only survives if $j=1$ and hence $p_j = p_1 =
\begin{bsmallmatrix}
  1
\end{bsmallmatrix}
$. The
integers $t_i$ in the $(*)$-sum range such that $\sum_i t_i = N + 2 -
j = N + 1$ \textbf{(}see Eq.~(\ref{eq:star})\textbf{)}. Furthermore,
in the first argument of the function $\mathcal{S}$, $m\ominus p_j =
\begin{bsmallmatrix}
  N-1 \\ 2
\end{bsmallmatrix}
$. For the inhomogeneous part of the differential equation, this
implies that $m_i =
\begin{bsmallmatrix}
  t_i
\end{bsmallmatrix}
$ $\forall i \leq k$ such that $\bigoplus_i m_i =
\begin{bsmallmatrix}
  N+1
\end{bsmallmatrix}
$. Then,
\begin{equation}
  \mathcal{S} \left(
    \begin{bmatrix}
      N-1 \\ 2
    \end{bmatrix},
    \begin{bmatrix}
      N+1
    \end{bmatrix}
  \right) =
  \begin{pmatrix}
    N+1 \\ 2
  \end{pmatrix} = \frac{(N+1)N}{2}.
\end{equation}
All together, we obtain
\begin{align}
  \begin{bmatrix}
    \bullet \\ 2 
  \end{bmatrix}'\bigg|_\mathrm{inhom.} =& \sum_{N = 0}^\infty \Bigg(
  -\frac{\mathrm{d}}{\mathrm{d}z} z^{(N+1)-(N+2)} \frac{(N+2)(N+1)}{2}
  \begin{bmatrix}
    N+2
  \end{bmatrix}
  z^{N+2} \nonumber\\
  &+ \sum_{k\geq1} \sum_{\sum_i t_i = N+1} z^{\left((N + 1) - 1 -
      \sum_i t_i\right)} \frac{(N+1)N}{2}
  \begin{bmatrix}
    t_1
  \end{bmatrix}
  z^{t_1} \ldots
  \begin{bmatrix}
    t_k
  \end{bmatrix}
  z^{t_k} \Bigg) \nonumber\\
  =& \sum_{N = 0}^\infty \left( -\frac{\mathrm{d}}{\mathrm{d}z}
    \frac{1}{z} \frac{1}{2} z^2 \frac{\mathrm{d}^2}{\mathrm{d}z^2}
    \begin{bmatrix}
      N+2
    \end{bmatrix} z^{N+2} + \sum_{k\geq1} \sum_{\sum_i t_i = N+1}
    \frac{1}{z} \frac{1}{2} z^2 \frac{\mathrm{d^2}}{\mathrm{d}z^2}
    \begin{bmatrix}
      t_1
    \end{bmatrix}z^{t_1}
    \ldots \begin{bmatrix}
      t_k
    \end{bmatrix} z^{t_k} \right) \nonumber\\
  =& - \frac{\mathrm{d}}{\mathrm{d}z} \frac{z}{2}
  \frac{\mathrm{d}^2}{\mathrm{d}z^2} \begin{bmatrix}\bullet\end{bmatrix}
  + \sum_{k\geq1} \frac{z}{2}
  \frac{\mathrm{d}^2}{\mathrm{d}z^2} \begin{bmatrix}\bullet\end{bmatrix}^k
  \nonumber\\
  =& - \frac{1}{2} \frac{\mathrm{d}^2}{\mathrm{d}z^2} \left( 1 -
    \sqrt{1 - 2z} \right) \nonumber\\
  =& -\frac{1}{2\sqrt{1-2z}^3}.
\end{align}
In the third line, we used the explicit expression for the generating
function $\begin{bsmallmatrix}\bullet\end{bsmallmatrix}$
\textbf{(}Eq.~(\ref{eq:bullet1})\textbf{)}.

Inserting this result into Eq.~(\ref{eq:generating-function-final}),
we finally obtain
\begin{equation}
  \label{eq:bullet5}
  \begin{bmatrix}
    \bullet \\ 2
  \end{bmatrix} = -\frac{1}{2\sqrt{1-2z}} \log \left(
    \frac{1}{\sqrt{1-2z}} \right).
\end{equation}

$
\begin{bsmallmatrix}
  \bullet & 1
\end{bsmallmatrix}
$ \textbf{(}Eq.~(\ref{eq:bullet2})\textbf{)} and $
\begin{bsmallmatrix}
  \bullet \\ 2
\end{bsmallmatrix}
$ are the only necessary generating functions to derive relations for
the next-to-leading log order. It is surprising that they are related
by a factor of $-1/2$,
\begin{equation}
  \label{eq:bullet5bullet2}
  \begin{bmatrix}
    \bullet \\ 2
  \end{bmatrix} = - \frac{1}{2} \begin{bmatrix} \bullet & 1
  \end{bmatrix}.
\end{equation}

\subsubsection{The generating function $ \protect\begin{bsmallmatrix}
    \bullet \protect\\ 3 \protect\end{bsmallmatrix} $}

Now, consider the next-to-next-to-leading log generating function $
\mathcal{M}(z) =
\begin{bsmallmatrix}
  \bullet \\ 3
\end{bsmallmatrix}
$. Matrices $m\sim\mathcal{M}$ belong to the shuffle products
$a_1^{\shuffle_\Theta\,N}\shuffle_\Theta \Theta(a_1,a_1,a_1)$ for
$N\in\mathbb{N}$.

Again in Eq.~(\ref{eq:master}), we replace the sum over
$m\sim\mathcal{M}$ by a sum over $N\in\mathbb{N}$ such that $m
= \begin{bsmallmatrix} N \\ 3
\end{bsmallmatrix}
$. The first term consists of two parts, $m' =
\begin{bsmallmatrix}
  N+3
\end{bsmallmatrix}
$ and $ m' =
\begin{bsmallmatrix}
  N + 1 \\ 2
\end{bsmallmatrix}
$. For other $m'$, $\mathcal{S}(m,m')$ vanishes. In particular,
\begin{equation}
  \mathcal{S}\left(
    \begin{bmatrix}
      N \\ 3
    \end{bmatrix},
    \begin{bmatrix}
      N+3
    \end{bmatrix}
  \right) =
  \begin{pmatrix}
    N+3 \\ 3
  \end{pmatrix} = \frac{(N+3)(N+2)(N+1)}{6}, \qquad \mathcal{S}\left(
    \begin{bmatrix}
      N \\ 3
    \end{bmatrix},
    \begin{bmatrix}
      N + 1 \\ 2
    \end{bmatrix}
  \right) =
  \begin{pmatrix}
    N+1 \\ 1
  \end{pmatrix} = N+1.
\end{equation}
The sum over $j$ only gives a non-zero value for $j=1$. Hence $p_j =
p_1 =
\begin{bsmallmatrix}
  1
\end{bsmallmatrix}
$ and the integers $t_i$ in the $(*)$-sum require $\sum_i t_i
= N + 3 - j = N + 2$ \textbf{(}see Eq.~(\ref{eq:star})\textbf{)}. The
first argument of the function $\mathcal{S}$ becomes $m\ominus p_j =
\begin{bsmallmatrix}
  N-1 \\ 3
\end{bsmallmatrix}
$. For the inhomogeneous part of the differential equation, this
implies that either $\bigoplus_i m_i =
\begin{bsmallmatrix}
  N+2
\end{bsmallmatrix}
$ or $\bigoplus_i m_i =
\begin{bsmallmatrix}
  N \\ 2
\end{bsmallmatrix}
$. The first case is realized if $m_i =
\begin{bsmallmatrix}
  t_i
\end{bsmallmatrix}
$ $\forall i \leq k$. The
second case implies that one of the matrices $m_i$ is equal to $
\begin{bsmallmatrix}
  t_i - 2 \\ 2
\end{bsmallmatrix}
$ and the rest of the matrices $m_i =
\begin{bsmallmatrix}
  t_i
\end{bsmallmatrix}
$. We compute
\begin{equation}
  \mathcal{S} \left(
    \begin{bmatrix}
      N-1 \\ 3
    \end{bmatrix},
    \begin{bmatrix}
      N+2
    \end{bmatrix}
  \right) =
  \begin{pmatrix}
    N+2 \\ 3
  \end{pmatrix} = \frac{(N+2)(N+1)N}{6}, \qquad 
  \mathcal{S}\left(
    \begin{bmatrix}
      N-1 \\ 3
    \end{bmatrix},
    \begin{bmatrix}
      N \\ 2
    \end{bmatrix}
  \right) =
  \begin{pmatrix}
    N \\ 1
  \end{pmatrix} = N.
\end{equation}
We use all these observations to obtain the inhomogeneous part of the
differential equation Eq.~(\ref{eq:master}),
\begin{align}
  \label{eq:bullet6-pre}
  \begin{bmatrix}
    \bullet \\ 3
  \end{bmatrix}'\bigg|_\mathrm{inhom.} =& \sum_{N = 0}^\infty \Bigg(
  -\frac{\mathrm{d}}{\mathrm{d}z} z^{(N+1) - (N+3)}
  \frac{(N+3)(N+2)(N+1)}{6}
  \begin{bmatrix}
    N+3
  \end{bmatrix}
  z^{N+3} \nonumber\\
  &-\frac{\mathrm{d}}{\mathrm{d}z} z^{(N+1) - (N+2)} (N+1)
    \begin{bmatrix}
      N + 1 \\ 2
    \end{bmatrix}
    z^{N+2}
    \nonumber\\
    &+ \sum_{k\geq1} \sum_{\sum_i t_i = N+2} z^{(N+1) - 1 - \sum_i
      t_i} \frac{(N+2)(N+1)N}{6}
    \begin{bmatrix}
      t_1
    \end{bmatrix}
    z^{t_1} \ldots
    \begin{bmatrix}
      t_k
    \end{bmatrix}
    z^{t_k}
    \nonumber\\
    &+ \sum_{k\geq1} \sum_{\sum_i t_i = N+2} z^{(N+1) - 1 - (\sum_i
      t_i - 1)} N \,k \,
    \begin{bmatrix}
      t_1
    \end{bmatrix}
    z^{t_1} \ldots
    \begin{bmatrix}
      t_{k-1}
    \end{bmatrix}
    z^{t_{k-1}}
  \begin{bmatrix}
    t_k - 2 \\ 2
  \end{bmatrix} z^{t_k - 1} \Bigg)
  \nonumber\\
  =& \sum_{N=0}^\infty \Bigg( - \frac{\mathrm{d}}{\mathrm{d}z}
  \frac{1}{z^2} \frac{1}{6} z^3 \frac{\mathrm{d}^3}{\mathrm{d}z^3}
  \begin{bmatrix}
    N+3
  \end{bmatrix}
  z^{N+3} - \frac{\mathrm{d}}{\mathrm{d}z} \frac{1}{z} z^2
  \frac{\mathrm{d}}{\mathrm{d}z} \frac{1}{z}
  \begin{bmatrix}
    N + 1 \\ 2
  \end{bmatrix}
  z^{N+2} \nonumber\\
  &+ \sum_{k\geq1} \sum_{\sum_i t_i = N+2} \frac{1}{z^2} \frac{1}{6}
  z^3 \frac{\mathrm{d}^3}{\mathrm{d}z^3}
  \begin{bmatrix}
    t_1
  \end{bmatrix}
  z^{t_1} \ldots
  \begin{bmatrix}
    t_k
  \end{bmatrix}
  z^{t_k} \nonumber\\
  &+ \sum_{k\geq1} \sum_{\sum_i t_i = N + 2} \frac{1}{z} z^2
  \frac{\mathrm{d}}{\mathrm{d}z} \frac{1}{z} \,k \,
  \begin{bmatrix}
    t_1
  \end{bmatrix}
  z^{t_1}
  \ldots
  \begin{bmatrix}
    t_{k-1}
  \end{bmatrix}
  z^{t_{k-1}}
  \begin{bmatrix}
    t_k - 2 \\ 2
  \end{bmatrix} z^{t_k - 1} \Bigg) \nonumber\\
  =& - \frac{\mathrm{d}}{\mathrm{d}z} \frac{z}{6}
  \frac{\mathrm{d}^3}{\mathrm{d}z^3} \begin{bmatrix}\bullet\end{bmatrix} -
  \frac{\mathrm{d}}{\mathrm{d}z} z \frac{\mathrm{d}}{\mathrm{d}z}
  \frac{1}{z}
  \begin{bmatrix}
    \bullet \\ 2
  \end{bmatrix}
  + \sum_{k\geq1} \frac{z}{6} \frac{\mathrm{d}^3}{\mathrm{d}z^3}
  \begin{bmatrix}\bullet\end{bmatrix}^k + \sum_{k\geq1} z
  \frac{\mathrm{d}}{\mathrm{d}z} 
  \frac{1}{z} \,k \,\begin{bmatrix}\bullet\end{bmatrix}^{k-1}
  \begin{bmatrix}
    \bullet \\ 2
  \end{bmatrix} \nonumber\\
  =& - \frac{\mathrm{d}}{\mathrm{d}z} \frac{z}{6}
  \frac{\mathrm{d}^3}{\mathrm{d}z^3} \begin{bmatrix}\bullet\end{bmatrix} -
  \frac{\mathrm{d}}{\mathrm{d}z} z \frac{\mathrm{d}}{\mathrm{d}z}
  \frac{1}{z}
  \begin{bmatrix}
    \bullet \\ 2
  \end{bmatrix}
  + \frac{z}{6} \frac{\mathrm{d}^3}{\mathrm{d}z^3}
  \frac{1}{1-\begin{bmatrix}\bullet\end{bmatrix}} + z
  \frac{\mathrm{d}}{\mathrm{d}z} \frac{1}{z} 
  \frac{1}{\left(1-\begin{bmatrix}\bullet\end{bmatrix}\right)^2}
  \begin{bmatrix}
    \bullet \\ 2
  \end{bmatrix} \nonumber\\
  =& \frac{1}{\sqrt{1-2z}^5} - \frac{1}{2\sqrt{1-2z}}
  \frac{\mathrm{d}}{\mathrm{d}z}\left( \frac{1}{z} \log \left(
      \frac{1}{\sqrt{1-2z}} \right) \right).
\end{align}
In the last line, we used the explicit formulas for the generating
functions $\begin{bsmallmatrix}\bullet\end{bsmallmatrix}$ and $
\begin{bsmallmatrix}
  \bullet \\ 2
\end{bsmallmatrix}
$ \textbf{(}Eqs.~(\ref{eq:bullet1},\ref{eq:bullet5})\textbf{)}. 

Inserting Eq.~(\ref{eq:bullet6-pre}) into the integration in
Eq.~(\ref{eq:generating-function-final}) finally results in
\begin{equation}
  \label{eq:bullet6}
  \begin{bmatrix}
    \bullet \\ 3
  \end{bmatrix} = \frac{1}{2\sqrt{1-2z}^3} - \frac{1}{2\sqrt{1-2z}}
  \frac{1}{z} \log \left( \frac{1}{\sqrt{1-2z}} \right).
\end{equation}

\subsubsection{The generating function $ \protect\begin{bsmallmatrix}
    \bullet & 0 \protect\\ 1 & 1 \protect\end{bsmallmatrix}$}

Now, consider $ \mathcal{M}(z) =
\begin{bsmallmatrix}
  \bullet & 0 \\ 1 & 1
\end{bsmallmatrix}
$. The respective shuffle products to the matrices $m\sim\mathcal{M}$
are $a_1^{\shuffle_\Theta\,N}\shuffle_\Theta \Theta(a_1,a_2)$ for
$N\in\mathbb{N}$. 

In Eq.~(\ref{eq:master}), we replace the sum over $m\sim\mathcal{M}$
by a sum over $N\in\mathbb{N}$ such that $m = \begin{bsmallmatrix} N & 0 \\
  1 & 1
\end{bsmallmatrix}
$. Here, only $m' =
\begin{bsmallmatrix}
  N + 1 & 1
\end{bsmallmatrix}$ contributes to the first sum and we calculate
\begin{equation}
  \mathcal{S}\left(
    \begin{bmatrix}
      N & 0 \\ 1 & 1
    \end{bmatrix}
    ,
    \begin{bmatrix}
      N + 1 & 1
    \end{bmatrix}
  \right) =
  \begin{pmatrix}
    N+1 \\ 1
  \end{pmatrix} = N+1.
\end{equation}
In the second sum, $j=1$ and $p_j = p_1 =
\begin{bsmallmatrix}
  1
\end{bsmallmatrix}
$. The integers $t_i$ in the $(*)$-sum are constraint by $\sum_i t_i =
N + 3 - j = N + 2$ \textbf{(}see Eq.~(\ref{eq:star})\textbf{)}. The
first argument of the function $\mathcal{S}$ becomes $m\ominus p_j =
\begin{bsmallmatrix}
  N-1 & 0 \\ 1 & 1
\end{bsmallmatrix}
$. The second argument must be $\bigoplus_i m_i =
\begin{bsmallmatrix}
  N & 1
\end{bsmallmatrix}
$ since we only consider the inhomogeneous part of the differential
equation. Thus, one of the matrices $m_i$ is equal to $
\begin{bsmallmatrix}
  t_i - 2 & 1
\end{bsmallmatrix}
$ and the rest of the matrices $m_i =
\begin{bsmallmatrix}
  t_i
\end{bsmallmatrix}
$. In particular,
\begin{equation}
  \mathcal{S} \left(
    \begin{bmatrix}
      N-1 & 0 \\ 1 & 1
    \end{bmatrix},
    \begin{bmatrix}
      N & 1
    \end{bmatrix}
  \right) =
  \begin{pmatrix}
    N \\ 1
  \end{pmatrix} = N.
\end{equation}
We deduce the inhomogeneous part of the differential equation
Eq.~(\ref{eq:master}),
\begin{align}
  \label{eq:bullet7-pre}
  \begin{bmatrix}
    \bullet & 0 \\ 1 & 1
  \end{bmatrix}'\bigg|_\mathrm{inhom.} =&
  \sum_{N=0}^\infty
  \Bigg(
  -\frac{\mathrm{d}}{\mathrm{d}z} z^{(N+1) - (N+2)} (N+1)
  \begin{bmatrix}
    N + 1 & 1
  \end{bmatrix} z^{N+2} \nonumber\\
  &+ \sum_{k\geq1} \sum_{\sum_i t_i = N + 2} z^{(N+1) - 1 - (\sum_i
    t_i - 1)} N \,k \,
  \begin{bmatrix}
    t_1
  \end{bmatrix}
  z^{t_1} \ldots
  \begin{bmatrix}
    t_{k-1}
  \end{bmatrix}
  z^{t_{k-1}}
  \begin{bmatrix}
    t_k - 2 & 1
  \end{bmatrix} z^{t_k - 1}
  \Bigg) \nonumber\\
  =& \sum_{N=0}^\infty
  \Bigg(
    - \frac{\mathrm{d}}{\mathrm{d}z} \frac{1}{z} z^2
    \frac{\mathrm{d}}{\mathrm{d}z} \frac{1}{z} 
    \begin{bmatrix}
      N + 1 & 1
    \end{bmatrix} z^{N+2} \nonumber\\
    & + \sum_{k\geq1} \sum_{\sum_i t_i = N + 2}
    \frac{1}{z} z^2 \frac{\mathrm{d}}{\mathrm{d}z} \frac{1}{z} k\,
    \begin{bmatrix}
      t_1
    \end{bmatrix}
    z^{t_1} \ldots
    \begin{bmatrix}
      t_{k-1}
    \end{bmatrix}
    z^{t_{k-1}}
    \begin{bmatrix}
      t_k - 2 & 1
    \end{bmatrix} z^{t_k - 1}.
  \Bigg) \nonumber\\
  =& - \frac{\mathrm{d}}{\mathrm{d}z} z \frac{\mathrm{d}}{\mathrm{d}z}
  \frac{1}{z}
  \begin{bmatrix}
    \bullet & 1
  \end{bmatrix} + \sum_{k\geq1} z \frac{\mathrm{d}}{\mathrm{d}z}
  \frac{1}{z} k\, \begin{bmatrix}\bullet\end{bmatrix}^{k-1}
  \begin{bmatrix}
    \bullet & 1
  \end{bmatrix} \nonumber\\
  =& - \frac{\mathrm{d}}{\mathrm{d}z} z \frac{\mathrm{d}}{\mathrm{d}z}
  \frac{1}{z}
  \begin{bmatrix}
    \bullet & 1
  \end{bmatrix} + z \frac{\mathrm{d}}{\mathrm{d}z} \frac{1}{z}
  \frac{1}{\left(1- \begin{bmatrix}\bullet\end{bmatrix}\right)^2}
  \begin{bmatrix}
    \bullet & 1
  \end{bmatrix} \nonumber\\
  =& - \frac{3}{\sqrt{1-2z}^5} + \frac{1}{\sqrt{1-2z}}
  \frac{\mathrm{d}}{\mathrm{d}z} \frac{1}{z} \log \left(
    \frac{1}{\sqrt{1-2z}} \right).
\end{align}
In the last line, we used the explicit formulas for the generating
functions $\begin{bsmallmatrix}\bullet\end{bsmallmatrix}$ and $
\begin{bsmallmatrix}
  \bullet & 1
\end{bsmallmatrix}
$ \textbf{(}Eqs.~(\ref{eq:bullet1},\ref{eq:bullet2})\textbf{)}.

We plug Eq.~(\ref{eq:bullet7-pre}) into
Eq.~(\ref{eq:generating-function-final}) and obtain
\begin{equation}
  \label{eq:bullet7}
  \begin{bmatrix}
    \bullet & 0 \\ 1 & 1
  \end{bmatrix} = \frac{1}{2\sqrt{1-2z}} - \frac{3}{2\sqrt{1-2z}^3} +
  \frac{1}{\sqrt{1-2z}} \frac{1}{z} \log \left( \frac{1}{\sqrt{1-2z}}
  \right).
\end{equation}

\subsubsection{The generating function $ \protect\begin{bsmallmatrix}
    \bullet \protect\\ 2 \protect\\ 2 \protect\end{bsmallmatrix}$}

The function $ \mathcal{M}(z) =
\begin{bsmallmatrix}
  \bullet \\ 2 \\ 2
\end{bsmallmatrix}
$ generates the rationals $m \sim \mathcal{M}$ to the respective
shuffle products $a_1^{\shuffle_\Theta\,N}\shuffle_\Theta
\Theta(a_1,a_1)\shuffle_\Theta \Theta(a_1,a_1)$ for $N\in\mathbb{N}$.

Consider Eq.~(\ref{eq:master}). We replace the sum over
$m\sim\mathcal{M}$ by a sum over $N\in\mathbb{N}$ such that $m
= \begin{bsmallmatrix} N \\ 2 \\ 2
\end{bsmallmatrix}
$. Here, only the matrices $m' =
\begin{bsmallmatrix}
  N+4
\end{bsmallmatrix}
$ and $m' =
\begin{bsmallmatrix}
  N + 2 \\ 2
\end{bsmallmatrix}$ contribute to the first sum and we find
\begin{gather}
  \mathcal{S}\left(
    \begin{bmatrix}
      N \\ 2 \\ 2
    \end{bmatrix}
    ,
    \begin{bmatrix}
      N+4
    \end{bmatrix}
  \right) = \frac{1}{2}
  \begin{pmatrix}
    N + 4 \\ 2
  \end{pmatrix}
  \begin{pmatrix}
    N + 2 \\ 2
  \end{pmatrix} = \frac{(N+4)(N+3)(N+2)(N+1)}{8}, \\
  \mathcal{S}\left(
    \begin{bmatrix}
      N \\ 2 \\ 2
    \end{bmatrix}
    ,
    \begin{bmatrix}
      N + 2 \\ 2
    \end{bmatrix}
  \right) =
  \begin{pmatrix}
    N+2 \\ 2
  \end{pmatrix} = \frac{(N+2)(N+1)}{2}.
\end{gather}
In the second sum, $j=1$ and $p_j = p_1 =
\begin{bsmallmatrix}
  1
\end{bsmallmatrix}
$ as before. The integers $t_i$ in the $(*)$-sum range over $\sum_i
t_i = N + 4 - j = N + 3$ \textbf{(}see Eq.~(\ref{eq:star})\textbf{)}.
The first argument of the function $\mathcal{S}$ is $m\ominus p_j =
\begin{bsmallmatrix}
  N-1 \\ 2 \\ 2
\end{bsmallmatrix}
$. Here, there are three possibilities for $\bigoplus_i m_i$ in the
second sum of Eq.~(\ref{eq:master}). First, $\bigoplus_i m_i =
\begin{bsmallmatrix}
  N+3
\end{bsmallmatrix}
$ and $m_i =
\begin{bsmallmatrix}
  t_i
\end{bsmallmatrix}
$ $\forall i \leq k$. Secondly, $\bigoplus_i m_i =
\begin{bsmallmatrix}
  N + 1 \\ 2
\end{bsmallmatrix}
$, which implies that one of the matrices $m_i$ is equal to $
\begin{bsmallmatrix}
  t_i - 2 \\ 2
\end{bsmallmatrix}
$ and the rest of the matrices $m_i =
\begin{bsmallmatrix}
  t_i
\end{bsmallmatrix}
$. In the third case,
$\bigoplus_i m_i =
\begin{bsmallmatrix}
  N - 1 \\ 2 \\ 2
\end{bsmallmatrix}
$. Note that $\bigoplus_i m_i \sim \mathcal{M}$ did not occur in the
previous examples because it was part of the homogeneous differential
equation Eq.~(\ref{eq:master}). Here, we realize $\bigoplus_i m_i =
\begin{bsmallmatrix}
  N - 1 \\ 2 \\ 2
\end{bsmallmatrix}
$ within the inhomogeneous part of the differential equation: two of
the matrices $m_i$ are equal to $
\begin{bsmallmatrix}
  t_i - 2 \\ 2
\end{bsmallmatrix}
$ and the rest of the matrices $m_i =
\begin{bsmallmatrix}
  t_i
\end{bsmallmatrix}
$. The corresponding
functions $\mathcal{S}(\cdot,\cdot)$ give
\begin{gather}
  \mathcal{S} \left(
    \begin{bmatrix}
      N-1 \\ 2 \\ 2
    \end{bmatrix},
    \begin{bmatrix}
      N+3
    \end{bmatrix}
  \right) = \frac{1}{2}
  \begin{pmatrix}
    N+3 \\ 2
  \end{pmatrix}
  \begin{pmatrix}
    N+1 \\ 2
  \end{pmatrix} = \frac{(N+3)(N+2)(N+1)N}{8}, \\
  \mathcal{S} \left(
    \begin{bmatrix}
      N-1 \\ 2 \\ 2
    \end{bmatrix},
    \begin{bmatrix}
      N + 1 \\ 2
    \end{bmatrix}
  \right) =
  \begin{pmatrix}
    N + 1 \\ 2
  \end{pmatrix} = \frac{(N+1)N}{2},\\
  \mathcal{S} \left(
    \begin{bmatrix}
      N-1 \\ 2 \\ 2
    \end{bmatrix},
    \begin{bmatrix}
      N-1 \\ 2 \\ 2
    \end{bmatrix}
  \right) = 1.
\end{gather}
We simplify the inhomogeneous part of the differential equation
Eq.~(\ref{eq:master}) as follows:
\begin{align}
  \begin{bmatrix}
    \bullet \\ 2 \\ 2
  \end{bmatrix}' \Bigg|_\mathrm{inhom.} =& \sum_{N = 0}^\infty \Bigg( -
  \frac{\mathrm{d}}{\mathrm{d}z} z^{(N+2) - (N+4)}
  \frac{(N+4)(N+3)(N+2)(N+1)}{8}
  \begin{bmatrix}
    N+4
  \end{bmatrix}
  z^{N+4} \nonumber\\
  &- \frac{\mathrm{d}}{\mathrm{d}z} z^{(N+2) - (N+3)}
  \frac{(N+2)(N+1)}{2}
  \begin{bmatrix}
    N + 2 \\ 2
  \end{bmatrix} z^{N+3} \nonumber\\
  & + \sum_{k\geq1} \sum_{\sum_i t_i = N+3} z^{(N+2) - 1 - \sum_i t_i}
  \frac{(N+3)(N+2)(N+1)N}{8}
  \begin{bmatrix}
    t_1
  \end{bmatrix}
  z^{t_1} \ldots
  \begin{bmatrix}
    t_k
  \end{bmatrix}
  z^{t_k}
  \nonumber\\
  &+ \sum_{k\geq1} \sum_{\sum_i t_i = N + 3} z^{(N+2) - 1 - (\sum_i
    t_i - 1)} \frac{(N+1)N}{2} k
  \begin{bmatrix}
    t_1
  \end{bmatrix}
  z^{t_1} \ldots
  \begin{bmatrix}
    t_{k-1}
  \end{bmatrix}
  z^{t_{k-1}}
  \begin{bmatrix}
    t_k - 2 \\ 2
  \end{bmatrix} z^{t_k - 1} \nonumber\\
  & + \sum_{k\geq1} \sum_{\sum_i t_i = N + 3} z^{(N+2) - 1 - (\sum_i
    t_i - 2)}
  \begin{pmatrix}
    k \\ 2
  \end{pmatrix}
  \begin{bmatrix}
    t_1
  \end{bmatrix}
  z^{t_1} \ldots
  \begin{bmatrix}
    t_{k-2}
  \end{bmatrix}
  z^{t_{k-2}}
  \begin{bmatrix}
    t_{k-1} - 2 \\ 2
  \end{bmatrix} z^{t_{k-1} - 1}
  \begin{bmatrix}
    t_k - 2 \\ 2
  \end{bmatrix} z^{t_k - 1} \Bigg)\nonumber\\
  =& \sum_{N=0}^\infty \Bigg( - \frac{\mathrm{d}}{\mathrm{d}z}
  \frac{1}{z^2} \frac{1}{8} z^4
  \frac{\mathrm{d}^4}{\mathrm{d}z^4}
  \begin{bmatrix}
    N+4
  \end{bmatrix}
  z^{N+4} -
  \frac{\mathrm{d}}{\mathrm{d}z} \frac{1}{z}\frac{1}{2} z^3
  \frac{\mathrm{d}^2}{\mathrm{d}z^2} \frac{1}{z}
    \begin{bmatrix}
      N + 2 \\ 2
    \end{bmatrix} z^{N+3} \nonumber\\
    &+ \sum_{k\geq1} \sum_{\sum_{t_i} = N+3} \frac{1}{z^2} \frac{1}{8}
    z^4 \frac{\mathrm{d}^4}{\mathrm{d}z^4}
    \begin{bmatrix}
      t_1
    \end{bmatrix}
    z^{t_1} \ldots
    \begin{bmatrix}
      t_k
    \end{bmatrix}
    z^{t_k} \nonumber\\
    & + \sum_{k\geq1} \sum_{\sum_{t_i} = N+3} \frac{1}{z}
    \frac{1}{2} z^3 \frac{\mathrm{d}^2}{\mathrm{d}z^2} \frac{1}{z} k
    \begin{bmatrix}
      t_1
    \end{bmatrix}
    z^{t_1} \ldots
    \begin{bmatrix}
      t_{k-1}
    \end{bmatrix}
    z^{t_{k-1}}
  \begin{bmatrix}
    t_k - 2 \\ 2
  \end{bmatrix} z^{t_k - 1} \nonumber\\
  & + \sum_{k\geq1} \sum_{\sum_{t_i} = N+3} \frac{k(k-1)}{2}
  \begin{bmatrix}
    t_1
  \end{bmatrix}
  z^{t_1} \ldots
  \begin{bmatrix}
    t_{k-2}
  \end{bmatrix}
  z^{t_{k-2}}
  \begin{bmatrix}
    t_{k-1} - 2 \\ 2
  \end{bmatrix} z^{t_{k-1} - 1}
  \begin{bmatrix}
    t_k - 2 \\ 2
  \end{bmatrix} z^{t_k - 1} \Bigg) \nonumber\\ =& -
  \frac{\mathrm{d}}{\mathrm{d}z} \frac{z^2}{8}
  \frac{\mathrm{d}^4}{\mathrm{d}z^4} \begin{bmatrix}\bullet\end{bmatrix}
  - \frac{\mathrm{d}}{\mathrm{d}z} \frac{z^2}{2}
  \frac{\mathrm{d}^2}{\mathrm{d}z^2} \frac{1}{z}
  \begin{bmatrix}
    \bullet \\ 2
  \end{bmatrix} + \sum_{k\geq1} \frac{z^2}{8}
  \frac{\mathrm{d}^4}{\mathrm{d}z^4} \begin{bmatrix}\bullet\end{bmatrix}^k
  + \sum_{k\geq1} \frac{z^2}{2} \frac{\mathrm{d}^2}{\mathrm{d}z^2}
  \frac{1}{z} k \begin{bmatrix}\bullet\end{bmatrix}^{k-1}
  \begin{bmatrix}
    \bullet \\ 2
  \end{bmatrix}  \nonumber\\
  &+ \sum_{k\geq1}
  \frac{k(k-1)}{2} \begin{bmatrix}\bullet\end{bmatrix}^{k-2}
  \begin{bmatrix}
    \bullet \\ 2
  \end{bmatrix}^2 \nonumber\\
  =& -
  \frac{\mathrm{d}}{\mathrm{d}z} \frac{z^2}{8}
  \frac{\mathrm{d}^4}{\mathrm{d}z^4} \begin{bmatrix}\bullet\end{bmatrix}
  - 
  \frac{\mathrm{d}}{\mathrm{d}z} \frac{z^2}{2}
  \frac{\mathrm{d}^2}{\mathrm{d}z^2} \frac{1}{z}
  \begin{bmatrix}
    \bullet \\ 2
  \end{bmatrix} + \frac{z^2}{8} \frac{\mathrm{d}^4}{\mathrm{d}z^4}
  \frac{1}{1 - \begin{bmatrix}\bullet\end{bmatrix}} + \frac{z^2}{2}
  \frac{\mathrm{d}^2}{\mathrm{d}z^2} \frac{1}{z} \frac{1}{\left(1
    - \begin{bmatrix}\bullet\end{bmatrix}\right)^2}  
  \begin{bmatrix}
    \bullet \\ 2
  \end{bmatrix} + \frac{1}{\left(1
      - \begin{bmatrix}\bullet\end{bmatrix}\right)^3}
  \begin{bmatrix}
    \bullet \\ 2
  \end{bmatrix}^2.
\end{align}

We use the previous results in
Eqs.~(\ref{eq:bullet1},\ref{eq:bullet5}) and insert the resulting
expression into Eq.~(\ref{eq:generating-function-final}).

A little calculation yields
\begin{align}
  \label{eq:bullet8}
  \begin{bmatrix}
    \bullet \\ 2 \\ 2
  \end{bmatrix} =& -\frac{1}{8\sqrt{1-2z}} - \frac{3}{8\sqrt{1-2z}^3}
  + \frac{1}{4\sqrt{1-2z}^3} \log \left( \frac{1}{\sqrt{1-2z}} \right)
  + \frac{1}{2\sqrt{1-2z}} \frac{1}{z} \log \left(
    \frac{1}{\sqrt{1-2z}} \right) \nonumber\\
  &+ \frac{1}{8\sqrt{1-2z}^3} \log^2 \left( \frac{1}{\sqrt{1-2z}}
  \right).
\end{align}

\subsubsection{The generating function $ \protect\begin{bsmallmatrix}
    \bullet & 1 \protect\\ 2 & 0 \protect\end{bsmallmatrix}$}

The only next-to-next-to-leading log generating function left is $
\mathcal{M}(z) =
\begin{bsmallmatrix}
  \bullet & 1 \\ 2 & 0
\end{bsmallmatrix}
$. It generates the rationals $m \sim \mathcal{M}$ with the respective
shuffle products $a_1^{\shuffle_\Theta\,N}\shuffle_\Theta a_2
\shuffle_\Theta \Theta(a_1,a_1)$ for $N\in\mathbb{N}$.

In Eq.~(\ref{eq:master}), we replace the sum over $m\sim\mathcal{M}$
by a sum over $N\in\mathbb{N}$ such that $m = \begin{bsmallmatrix} N & 1 \\
  2 & 0
\end{bsmallmatrix}
$. The function $\mathcal{S}$ in the first sum vanishes except for $m'
=
\begin{bsmallmatrix}
  N + 2 & 1
\end{bsmallmatrix}
$:
\begin{equation}
  \mathcal{S}\left(
    \begin{bmatrix}
      N & 1 \\ 2 & 0
    \end{bmatrix}
    ,
    \begin{bmatrix}
      N + 2 & 1
    \end{bmatrix}
  \right) =
  \begin{pmatrix}
    N+2 \\ 2
  \end{pmatrix} = \frac{(N+2)(N+1)}{2}.
\end{equation}
In the second sum, either $j = 1$ or $j = 2$. This is the main
difference to the previous examples. The sum does not vanish for $j=2$
because $\mathcal{M}_{1\,2} = 1$. For $j=1$, $p_j = p_1 =
\begin{bsmallmatrix}
  1
\end{bsmallmatrix}
$ as
before. Then, the integers $t_i$ in the $(*)$-sum range over $\sum_i
t_i = N + 4 - j = N + 3$ \textbf{(}see Eq.~(\ref{eq:star})\textbf{)}.
Furthermore, $m\ominus p_j =
\begin{bsmallmatrix}
  N-1 & 1 \\ 2 & 0
\end{bsmallmatrix}
$ in the first argument of the function $\mathcal{S}$, which implies
that either $\bigoplus_i m_i =
\begin{bsmallmatrix}
  N + 1 & 1
\end{bsmallmatrix}
$ or $\bigoplus_i m_i =
\begin{bsmallmatrix}
  N - 1 & 1 \\ 2 & 0
\end{bsmallmatrix}
$. In the first case, there is one $ i \leq k$ such that $m_i =
\begin{bsmallmatrix}
  t_i - 2 & 1
\end{bsmallmatrix}
$ and for all other $ i \leq k$, $m_i =
\begin{bsmallmatrix}
  t_i
\end{bsmallmatrix}
$. In the second case,
one of the matrices $m_i$ must be equal to $\begin{bsmallmatrix} t_i - 2 &
  1
\end{bsmallmatrix}
$ and another one must be equal to $
\begin{bsmallmatrix}
  t_i - 2 \\ 2
\end{bsmallmatrix}
$. The rest of the matrices $m_i =
\begin{bsmallmatrix}
  t_i
\end{bsmallmatrix}
$. The corresponding
$\mathcal{S}$-function factors are
\begin{gather}
  \mathcal{S} \left(
    \begin{bmatrix}
      N-1 & 1 \\ 2 & 0
    \end{bmatrix},
    \begin{bmatrix}
      N + 1 & 1
    \end{bmatrix}
  \right) =
  \begin{pmatrix}
    N + 1 \\ 2
  \end{pmatrix} = \frac
  {(N+1)N}{2},\\
  \mathcal{S} \left(
    \begin{bmatrix}
      N-1 & 1 \\ 2 & 0
    \end{bmatrix},
    \begin{bmatrix}
      N-1 & 1 \\ 2 & 0
    \end{bmatrix}
  \right) = 1.
\end{gather}
For $j=2$, $p_j = p_2 =
\begin{bsmallmatrix}
  0 & 1
\end{bsmallmatrix}
$. The integers $t_i$ in the $(*)$-sum then require $\sum_i t_i = N +
4 - j = N + 2$. Furthermore, $m\ominus p_j =
\begin{bsmallmatrix}
  N \\ 2
\end{bsmallmatrix}
$ in the first argument of the function $\mathcal{S}$. Thus, either
$\bigoplus_i m_i =
\begin{bsmallmatrix}
  N+2
\end{bsmallmatrix}
$ or $\bigoplus_i m_i =
\begin{bsmallmatrix}
  N \\ 2
\end{bsmallmatrix}
$. In the former case, $m_i =
\begin{bsmallmatrix}
  t_i
\end{bsmallmatrix}
$ $\forall i \leq k$. In the latter
case, one of the matrices $m_i$ must be equal to $
\begin{bsmallmatrix}
  t_i - 2 \\ 2
\end{bsmallmatrix}
$ and the other matrices $m_i =
\begin{bsmallmatrix}
  t_i
\end{bsmallmatrix}
$. The corresponding functions
$\mathcal{S}(\cdot,\cdot)$ give
\begin{gather}
  \mathcal{S} \left(
    \begin{bmatrix}
      N \\ 2
    \end{bmatrix},
    \begin{bmatrix}
      N+2
    \end{bmatrix}
  \right) =
  \begin{pmatrix}
    N +2 \\ 2
  \end{pmatrix} = \frac
  {(N+2)(N+1)}{2},\\
  \mathcal{S} \left(
    \begin{bmatrix}
      N \\ 2
    \end{bmatrix},
    \begin{bmatrix}
      N \\ 2
    \end{bmatrix}
  \right) = 1.
\end{gather}
Using these observations yields the inhomogeneous part of the
differential equation Eq.~(\ref{eq:master}),
\begin{align}
  \begin{bmatrix}
    \bullet & 1 \\ 2 & 0
  \end{bmatrix}' \bigg|_\mathrm{inhom.} =& \sum_{N=0}^\infty \Bigg( -
  \frac{\mathrm{d}}{\mathrm{d}z} z^{(N+2) - (N+3)}
  \frac{(N+2)(N+1)}{2}
  \begin{bmatrix}
    N + 2 & 1
  \end{bmatrix}
  z^{N+3} \nonumber\\
  &+ \sum_{k\geq1} \sum_{\sum_i t_i = N + 3} z^{(N+2) - 1 - (\sum_i
    t_i - 1)} \frac {(N+1)N}{2}\, k\,
  \begin{bmatrix}
    t_1
  \end{bmatrix}
  z^{t_1} \ldots
  \begin{bmatrix}
    t_{k-1}
  \end{bmatrix}
  z^{t_{k-1}}
  \begin{bmatrix}
    t_k - 2 & 1
  \end{bmatrix}
  z^{t_k - 1} \nonumber\\
  &+ \sum_{k\geq1} \sum_{\sum_i t_i = N+3} \!\!\! z^{(N+2) - 1 -
    (\sum_i t_i - 2)} k(k-1) \,
  \!\!\begin{bmatrix}
    t_1
  \end{bmatrix}
  z^{t_1} \!\ldots
  \begin{bmatrix}
    t_{k-2}
  \end{bmatrix}
  z^{t_{k-2}}
  \begin{bmatrix}
    t_{k-1} - 2 & 1
  \end{bmatrix}
  z^{t_{k-1} - 1} \!\!
  \begin{bmatrix}
    t_k - 2 \\ 2 
  \end{bmatrix}\!
  z^{t_k - 1}
 \nonumber\\
  &+ \sum_{k\geq1}
  \begin{pmatrix}
    2 + k \\ k
  \end{pmatrix}
  \sum_{\sum_i t_i = N+2} z^{(N+2) - 1 - \sum_i t_i} \, \frac
  {(N+2)(N+1)}{2} \,
  \begin{bmatrix}
    t_1
  \end{bmatrix}
  z^{t_1} \ldots
  \begin{bmatrix}
    t_k
  \end{bmatrix}
  z^{t_{k}}
  \nonumber\\
  & + \sum_{k\geq1}
  \begin{pmatrix}
    2 + k \\ k
  \end{pmatrix}
  \sum_{\sum_i t_i = N + 2} z^{(N+2) - 1 - (\sum_i t_i - 1)} \,k\,
  \begin{bmatrix}
    t_1
  \end{bmatrix}
  z^{t_1} \ldots
  \begin{bmatrix}
    t_{k-1}
  \end{bmatrix}
  z^{t_{k-1}}
  \begin{bmatrix}
    t_k - 2 \\ 2
  \end{bmatrix} z^{t_k - 1} \Bigg) \nonumber\\
  =& \sum_{N=0}^\infty \Bigg( -
  \frac{\mathrm{d}}{\mathrm{d}z} \frac{1}{z} \frac{1}{2} z^3
  \frac{\mathrm{d}^2}{\mathrm{d}z^2} \frac{1}{z} 
  \begin{bmatrix}
    N + 2 & 1
  \end{bmatrix}
  z^{N+3} \nonumber\\
  &+ \sum_{k\geq1} \sum_{\sum_i t_i = N+3} \frac{1}{z} \frac{1}{2} z^3
  \frac{\mathrm{d}^2}{\mathrm{d}z^2} \frac{1}{z} k
  \begin{bmatrix}
    t_1
  \end{bmatrix}
  z^{t_1}
  \ldots
  \begin{bmatrix}
    t_{k-1}
  \end{bmatrix}
  z^{t_{k-1}}
  \begin{bmatrix}
    t_k - 2 & 1
  \end{bmatrix}
  z^{t_k - 1} \nonumber\\
  &+ \sum_{k\geq1} \sum_{\sum_i t_i = N + 3} k(k-1)
  \begin{bmatrix}
    t_1
  \end{bmatrix}
  z^{t_1}
  \ldots
  \begin{bmatrix}
    t_{k-2}
  \end{bmatrix}
  z^{t_{k-2}}
  \begin{bmatrix}
    t_{k-1} - 2 & 1
  \end{bmatrix}
  z^{t_{k-1} - 1}
  \begin{bmatrix}
    t_k - 2 \\ 2 
  \end{bmatrix}
  z^{t_k - 1}
  \nonumber\\
  &+ \sum_{k\geq1}
  \begin{pmatrix}
    2 + k \\ k
  \end{pmatrix}
  \sum_{\sum_i t_i = N+2} \frac{1}{z} \frac{1}{2} z^2
  \frac{\mathrm{d}^2}{\mathrm{d}z^2}
  \begin{bmatrix}
    t_1
  \end{bmatrix}
  z^{t_1} \ldots
  \begin{bmatrix}
    t_k
  \end{bmatrix}
  z^{t_{k}} \nonumber\\
  &+ \sum_{k\geq1}
  \begin{pmatrix}
    2 + k \\ k
  \end{pmatrix}
  \sum_{\sum_i t_i = N + 2} k
  \begin{bmatrix}
    t_1
  \end{bmatrix}
  z^{t_1} \ldots
  \begin{bmatrix}
    t_{k-1}
  \end{bmatrix}
  z^{t_{k-1}}
  \begin{bmatrix}
    t_k - 2 \\ 2
  \end{bmatrix} z^{t_k - 1} \Bigg) \nonumber\\
  =& - \frac{\mathrm{d}}{\mathrm{d}z} \frac{z^2}{2}
  \frac{\mathrm{d}^2}{\mathrm{d}z^2} \frac{1}{z}
  \begin{bmatrix}
    \bullet & 1
  \end{bmatrix}
  + \sum_{k\geq1} \frac{z^2}{2} \frac{\mathrm{d}^2}{\mathrm{d}z^2}
  \frac{1}{z} k \begin{bmatrix}\bullet\end{bmatrix}^{k-1}
  \begin{bmatrix}
    \bullet & 1
  \end{bmatrix}
  + \sum_{k\geq1} k(k-1) \begin{bmatrix}\bullet\end{bmatrix}^{k-2}
  \begin{bmatrix}
    \bullet & 1
  \end{bmatrix}
  \begin{bmatrix}
    \bullet \\ 2
  \end{bmatrix} \nonumber\\
  &+ \sum_{k\geq1}
  \begin{pmatrix}
    2 + k \\ k
  \end{pmatrix}
  \frac{z}{2}
  \frac{\mathrm{d}^2}{\mathrm{d}z^2} \begin{bmatrix}\bullet\end{bmatrix}^k
  + 
  \sum_{k\geq1}
  \begin{pmatrix}
    2 + k \\ k
  \end{pmatrix}
  k \begin{bmatrix}\bullet\end{bmatrix}^{k-1}
  \begin{bmatrix}
    \bullet \\ 2
  \end{bmatrix} \nonumber\\
  =& - \frac{\mathrm{d}}{\mathrm{d}z} \frac{z^2}{2}
  \frac{\mathrm{d}^2}{\mathrm{d}z^2} \frac{1}{z}
  \begin{bmatrix}
    \bullet & 1
  \end{bmatrix}
  + \frac{z^2}{2} \frac{\mathrm{d}^2}{\mathrm{d}z^2} \frac{1}{z}
  \frac{1}{\left(1-\begin{bmatrix}\bullet\end{bmatrix}\right)^2}
  \begin{bmatrix}
    \bullet & 1
  \end{bmatrix}
  + \frac{2}{\left(1 - \begin{bmatrix}\bullet\end{bmatrix}\right)^3}
  \begin{bmatrix}
    \bullet & 1
  \end{bmatrix}
  \begin{bmatrix}
    \bullet \\ 2
  \end{bmatrix} + \frac{z}{2} \frac{\mathrm{d}^2}{\mathrm{d}z^2}
  \frac{1}{\left(1 - \begin{bmatrix}\bullet\end{bmatrix}\right)^3} \nonumber\\
  &+ \frac{3}{\left(1 - \begin{bmatrix}\bullet\end{bmatrix}\right)^4}
  \begin{bmatrix}
    \bullet \\ 2
  \end{bmatrix}.
\end{align}

We insert this into the integration in
Eq.~(\ref{eq:generating-function-final}). We also need
Eqs.~(\ref{eq:bullet1},\ref{eq:bullet2},\ref{eq:bullet5}) and after
some calculation, we obtain
\begin{equation}
  \label{eq:bullet9}
  \begin{bmatrix}
    \bullet & 1 \\ 2 & 0
  \end{bmatrix} = \frac{1}{\sqrt{1-2z}^3} -
  \frac{1}{\sqrt{1-2z}^3}\log \left( \frac{1}{\sqrt{1-2z}} \right) -
  \frac{1}{\sqrt{1-2z}} \frac{1}{z} \log \left( \frac{1}{\sqrt{1-2z}}
  \right) - \frac{1}{2\sqrt{1-2z}^3}\log^2 \left(
    \frac{1}{\sqrt{1-2z}} \right).
\end{equation}

\subsection{Generating functions for indexed matrices}
\label{sec:commutators}

In this paper, we do not give a general method to obtain the
generating functions for indexed matrices. In this section however, we
derive the two `indexed generating functions'
\begin{equation}
  \begin{bmatrix}\bullet\end{bmatrix}_{[a_1,a_2]} = \sum_{N\geq0}
  \begin{bmatrix}
    N
  \end{bmatrix}
  _{[a_1,a_2]} z^{N+1}, \qquad
  \begin{bmatrix}\bullet\end{bmatrix}_{[a_1,\Theta(a_1,a_1)]} = \sum_{N\geq0}
  \begin{bmatrix}
    N
  \end{bmatrix}
  _{[a_1,\Theta(a_1,a_1)]} z^{N+1}.
\end{equation}
These generate the indexed matrices $
\begin{bsmallmatrix}
  N
\end{bsmallmatrix}
_{[a_1,a_2]}$ and $
\begin{bsmallmatrix}
  N
\end{bsmallmatrix}
_{[a_1,\Theta(a_1,a_1)]}$ that belong to the full shuffle products
\begin{equation}
  a_1^{\shuffle_\Theta\,N} \shuffle_\Theta[a_1,a_2],\qquad
  a_1^{\shuffle_\Theta\,N} \shuffle_\Theta[a_1,\Theta(a_1,a_1)]
\end{equation}
respectively. These shuffles make part of the filtered words $w_{N+3}$
\textbf{(}Eq.~(\ref{eq:DSE-solution-Yuk})\textbf{)} and map to the
next-to-next-to-leading log order in the log-expansion. In particular,
$\begin{bsmallmatrix}\bullet\end{bsmallmatrix}_{[a_1,a_2]}$ and
$\begin{bsmallmatrix}\bullet\end{bsmallmatrix}_{[a_1,\Theta(a_1,a_1)]}$
complete the set of generating functions that are necessary to obtain
the next-to-next-to-leading log expansion. We will work on a general
method to derive indexed generating functions in future work.

\subsubsection{The generating function
  $\protect\begin{bsmallmatrix}\bullet\protect\end{bsmallmatrix}_{[a_1,a_2]}$}

Only words with $(N-2)$ letters $a_1$ and one letter $a_2$ in the {\it
  unfiltered word $w_N$} can contribute to the term
\begin{equation}
  \label{eq:cterm}
  \begin{bmatrix}
    N-3
  \end{bmatrix}
  _{[a_1,a_2]} a_1^{\shuffle_\Theta\,N-3}
  \shuffle_\Theta[a_1,a_2]
\end{equation}
in the {\it filtered word $w_N$}. Consider
Eq.~(\ref{eq:DSE-solution-Yuk}) and let the words $w_{t_i}$ on the rhs
be filtered. Then,
\begin{align}
  \label{eq:wn-contribution-1}
  w_N =& B_+^{a_1} \sum_{k\geq 1} \sum_{\sum_i t_i = N-1} k
  \begin{bmatrix}
    t_1
  \end{bmatrix}
  a_1^{\shuffle_\Theta\,t_1} \shuffle_\Theta \ldots
  \shuffle_\Theta
  \begin{bmatrix}
    t_{k-1}
  \end{bmatrix}
  a_1^{\shuffle_\Theta\,t_{k-1}}
  \shuffle_\Theta \left(
    \begin{bmatrix}
      t_k - 3
    \end{bmatrix}
    _{[a_1,a_2]}
    a_1^{\shuffle_\Theta\,t_k-3} \shuffle_\Theta [a_1,a_2] \right)
  \nonumber\\
  &+ B_+^{a_1} \sum_{k\geq 1} \sum_{\sum_i t_i = N-1} k
  \begin{bmatrix}
    t_1
  \end{bmatrix}
  a_1^{\shuffle_\Theta\,t_1} \shuffle_\Theta \ldots
  \shuffle_\Theta
  \begin{bmatrix}
    t_{k-1}
  \end{bmatrix}
  a_1^{\shuffle_\Theta\,t_{k-1}}
  \shuffle_\Theta
  \begin{bmatrix}
    t_k-2 & 1
  \end{bmatrix} a_1^{\shuffle_\Theta\,t_k-2} \shuffle_\Theta a_2
  \nonumber\\ 
  &+ B_+^{a_2} \sum_{k\geq1}
  \begin{pmatrix}
    2 + k \\ k 
  \end{pmatrix} \sum_{\sum_i t_i = N-2}
  \begin{bmatrix}
    t_1
  \end{bmatrix}
  a_1^{\shuffle_\Theta\,t_1} \shuffle_\Theta \ldots
  \shuffle_\Theta
  \begin{bmatrix}
    t_k
  \end{bmatrix}
  a_1^{\shuffle_\Theta\,t_k} + \ldots,
\end{align}
where the dots represent all missing terms in
Eq.~(\ref{eq:DSE-solution-Yuk}), for example
\begin{equation}
  \label{eq:multi-commutator-term}
  B_+^{a_1} \sum_{k\geq 1} \sum_{\sum_i t_i = N-1} k
  \begin{bmatrix}
    t_1
  \end{bmatrix}
  a_1^{\shuffle_\Theta\,t_1} \shuffle_\Theta \ldots
  \shuffle_\Theta
  \begin{bmatrix}
    t_{k-1}
  \end{bmatrix}
  a_1^{\shuffle_\Theta\,t_{k-1}}
  \shuffle_\Theta \left( \begin{bmatrix}t_k - 4\end{bmatrix}_{[a_1,[a_1,a_2]]}
    a_1^{\shuffle_\Theta\,t_k-4} \shuffle_\Theta [a_1,[a_1,a_2]]
  \right). 
\end{equation}
All these other terms do not contribute to $
\begin{bsmallmatrix}
  N-3
\end{bsmallmatrix}
_{[a_1,a_2]} a_1^{\shuffle_\Theta\,N-3} \shuffle_\Theta[a_1,a_2]$ in
the filtered word. Indeed, the only missing terms in
Eq.~(\ref{eq:wn-contribution-1}) that give words with $(N-2)$ letters
$a_1$ and one $a_2$ include one of the multi-commutator letters
$[a_1,\ldots[a_1,a_2]\ldots]$ in the shuffle product. Computing all
these shuffles and filtrating the resulting words will not give words
with $(N-3)$ letters $a_1$ and one $[a_1,a_2]$ but words with
multi-commutator letters.

Given a filtered word $w_N$, one regains the original unfiltered word
by {\it first computing all shuffle products in $w_N$ and then,
  computing all (multi-)commutators}. This can be seen by a look at
the filtration algorithm in Section \ref{sec:filtration-algorithm}.
For the first term in Eq.~(\ref{eq:wn-contribution-1}), this implies
that we must first compute the shuffle products in the bracket.
Secondly, we have to replace the commutator letter $[a_1,a_2]$ by
$a_1a_2-a_2a_1$ and finally, we must compute all remaining shuffle
products to obtain the respective terms in the unfiltered word $w_N$
\textbf{(}Eq.~(\ref{eq:wn-contribution-1})\textbf{)}.

Let us calculate the unfiltered $w_N$. We are only interested in words
that contribute to
$\begin{bsmallmatrix}\bullet\end{bsmallmatrix}_{[a_1,a_2]}$. Hence,
when computing the shuffle products in
Eq.~(\ref{eq:wn-contribution-1}), we shift all words with
$\Theta(\cdot,\cdot)$ letters to the $\ldots$ terms. In the following,
it is convenient to define the words
\begin{align}
  A(p,q) :=& \underbrace{a_1 \ldots a_1}_{p\times} a_2
  \underbrace{a_1\ldots a_1}_{q\times},\\
  B(p,q) :=& \underbrace{a_1 \ldots a_1}_{p\times} [a_1,a_2]
  \underbrace{a_1\ldots a_1}_{q\times}
\end{align}
for $p,q\in \mathbb{N}$. We note that
\begin{align}
  a_1^{\shuffle_\Theta p} \shuffle_\Theta [a_1,a_2] =& p!
  \textbf{(}A(p+1,0) - A(0,p+1)\textbf{)} + \ldots, \\
  a_1^{\shuffle_\Theta p}\shuffle_\Theta A(q, 0) =& p!\sum_{r = 0}^{p}
  \begin{pmatrix}
    q + r \\ r
  \end{pmatrix} A(q + r, p - r) + \ldots,\\
  a_1^{\shuffle_\Theta p}\shuffle_\Theta A(0, q) =& p!\sum_{r =
    0}^{p}
  \begin{pmatrix}
    q + r \\ r
  \end{pmatrix} A( p -
  r,q + r) + \ldots, \\
  a_1^{\shuffle_\Theta p} \shuffle_\Theta a_2 =& p! \sum_{q = 0}^{p}
  A(q,p-q) + \ldots
\end{align}
The dots collect all words that consist of other letters than one
$a_2$ and some $a_1$. Using these relations in
Eq.~(\ref{eq:wn-contribution-1}), we obtain
\begin{align}
  \label{eq:wn-contribution-2}
  w_N =& \sum_{k\geq 1} \sum_{\sum_i t_i = N-1} k
  \begin{bmatrix}
    t_1
  \end{bmatrix}
  \ldots
  \begin{bmatrix}
    t_{k-1}
  \end{bmatrix}
  \begin{bmatrix}
    t_k - 3
  \end{bmatrix}
  _{[a_1,a_2]} (t_k - 3)! (N-1-t_k)!
  \sum_{p=0}^{N-1-t_k}
  \begin{pmatrix}
    t_k - 2 + p \\ p
  \end{pmatrix} \nonumber\\
  &\times \textbf{(} A(t_k - 1 + p, N - 1 - t_k - p) - A( N - t_k - p,t_k -
    2 + p)
  \textbf{)}\nonumber\\
  &+ \sum_{k\geq 1} \sum_{\sum_i t_i = N-1} k
  \begin{bmatrix}
    t_1
  \end{bmatrix}
  \ldots
  \begin{bmatrix}
    t_{k-1}
  \end{bmatrix}
  \begin{bmatrix}
    t_k-2 & 1
  \end{bmatrix} (N-3)! \sum_{p=0}^{N-3} A(p + 1, N - 3 - p) \nonumber\\
  &+ \sum_{k\geq1}
  \begin{pmatrix}
    2 + k \\ k 
  \end{pmatrix} \sum_{\sum_i t_i = N-2}
  \begin{bmatrix}
    t_1
  \end{bmatrix}
  \ldots
  \begin{bmatrix}
    t_k
  \end{bmatrix}
  (N-2)!
  A(0,N-2) + \ldots
\end{align}

$w_N$ is still unfiltered. Now, the filtration algorithm brings the
words $A(p,q)$ into lexicographical order using the concatenation
commutator. In our case, $a_2$ is sorted to the right,
\begin{equation}
  A(p,q) = -B(p,q-1) + A(p+1,q-1), \qquad \Rightarrow A(p,q) =
  A(p+q,0) -\sum_{r = 0}^{q-1} B(p+r,q-1-r).
\end{equation}
In this section, a filtration algorithm that sorts $a_2$ to the left
would be more convenient because the last term in
Eq.~(\ref{eq:wn-contribution-2}) would already be given in
lexicographical order. Since the resulting filtered words must not
depend on the lexicographical order of letters, we now assume
throughout this section that we work with a filtration algorithm that
sorts $a_2$ to the left,
\begin{equation}
  \label{eq:sorting}
  A(p,q) = B(p-1,q) + A(p-1,q+1), \qquad \Rightarrow A(p,q) =
  A(0,p+q) + \sum_{r = 0}^{p-1} B(p-1-r,q+r).
\end{equation}
This change does not effect the final generating functions
$\begin{bsmallmatrix}\bullet\end{bsmallmatrix}_{[a_1,a_2]}$ and the
respective generated matrices.
Inserting Eq.~(\ref{eq:sorting}) into Eq.~(\ref{eq:wn-contribution-2})
yields
\begin{align}
  \label{eq:wn-contribution-3}
  w_N =& \sum_{k\geq 1} \sum_{\sum_i t_i = N-1} k
  \begin{bmatrix}
    t_1
  \end{bmatrix}
  \ldots
  \begin{bmatrix}
    t_{k-1}  
  \end{bmatrix}
  \begin{bmatrix}
    t_k - 3
  \end{bmatrix}
  _{[a_1,a_2]} (t_k - 3)! (N-1-t_k)!
  \sum_{p=0}^{N-1-t_k}
  \begin{pmatrix}
    t_k - 2 + p \\ p
  \end{pmatrix} \nonumber\\
  &\times \left( \sum_{q=0}^{t_k - 2 + p}B(t_k - 2 + p - q, N - 1 -
    t_k - p + q) - \sum_{q=0}^{N - t_k - p - 1}B( N - t_k - p - 1 -
    q,t_k - 2 + p + q)
  \right)\nonumber\\
  &+ \sum_{k\geq 1} \sum_{\sum_i t_i = N-1} k
  \begin{bmatrix}
    t_1
  \end{bmatrix}
  \ldots
  \begin{bmatrix}
    t_{k-1}
  \end{bmatrix}
  \begin{bmatrix}
    t_k-2 & 1
  \end{bmatrix} (N-3)! \sum_{p=0}^{N-3} \sum_{q = 0}^{p}B(p - q , N -
  3 - p + q) \nonumber\\
  &+ \sum_{k\geq 1} \sum_{\sum_i t_i = N-1} k
  \begin{bmatrix}
    t_1
  \end{bmatrix}
  \ldots
  \begin{bmatrix}
    t_{k-1}
  \end{bmatrix}
  \begin{bmatrix}
    t_k-2 & 1
  \end{bmatrix} (N-2)! A(0, N - 2) \nonumber\\
  &+ \sum_{k\geq1}
  \begin{pmatrix}
    2 + k \\ k 
  \end{pmatrix} \sum_{\sum_i t_i = N-2}
  \begin{bmatrix}
    t_1
  \end{bmatrix}
  \ldots
  \begin{bmatrix}
    t_k
  \end{bmatrix}
  (N-2)! A(0,N-2) + \ldots
\end{align}
The last two terms together are equal to $
\begin{Bsmallmatrix}
  N - 2 & 1
\end{Bsmallmatrix} A(0,N-2) $, see Section \ref{sec:bullet2}.

Step 2 of the filtration loop in Section \ref{sec:pres-filtr-algor}
computes
\begin{equation}
  a_1^{\shuffle_\Theta N-2}\shuffle_\Theta a_2 = (N-2)!
  \sum_{p=0}^{N-2} A(p,N-2-p) + \ldots
\end{equation}
and brings all the words $A(p,N-2-p)$ on the rhs into lexicographical
order using Eq.~(\ref{eq:sorting}). Hence, in
Eq.~(\ref{eq:wn-contribution-3}), $A(0,N-2)$ is replaced by
\begin{equation}
  A(0,N-2) = \frac{1}{(N-1)!} a_1^{\shuffle_\Theta N-2}\shuffle_\Theta
  a_2 - \frac{1}{N-1} \sum_{p=1}^{N-2}\sum_{q = 0}^{p-1}
  B(p-1-q,N-2-p+q).
\end{equation}
Thus, after step 2 of the respective filtration loop,
\begin{align}
  \label{eq:wn-contribution-4}
  w_N =& \sum_{k\geq 1} \sum_{\sum_i t_i = N-1} k
  \begin{bmatrix}
    t_1
  \end{bmatrix}
  \ldots
  \begin{bmatrix}
    t_{k-1}
  \end{bmatrix}
  \begin{bmatrix}
    t_k - 3
  \end{bmatrix}
  _{[a_1,a_2]} (t_k - 3)! (N-1-t_k)!
  \sum_{p=0}^{N-1-t_k}
  \begin{pmatrix}
    t_k - 2 + p \\ p
  \end{pmatrix} \nonumber\\
  &\times \left( \sum_{q=0}^{t_k - 2 + p}B(t_k - 2 + p - q, N - 1 -
    t_k - p + q) - \sum_{q=0}^{N - t_k - p - 1}B( N - t_k - p - 1 -
    q,t_k - 2 + p + q)
  \right)\nonumber\\
  &+ \sum_{k\geq 1} \sum_{\sum_i t_i = N-1} k
  \begin{bmatrix}
    t_1
  \end{bmatrix}
  \ldots
  \begin{bmatrix}
    t_{k-1}
  \end{bmatrix}
  \begin{bmatrix}
    t_k-2 & 1
  \end{bmatrix} (N-3)! \sum_{p=0}^{N-3} \sum_{q = 0}^{p}B(p - q , N -
  3 - p + q) \nonumber\\
  &+
  \begin{bmatrix}
    N-2 & 1
  \end{bmatrix} a_1^{\shuffle_\Theta N-2}\shuffle_\Theta a_2 -
  \frac{1}{N-1}
  \begin{Bmatrix}
    N-2 & 1
  \end{Bmatrix} \sum_{p=1}^{N-2}\sum_{q = 0}^{p-1} B(p-1-q,N-2-p+q) +
  \ldots
\end{align}
Note that we do not take the filtered term $\begin{bsmallmatrix} N-2 & 1
\end{bsmallmatrix} a_1^{\shuffle_\Theta N-2}\shuffle_\Theta a_2$ into
account because it does not contribute to the term in
Eq.~(\ref{eq:cterm}).

We now proceed as in the derivation of the master differential
equation in Section \ref{sec:deriv-mast-equat}. We denote by $
\begin{Bsmallmatrix}
  N-3
\end{Bsmallmatrix}
_{[a_1,a_2]}$ the number of words on the rhs of
Eq.~(\ref{eq:wn-contribution-4}) with $(N-3)$ letters $a_1$ and one
letter $[a_1,a_2]$. As in Section \ref{sec:notation},
$
\begin{Bsmallmatrix}
  N-3
\end{Bsmallmatrix}
_{[a_1,a_2]}$ is related to the indexed matrix $
\begin{bsmallmatrix}
  N-3
\end{bsmallmatrix}
_{[a_1,a_2]}$ by
\begin{equation}
  \label{eq:curly-square-4}
  \begin{bmatrix}
    N-3
  \end{bmatrix}
  _{[a_1,a_2]} = \frac{1}{(N-2)!}
  \begin{Bmatrix}
    N-3
  \end{Bmatrix}
  _{[a_1,a_2]}.
\end{equation}
We obtain $
\begin{Bsmallmatrix}
  N-3
\end{Bsmallmatrix}
_{[a_1,a_2]}$ by setting all words $B(\cdot,\cdot)$ on the rhs of
Eq.~(\ref{eq:wn-contribution-4}) to 1, hence
\begin{align}
  \label{eq:wn-contribution-5}
  \begin{Bmatrix}
    N-3
  \end{Bmatrix}
  _{[a_1,a_2]} =& \sum_{k\geq 1} \sum_{\sum_i t_i = N-1} \!\!\!\!\!\!\! k
  \begin{bmatrix}
    t_1
  \end{bmatrix}
  \ldots
  \begin{bmatrix}
    t_{k-1}
  \end{bmatrix}\!
  \begin{bmatrix}
    t_k - 3
  \end{bmatrix}
  _{[a_1,a_2]} (t_k - 3)! (N-1-t_k)! \!\!\!
  \sum_{p=0}^{N-1-t_k}\!\!
  \begin{pmatrix}
    t_k - 2 + p \\ p
  \end{pmatrix} \! ( 2t_k + 2p - N - 1)\nonumber\\
  &+ \sum_{k\geq 1} \sum_{\sum_i t_i = N-1} \frac{(N-1)!}{2} k
  \begin{bmatrix}
    t_1
  \end{bmatrix}
  \ldots
  \begin{bmatrix}
    t_{k-1}
  \end{bmatrix}
  \begin{bmatrix}
    t_k-2 & 1
  \end{bmatrix}  - \frac{N-2}{2}
  \begin{Bmatrix}
    N-2 & 1
  \end{Bmatrix}.
\end{align}
Standard combinatorial calculation yields
\begin{equation}
  \label{eq:combinatorics}
  (t_k-3)!(N-1-t_k)! \sum_{p=0}^{N-1-t_k}
  \begin{pmatrix}
    t_k-2+p \\ p
  \end{pmatrix}(2t_k + 2p - N - 1) = \frac{(N-1)!}{t_k(t_k-1)}.
\end{equation}
Furthermore, we find $
\begin{Bsmallmatrix}
  N-2 & 1
\end{Bsmallmatrix} = (N-1)!
\begin{bsmallmatrix}
  N - 2 & 1
\end{bsmallmatrix}
$ from Eq.~(\ref{eq:curly-square-2}). We divide
Eq.~(\ref{eq:wn-contribution-5}) by $(N-1)!$ and insert
Eq.~(\ref{eq:curly-square-4}) on the lhs. This results in
\begin{align}
  \frac{1}{N-1}
  \begin{bmatrix}
    N-3
  \end{bmatrix}
  _{[a_1,a_2]} =& \sum_{k\geq 1} \sum_{\sum_i t_i =
    N-1} k
  \begin{bmatrix}
    t_1
  \end{bmatrix}
  \ldots
  \begin{bmatrix}
    t_{k-1}
  \end{bmatrix}
  \frac{1}{t_k(t_k-1)}
  \begin{bmatrix}
    t_k - 3
  \end{bmatrix}
  _{[a_1,a_2]} \nonumber\\
  &+ \sum_{k\geq 1} \sum_{\sum_i t_i = N-1} \frac{1}{2} k
  \begin{bmatrix}
    t_1
  \end{bmatrix}
  \ldots
  \begin{bmatrix}
    t_{k-1}
  \end{bmatrix}
  \begin{bmatrix}
    t_k-2 & 1
  \end{bmatrix} - \frac{N-2}{2}
  \begin{bmatrix}
    N-2 & 1
  \end{bmatrix}.
\end{align}
Finally, we multiply with $z^{N-1}$ and sum over all $N\in
\mathbb{N}$. This will give us
$\int \begin{bsmallmatrix}\bullet\end{bsmallmatrix}_{[a_1,a_2]}
\mathrm{d}z$ on the lhs with zero integration constant. On the rhs, we
obtain
\begin{align}
  \label{eq:dgl-commutator}
  \int \begin{bmatrix}\bullet\end{bmatrix}_{[a_1,a_2]} \mathrm{d}z \;=\; &
  \sum_{N=0}^\infty \Bigg( 
  \sum_{k\geq 1} \sum_{\sum_i t_i = N-1} k
  \begin{bmatrix}
    t_1
  \end{bmatrix}
  z^{t_1} \ldots
  \begin{bmatrix}
    t_{k-1}
  \end{bmatrix}
  z^{t_{k-1}}
  \frac{1}{t_k(t_k-1)}
  \begin{bmatrix}
    t_k - 3
  \end{bmatrix}
  _{[a_1,a_2]}z^{t_k} \nonumber\\
  &+ \sum_{k\geq 1} \sum_{\sum_i t_i = N-1} \frac{z}{2} k
  \begin{bmatrix}
    t_1
  \end{bmatrix}
  z^{t_1}\ldots
  \begin{bmatrix}
    t_{k-1}
  \end{bmatrix}
  z^{t_{k-1}}
  \begin{bmatrix}
    t_k-2 & 1
  \end{bmatrix} z^{t_k - 1} -
  \frac{z^2}{2}\frac{\mathrm{d}}{\mathrm{d}z}\frac{1}{z}
  \begin{bmatrix}
    N-2 & 1
  \end{bmatrix}z^{N-1} \Bigg) \nonumber\\
 \; =\; & \sum_{k\geq 1} k \begin{bmatrix}\bullet\end{bmatrix}^{k-1} \int
  \left(
    \int \begin{bmatrix}\bullet\end{bmatrix}_{[a_1,a_2]} \mathrm{d}z
  \right) \mathrm{d}z + \sum_{k\geq 1} \frac{z}{2} k
  \begin{bmatrix}\bullet\end{bmatrix}^{k-1}
  \begin{bmatrix}
    \bullet & 1
  \end{bmatrix} -
  \frac{z^2}{2}\frac{\mathrm{d}}{\mathrm{d}z}\frac{1}{z}
  \begin{bmatrix}
    \bullet & 1
  \end{bmatrix} \nonumber\\
 \;  = \; & \frac{1}{1-2z}\int \left(
    \int \begin{bmatrix}\bullet\end{bmatrix}_{[a_1,a_2]} \mathrm{d}z
  \right) \mathrm{d}z + \frac{1}{4\sqrt{1-2z}} -
  \frac{1}{4\sqrt{1-2z}^3} + \frac{1}{2\sqrt{1-2z}}\log \left(
    \frac{1}{\sqrt{1-2z}} \right),
\end{align}
where we used the explicit form of the generating functions
$\begin{bsmallmatrix}\bullet\end{bsmallmatrix}$ and $
\begin{bsmallmatrix}
  \bullet & 1
\end{bsmallmatrix}
$ in the last line
\textbf{(}Eqs.~(\ref{eq:bullet1},\ref{eq:bullet2})\textbf{)}. Again,
the integration constants are zero.

Eq.~(\ref{eq:dgl-commutator}) is an ordinary first order differential
equation for $\int \left(
  \int \begin{bsmallmatrix}\bullet\end{bsmallmatrix}_{[a_1,a_2]}
  \mathrm{d}z \right) \mathrm{d}z$ with the same homogeneous part as
in the case of index-free generating functions, see Section
\ref{sec:deriv-mast-equat}. We can thus, use
Eq.~(\ref{eq:generating-function-final}) to obtain
\begin{equation}
\int  \left( \int \begin{bmatrix}\bullet\end{bmatrix}_{[a_1,a_2]}
    \mathrm{d}z \right) 
  \mathrm{d}z = -\frac{\sqrt{1-2z}}{4} + \frac{1}{4\sqrt{1-2z}} -
  \frac{\sqrt{1-2z}}{4}\log \left( \frac{1}{\sqrt{1-2z}} \right) -
  \frac{1}{4\sqrt{1-2z}}\log \left( \frac{1}{\sqrt{1-2z}} \right).
\end{equation}
The second derivative finally results in the generating function
\begin{equation}
  \label{eq:bullet10}
  \begin{bmatrix}\bullet\end{bmatrix}_{[a_1,a_2]} =
  \frac{1}{4\sqrt{1-2z}^3} - 
  \frac{1}{4\sqrt{1-2z}^5} + \frac{1}{4\sqrt{1-2z}^3}\log \left(
    \frac{1}{\sqrt{1-2z}} \right) - \frac{3}{4\sqrt{1-2z}^5}\log
  \left( \frac{1}{\sqrt{1-2z}} \right).
\end{equation}

\subsubsection{The generating function
  $\protect \begin{bsmallmatrix} \bullet
    \protect\end{bsmallmatrix}_{[a_1,\Theta(a_1,a_1)]}$}

In this section, we derive the generating function
$\begin{bsmallmatrix}\bullet\end{bsmallmatrix}_{[a_1,\Theta(a_1,a_1)]}$
for the matrices $
\begin{bsmallmatrix}
  N-3
\end{bsmallmatrix}
_{[a_1,\Theta(a_1,a_1)]}$. These belong to the shuffle products
$a_1^{\shuffle_\Theta \, N-3} \shuffle_\Theta [a_1,\Theta(a_1,a_1)]$
in the filtered word $w_N$.

In full analogy to the previous section, we derive an equivalent to
Eq.~(\ref{eq:wn-contribution-5}). In the partly filtered $w_N$, we
denote the number of words with $(N-3)$ letters $a_1$ and one letter
$[a_1,\Theta(a_1,a_1)]$ by
\begin{equation}
  \label{eq:curly-square-5}
  \begin{Bmatrix}
    N-3
  \end{Bmatrix}
  _{[a_1,\Theta(a_1,a_1)]} = (N-2)!
  \begin{bmatrix}
    N-3
  \end{bmatrix}
  _{[a_1,\Theta(a_1,a_1)]}.
\end{equation}
We then find
\begin{align}
  \label{eq:wn-contribution-6}
  \begin{Bmatrix}
    N-3
  \end{Bmatrix}
  _{[a_1,\Theta(a_1,a_1)]} =& \sum_{k\geq 1} \sum_{\sum_i t_i =
    N-1} k
  \begin{bmatrix}
    t_1
  \end{bmatrix}
  \ldots
  \begin{bmatrix}
    t_{k-1}
  \end{bmatrix}
  \begin{bmatrix}
    t_k - 3
  \end{bmatrix}
  _{[a_1,\Theta(a_1,a_1)]}
  (t_k - 3)! (N-1-t_k)! \sum_{p=0}^{N-1-t_k}
  \begin{pmatrix}
    t_k - 2 + p \\ p
  \end{pmatrix} \nonumber\\
  &\times ( 2t_k + 2p - N - 1) + \sum_{k\geq 1} \sum_{\sum_i t_i =
    N-1} \frac{(N-1)!}{2} k
  \begin{bmatrix}
    t_1
  \end{bmatrix}
  \ldots
  \begin{bmatrix}
    t_{k-1}
  \end{bmatrix}
  \begin{bmatrix}
    t_k-2 \\ 2
  \end{bmatrix} \nonumber\\
  &+ \sum_{k\geq 1} \sum_{\sum_i t_i = N-1}
  \mathcal{S}
  \left(
    \begin{bmatrix}
      N-3 \\ 2
    \end{bmatrix},
    \begin{bmatrix}
      N-1
    \end{bmatrix}
  \right)\frac{(N-1)!}{2}
  \begin{bmatrix}
    t_1
  \end{bmatrix}
  \ldots
  \begin{bmatrix}
    t_k
  \end{bmatrix}
  -
  \frac{N-2}{2}
  \begin{Bmatrix}
    N-2 \\ 2
  \end{Bmatrix}.
\end{align}
Compared to Eq.~(\ref{eq:wn-contribution-5}), we made the obvious
replacements
\begin{equation}
  \begin{bmatrix}
    t_k - 3  
  \end{bmatrix}
  _{[a_1,a_2]} \to
  \begin{bmatrix}
    t_k - 3
  \end{bmatrix}
  _{[a_1,\Theta(a_1,a_1)]},
  \qquad \begin{bmatrix} 
    t_k-2 & 1
  \end{bmatrix} \to \begin{bmatrix} t_k-2 \\ 2
  \end{bmatrix}, \qquad \begin{Bmatrix} N-2 & 1
  \end{Bmatrix} \to
  \begin{Bmatrix}
    N-2 \\ 2
  \end{Bmatrix}.
\end{equation}
The only new term is the third one. It arises because on the rhs of
Eq.~(\ref{eq:wn-contribution-1}), one must also consider the term
\begin{equation}
  B_+^{a_1} \sum_{k\geq 1} \sum_{\sum_i t_i = N-1}
  \begin{bmatrix}
    t_1
  \end{bmatrix}
  a_1^{\shuffle_\Theta t_1} \ldots
  \begin{bmatrix}
    t_k
  \end{bmatrix}
  a_1^{\shuffle_\Theta t_k}
\end{equation}
and calculate
\begin{equation}
  a_1^{\shuffle_\Theta\,N-1} = (N-1)!\underbrace{a_1\ldots
    a_1}_{(N-1)\times} \,+\, \mathcal{S}
  \left(
    \begin{bmatrix}
      N-3 \\ 2
    \end{bmatrix},
    \begin{bmatrix}
      N-1
    \end{bmatrix}
  \right)
  a_1^{\shuffle_\Theta\,N-3}\shuffle_\Theta \Theta(a_1,a_1).
\end{equation}
We divide Eq.~(\ref{eq:wn-contribution-6}) by $(N-1)!$ and use
Eqs.~(\ref{eq:curly-square-2},\ref{eq:combinatorics},\ref{eq:curly-square-5})
to obtain
\begin{align}
  \frac{1}{N-1}
  \begin{bmatrix}
    N-3
  \end{bmatrix}
  _{[a_1,\Theta(a_1,a_1)]} =& \sum_{k\geq 1}
  \sum_{\sum_i t_i = N-1} k
  \begin{bmatrix}
    t_1
  \end{bmatrix}
  \ldots
  \begin{bmatrix}
    t_{k-1}
  \end{bmatrix}
  \frac{1}{t_k(t_k-1)}
  \begin{bmatrix}
    t_k - 3
  \end{bmatrix}
  _{[a_1,\Theta(a_1,a_1)]} \nonumber\\
  +& \sum_{k\geq 1} \sum_{\sum_i t_i = N-1} \frac{1}{2} k
  \begin{bmatrix}
    t_1
  \end{bmatrix}
  \ldots
  \begin{bmatrix}
    t_{k-1}
  \end{bmatrix}
  \begin{bmatrix}
    t_k-2 \\ 2
  \end{bmatrix}\nonumber\\
  +& \sum_{k\geq 1} \sum_{\sum_i t_i = N-1} \mathcal{S} \left(
    \begin{bmatrix}
      N-3 \\ 2
    \end{bmatrix},
    \begin{bmatrix}
      N-1
    \end{bmatrix}
  \right)\frac{1}{2}
  \begin{bmatrix}
    t_1
  \end{bmatrix}
  \ldots
  \begin{bmatrix}
    t_k
  \end{bmatrix}
  \nonumber\\
  -&
  \frac{N-2}{2}
  \begin{bmatrix}
    N-2 \\ 2
  \end{bmatrix} - \frac{N-2}{2} \mathcal{S} \left(
    \begin{bmatrix}
      N-2 \\ 2
    \end{bmatrix},
    \begin{bmatrix}
      N
    \end{bmatrix}
  \right)
  \begin{bmatrix}
    N
  \end{bmatrix}
  .
\end{align}

As in the previous section, we multiply with $z^{N-1}$ and sum over
all $N\in \mathbb{N}$. With 
\begin{equation}
  \mathcal{S} \left(
  \begin{bmatrix}
    N-2 \\ 2
  \end{bmatrix},
  \begin{bmatrix}
    N
  \end{bmatrix}
\right) = \frac{N(N-1)}{2},
\end{equation}
we find
\begin{align}
  \label{eq:dgl-commutator-2}
  \int \begin{bmatrix}\bullet\end{bmatrix}_{[a_1,\Theta(a_1,a_1)]}
  \mathrm{d}z =&\, 
  \sum_{N=0}^\infty \Bigg( \sum_{k\geq 1} \sum_{\sum_i t_i = N-1} k
  \begin{bmatrix}
    t_1
  \end{bmatrix}
  z^{t_1} \ldots
  \begin{bmatrix}
    t_{k-1}
  \end{bmatrix}
  z^{t_{k-1}} \frac{1}{t_k(t_k-1)}
  \begin{bmatrix}
    t_k - 3
  \end{bmatrix}
  _{[a_1,\Theta(a_1,a_1)]}z^{t_k}
  \nonumber\\
  &+ \sum_{k\geq 1} \sum_{\sum_i t_i = N-1} \frac{z}{2} k
  \begin{bmatrix}
    t_1
  \end{bmatrix}
  z^{t_1}\ldots
  \begin{bmatrix}
    t_{k-1}
  \end{bmatrix}
  z^{t_{k-1}}
  \begin{bmatrix}
    t_k-2 \\ 2
  \end{bmatrix} z^{t_k - 1} \nonumber\\
  & + \sum_{k\geq 1} \sum_{\sum_i t_i = N-1}
  \frac{z^2}{4}\frac{\mathrm{d^2}}{\mathrm{d}z^2}
  \begin{bmatrix}
    t_1
  \end{bmatrix}
  z^{t_1}\ldots
  \begin{bmatrix}
    t_k
  \end{bmatrix}
  z^{t_k} -
  \frac{z^2}{2}\frac{\mathrm{d}}{\mathrm{d}z}\frac{1}{z}
  \begin{bmatrix}
    N-2 \\ 2
  \end{bmatrix}z^{N-1} - \frac{z^2}{4}
  \frac{\mathrm{d}^3}{\mathrm{d}z^3}
  \begin{bmatrix}
    N
  \end{bmatrix}
  z^N \Bigg) \nonumber\\
  =&\, \sum_{k\geq 1} k \begin{bmatrix}\bullet\end{bmatrix}^{k-1} \int
  \left( \int
    \begin{bmatrix}\bullet\end{bmatrix}_{[a_1,\Theta(a_1,a_1)]}
    \mathrm{d}z \right) \mathrm{d}z + \sum_{k\geq 1} \frac{z}{2}
  k \begin{bmatrix}\bullet\end{bmatrix}^{k-1}
  \begin{bmatrix}
    \bullet \\ 2
  \end{bmatrix} + \sum_{k\geq1} \frac{z^2}{4}
  \frac{\mathrm{d}^2}{\mathrm{d}z^2} \begin{bmatrix}\bullet\end{bmatrix}^k
  \nonumber\\ 
  &- \frac{z^2}{2}\frac{\mathrm{d}}{\mathrm{d}z}\frac{1}{z}
  \begin{bmatrix}
    \bullet \\ 2
  \end{bmatrix} - \frac{z^2}{4}
  \frac{\mathrm{d}^3}{\mathrm{d}z^3}\begin{bmatrix}\bullet\end{bmatrix}
  \nonumber\\ 
  =&\, \frac{1}{1-2z}\int \left(
    \int \begin{bmatrix}\bullet\end{bmatrix}_{[a_1,\Theta(a_1,a_1)]}
    \mathrm{d}z \right) \mathrm{d}z - \frac{1}{8\sqrt{1-2z}} +
  \frac{1}{8\sqrt{1-2z}^3} - \frac{1}{4\sqrt{1-2z}}\log \left(
    \frac{1}{\sqrt{1-2z}} \right).
\end{align}
Note that the third and the fifth term on the rhs of the second
equation cancel. This is an interesting incidence. Because of
Eq.~(\ref{eq:bullet5bullet2}), the inhomogeneous parts of the
differential equations
Eqs.~(\ref{eq:dgl-commutator},\ref{eq:dgl-commutator-2}) only differ
by a factor of $-1/2$. We therefore obtain
\begin{equation}
  \label{eq:bullet11}
  \begin{bmatrix}\bullet\end{bmatrix}_{[a_1,\Theta(a_1,a_1)]} =
  -\frac{1}{2} 
  \begin{bmatrix}\bullet\end{bmatrix}_{[a_1,a_2]} =
  -\frac{1}{8\sqrt{1-2z}^3} + 
  \frac{1}{8\sqrt{1-2z}^5} - \frac{1}{8\sqrt{1-2z}^3}\log \left(
    \frac{1}{\sqrt{1-2z}} \right) + \frac{3}{8\sqrt{1-2z}^5}\log
  \left( \frac{1}{\sqrt{1-2z}} \right).
\end{equation}

\subsection{Results}
\label{sec:results}

We now demonstrate the power of the generating functions derived in
the previous sections: one can write the next-to$^{\{j\}}$-leading log
order as a function of terms up to $\mathcal{O}(\alpha^{j+1})$ in the
log-expansion \textbf{(}Eq.~(\ref{eq:log-expansion})\textbf{)}. We
will show this for the Yukawa fermion propagator
\textbf{(}Eq.~(\ref{eq:DSE-Yuk})\textbf{)} up to $j \leq 2$ using the
explicit generating functions obtained in the previous section. They are also collected in the second column of Table
\ref{tab:results}. We discuss our results for $j = 0,1,2$ separately.

\subsubsection{Leading log expansion}
\label{sec:leading-log-order}

Consider the filtered solution $W_\mathrm{Yuk}$ of the DSE
Eq.~(\ref{eq:DSE-Yuk-words}), see Eq.~(\ref{eq:DSE-solution}). The
leading log order is
\begin{equation}
  W_\mathrm{Yuk}|_\mathrm{l.l.} = \sum_{n\geq1} \left(\alpha^n
    w_n^\mathrm{Yuk}\right)\big|_\mathrm{l.l.}.
\end{equation}
Contributing terms in the filtered words $w_n$ map to $L^n$ under
renormalized Feynman rules since the leading log order is $\propto
\alpha^n L^n$. These are only the full shuffle products
$a_1^{\shuffle_\Theta\, n}$ \textbf{(}see Section
\ref{sec:renormalized-feynman-rules} that
$\Psi_R\left(a_1^{\shuffle_\Theta\, n}\right) \propto L^n$\textbf{)}.
Thus,
\begin{equation}
  W_\mathrm{Yuk}|_\mathrm{l.l.} = \sum_{n\geq0} \alpha^n
  \begin{bmatrix}
    n
  \end{bmatrix}
  a_1^{\shuffle_\Theta\,n} = \alpha a_1 + \frac{1}{2} \alpha^2 a_1
  \shuffle_\Theta a_1 + \frac{1}{2} \alpha^3 a_1^{\shuffle_\Theta 3} +
  \frac{5}{8} \alpha^4 a_1^{\shuffle_\Theta 4} + \frac{7}{8} \alpha^5
  a_1^{\shuffle_\Theta 5} + \ldots
\end{equation}
See the first row of Table \ref{tab:yukawa} for the explicit
multiplicities. Acting with renormalized Feynman rules $\Psi_R$ on
both sides results in
\begin{equation}
  \label{eq:psiW-ll}
  \Psi_R\left(W_\mathrm{Yuk}\right)\big|_\mathrm{l.l.} = \sum_{n\geq0}
  \begin{bmatrix}
    n
  \end{bmatrix}
  \alpha^n \Psi_R(a_1)^{n} = \begin{bmatrix}\bullet\end{bmatrix}\big|_{z
    \to \alpha\Psi_R(a_1)}. 
\end{equation}

We write this equation in terms of Feynman graphs. Therefore, set
$\Psi_R = \Phi_R \circ \Upsilon^{-1}$. We find that $\Psi_R(a_1) =
\Phi_R(\Gamma_1)$ on the rhs using the properties of the Hopf algebra
morphism $\Upsilon^{-1}$ in
Eqs.~(\ref{eq:morphism-begin}-\ref{eq:morphism-end}). On the lhs,
$\Psi_R(W_\mathrm{Yuk}) = \Phi_R(X_\mathrm{Yuk}) \equiv
G_R(X_\mathrm{Yuk})$, which is the full Green function of the fermion
propagator. Hence, Eq.~(\ref{eq:psiW-ll}) yields
\begin{equation}
  \label{eq:GR-ll1}
  G_R(X_\mathrm{Yuk}) \big|_{l.l.}
  = \begin{bmatrix}\bullet\end{bmatrix}\big|_{z \to
    \alpha\Phi_R(\Gamma_1)}. 
\end{equation}
Using Eq.~(\ref{eq:bullet1}) for the generating function
$\begin{bsmallmatrix}\bullet\end{bsmallmatrix}$, we finally obtain
\begin{equation}
  \label{eq:GR-ll2}
  G_R(X_\mathrm{Yuk}) \big|_{l.l.} = 1 - \sqrt{1-2\alpha\Phi_R(\Gamma_1)}.
\end{equation}
Without this result, the computation of $G_R(X_\mathrm{Yuk})
\big|_{l.l.}$ would be quite more complicated, even impossible.
Computing $G_R(X_\mathrm{Yuk}) \big|_{l.l.}$ the ordinary way includes
to calculate an infinite number of Feynman integrals with any number
of loops. For example, the graphs $B_+^{\Gamma_1} \left(
  B_+^{\Gamma_1} \left( B_+^{\Gamma_1}(\ldots) \right) \right)$
contribute to $G_R(X_\mathrm{Yuk}) \big|_{l.l.}$. Using our formula in
Eq.~(\ref{eq:GR-ll2}), {\it we only need to compute the one-loop Feynman
  integral $\Phi_R(\Gamma_1)$ to derive the full leading log order
  Green function $G_R(X_\mathrm{Yuk}) \big|_{l.l.}$.}

\subsubsection{Next-to-leading log expansion}
\label{sec:next-leading-log}

The next-to-leading log order of Eq.~(\ref{eq:DSE-solution}) is
\begin{equation}
  W_\mathrm{Yuk}|_\mathrm{n.l.l.} = \sum_{n\geq1} (\alpha^n
  w_n^\mathrm{Yuk})\big|_\mathrm{n.l.l.}.
\end{equation}
Contributing terms of the filtered words $w_n$ map to $L^{n-1}$ under
renormalized Feynman rules since the next-to-leading log order is
$\propto \alpha^n L^{n-1}$. These are the full shuffle products
$a_1^{\shuffle_\Theta\, (n-2)}\shuffle_\Theta a_2$ and
$a_1^{\shuffle_\Theta\, (n-2)}\shuffle_\Theta \Theta(a_1,a_1)$.
Indeed, renormalized Feynman rules are character-like. $\Psi_R$ acting
on a full shuffle product of $n-1$ letters is $\propto L^{n-1}$ in the
log-expansion, see Section \ref{sec:renormalized-feynman-rules}. Thus,
\begin{equation}
  W_\mathrm{Yuk}\big|_\mathrm{n.l.l.} = \sum_{n\geq2} \left(
  \begin{bmatrix}
    n - 2 & 1
  \end{bmatrix} (\alpha a_1)^{\shuffle_\Theta\, (n-2)}\shuffle_\Theta
  \left(\alpha^2 a_2\right) +
  \begin{bmatrix}
    n - 2 \\ 2
  \end{bmatrix} (\alpha a_1)^{\shuffle_\Theta\, (n-2)}\shuffle_\Theta
  \left(\alpha^2 \Theta(a_1,a_1)\right) \right).
\end{equation}
Acting with renormalized Feynman rules $\Psi_R$ on both sides results
in
\begin{align}
  \Psi_R\left(W_\mathrm{Yuk}\right)\big|_\mathrm{n.l.l.} =&
  \sum_{n\geq2} \left(
    \begin{bmatrix}
      n - 2 & 1
    \end{bmatrix}
    \alpha^{n-2} \Psi_R(a_1)^{n-2} \alpha^2\Psi_R(a_2)
    + \begin{bmatrix} n - 2 \\ 2
    \end{bmatrix}
    \alpha^{n-2}\Psi_R(a_1)^{n-2}
    \alpha^2\Psi_R\textbf{(}\Theta(a_1,a_1)\textbf{)}
  \right) \nonumber\\
  =&
  \begin{bmatrix} \bullet & 1
  \end{bmatrix}\big|_{z \to \alpha\Psi_R(a_1)} \alpha^2\Psi_R(a_2) +
  \begin{bmatrix}
    \bullet \\ 2
  \end{bmatrix}\bigg|_{z \to \alpha\Psi_R(a_1)}
  \alpha^2\Psi_R\textbf{(}\Theta(a_1,a_1)\textbf{)}.
\end{align}

Again, we write $\Psi_R = \Phi_R \circ \Upsilon^{-1}$ and obtain the
full next-to-leading log order renormalized Green function of the
Yukawa fermion propagator on the lhs. On the rhs, the only subtle
point is that $\Theta(a_1,a_1)$ has no single corresponding Feynman
graph. However, we find the period
\begin{equation}
  \label{eq:PsiTheta1}
  \Psi_R\textbf{(}\Theta(a_1,a_1)\textbf{)} = \Phi_R \circ
  \Upsilon^{-1}\textbf{(}\Theta(a_1,a_1)\textbf{)} =
  \Phi_R \circ \Upsilon^{-1} \left( (a_1\shuffle_\Theta a_1) - 2
    B_+^{a_1}(a_1)\right) = \Phi_R
  (\Gamma_1)^2 - 2 \Phi_R
  \left(B_+^{\Gamma_1}(\Gamma_1)\right).
\end{equation}
Thus,
\begin{equation}
  \label{eq:GR-nll1}
  G_R(X_\mathrm{Yuk}) \big|_{n.l.l.} = \begin{bmatrix} \bullet & 1
  \end{bmatrix}\big|_{z \to \alpha\Phi_R(\Gamma_1)}
  \alpha^2\Phi_R(\Gamma_2) +
    \begin{bmatrix}
      \bullet \\ 2
    \end{bmatrix}\bigg|_{z \to \alpha\Phi_R(\Gamma_1)}
    \alpha^2 \left( \Phi_R (\Gamma_1)^2 - 2 \Phi_R
      \left(B_+^{\Gamma_1}(\Gamma_1)\right)\right).
\end{equation}
Using the generating functions in
Eqs.~(\ref{eq:bullet2},\ref{eq:bullet5}), we finally derive
\begin{equation}
  G_R(X_\mathrm{Yuk}) \big|_{n.l.l.} =
  \frac{\alpha^2}{\sqrt{1-2\alpha\Phi_R(\Gamma_1)}} \log \left(
    \frac{1}{\sqrt{1-2\alpha\Phi_R(\Gamma_1)}} \right) \left(
    \Phi_R(\Gamma_2) + \Phi_R \left(B_+^{\Gamma_1}(\Gamma_1)\right) -
    \frac{1}{2}\Phi_R(\Gamma_1)^2 \right).
\end{equation}
This is an enormous simplification: {\it we only need to compute the
  one-loop Feynman integral $\Phi_R(\Gamma_1)$ as well as the two-loop
  integrals $\Phi_R(\Gamma_2)$ and $\Phi_R
  \left(B_+^{\Gamma_1}(\Gamma_1)\right)$ to calculate the full
  next-to-leading log order Green function $G_R(X_\mathrm{Yuk})
  \big|_{n.l.l.}$.}

\subsubsection{Next-to-next-to-leading log expansion}
\label{sec:next-tonext-leading}

The next-to-next-to-leading log order of Eq.~(\ref{eq:DSE-solution})
is
\begin{equation}
  \label{eq:W-nnll}
  W_\mathrm{Yuk}|_\mathrm{n.n.l.l.} = \sum_{n\geq1} (\alpha^n
  w_n^\mathrm{Yuk})\big|_\mathrm{n.n.l.l.}.
\end{equation}
Contributing terms of the filtered words $w_n$ must map to $L^{n-2}$
under renormalized Feynman rules since the next-next-to-leading log
order is $\propto \alpha^n L^{n-2}$. These are the full shuffle
products
\begin{gather}
  a_1^{\shuffle_\Theta\, (n-3)}\shuffle_\Theta a_3, \qquad
  a_1^{\shuffle_\Theta\, (n-4)}\shuffle_\Theta
  a_2^{\shuffle_\Theta\, 2},\\
  a_1^{\shuffle_\Theta\, (n-3)}\shuffle_\Theta \Theta(a_1,a_1,a_1),
  \qquad a_1^{\shuffle_\Theta\, (n-3)}\shuffle_\Theta
  \Theta(a_1,a_2),\\ a_1^{\shuffle_\Theta\, (n-4)}\shuffle_\Theta
  \Theta(a_1,a_1)^{\shuffle_\Theta \, 2}, \qquad
  a_1^{\shuffle_\Theta\, (n-4)}\shuffle_\Theta a_2 \shuffle_\Theta
  \Theta(a_1,a_1), \\
  a_1^{\shuffle_\Theta\, (n-3)}\shuffle_\Theta [a_1,a_2], \qquad
  a_1^{\shuffle_\Theta\, (n-3)}\shuffle_\Theta [a_1,\Theta(a_1,a_1)].
\end{gather}
>From Eq.~(\ref{eq:W-nnll}),
\begin{align}
  W_\mathrm{Yuk}|_\mathrm{n.n.l.l.} =& \sum_{n\geq0} \Bigg(
  \begin{bmatrix}
    n - 3 & 0 & 1
  \end{bmatrix}
  (\alpha a_1)^{\shuffle_\Theta\, (n-3)}\shuffle_\Theta \left(
    \alpha^3 a_3 \right) +
  \begin{bmatrix}
    n - 4 & 2
  \end{bmatrix} (\alpha a_1)^{\shuffle_\Theta\, (n-4)}\shuffle_\Theta
  \left( \alpha^2 a_2 \right)^{\shuffle_\Theta\, 2} \nonumber\\ 
  +&
  \begin{bmatrix}
    n - 3 \\ 3
  \end{bmatrix} ( \alpha a_1)^{\shuffle_\Theta\, (n-3)}\shuffle_\Theta
  \left(\alpha^3 \Theta(a_1,a_1,a_1)\right) +
  \begin{bmatrix}
    n - 3 & 0 \\ 1 & 1
  \end{bmatrix} (\alpha a_1)^{\shuffle_\Theta\, (n-3)}\shuffle_\Theta
  \left( \alpha^3 \Theta(a_1,a_2) \right) \nonumber\\ 
  +&
  \begin{bmatrix}
    n - 4 \\ 2 \\ 2
  \end{bmatrix}
  (\alpha a_1)^{\shuffle_\Theta\, (n-4)}\shuffle_\Theta \left(
    \alpha^2 \Theta(a_1,a_1) \right)^{\shuffle_\Theta \, 2} +
  \begin{bmatrix}
    n - 4 & 1 \\ 2 & 0
  \end{bmatrix} ( \alpha a_1)^{\shuffle_\Theta\, (n-4)}\shuffle_\Theta
  \left( \alpha^2 a_2 \right) \shuffle_\Theta \left( \alpha^2
    \Theta(a_1,a_1) \right) \nonumber\\
  +&
  \begin{bmatrix}
    n-3
  \end{bmatrix}
  _{[a_1,a_2]} (\alpha a_1)^{\shuffle_\Theta\,
    (n-3)}\shuffle_\Theta \left(\alpha^3 [a_1,a_2]\right) +
  \begin{bmatrix}
    n-3
  \end{bmatrix}
  _{[a_1,\Theta(a_1,a_1)]} (\alpha a_1)^{\shuffle_\Theta\,
    (n-3)}\shuffle_\Theta \left( \alpha^3 [a_1,\Theta(a_1,a_1)]
  \right)\!\! \Bigg).
\end{align}
Acting with renormalized Feynman rules $\Psi_R$ on both sides results
in
\begin{align}
  \label{eq:PsiW-nnll}
  \Psi_R\left(W_\mathrm{Yuk}\right)\big|_\mathrm{n.n.l.l.} =&
  \Bigg( \begin{bmatrix} \bullet & 0 & 1
  \end{bmatrix} \alpha^3 \Psi_R(a_3) +
  \begin{bmatrix}
    \bullet & 2
  \end{bmatrix} \alpha^4 \Psi_R(a_2)^2 +
  \begin{bmatrix}
    \bullet \\ 3
  \end{bmatrix} \alpha^3 \Psi_R\textbf{(}
  \Theta(a_1,a_1,a_1)\textbf{)} +
  \begin{bmatrix}
    \bullet & 0 \\ 1 & 1
  \end{bmatrix} \alpha^3 \Psi_R\textbf{(} \Theta(a_1,a_2)\textbf{)}
  \nonumber\\ 
  &+
  \begin{bmatrix}
    \bullet \\ 2 \\ 2
  \end{bmatrix} \alpha^4 \Psi_R\textbf{(}\Theta(a_1,a_1)\textbf{)}^2 + 
  \begin{bmatrix}
    \bullet & 1 \\ 2 & 0
  \end{bmatrix} \alpha^4 \Psi_R(a_2)
  \Psi_R\textbf{(}\Theta(a_1,a_1)\textbf{)}
  \nonumber\\
  &+
  \begin{bmatrix}\bullet\end{bmatrix}_{[a_1,a_2]} \alpha^3 \Psi_R(
  [a_1,a_2] ) + 
  \begin{bmatrix}\bullet\end{bmatrix}_{[a_1,\Theta(a_1,a_1)]} \alpha^3
  \Psi_R( 
  [a_1,\Theta(a_1,a_1)]) \Bigg)\Bigg|_{z \to \alpha \Psi_R(a_1)}.
\end{align}

We write $\Psi_R = \Phi_R \circ \Upsilon^{-1}$ and obtain the full
next-to-next-to-leading log order Green function on the lhs. On the
rhs, for example $\Psi_R(a_3) = \Phi_R(\Gamma_3)$. However, the
letters $\Theta(a_1,a_1,a_1)$, $\Theta(a_1,a_2)$, $[a_1,a_2]$,
$[a_1,\Theta(a_1,a_1)]$ have no obvious corresponding Feynman graphs.
We therefore write
\begin{align}
  \Theta(a_1,a_1,a_1) =&\, 3
  B_+^{a_1}\left(B_+^{a_1}\left(a_1\right)\right) + \frac{3}{2} a_1
  \shuffle_\Theta \Theta(a_1,a_1) - \frac{1}{2} a_1\shuffle_\Theta a_1
  \shuffle_\Theta a_1, \\
  \Theta(a_1,a_2) =&\, - B_+^{a_1}(a_2) -
  B_+^{a_2}(a_1) + a_1 \shuffle_\Theta a_2, \\
  [a_1,a_2] =&\, B_+^{a_1}(a_2) - B_+^{a_2}(a_1), \\
  [a_1, \Theta(a_1,a_1)] =&\, 2B_+^{a_1} \left( a_1 \shuffle_\Theta
    a_1 \right) - B_+^{a_1}\left(B_+^{a_1}\left(a_1\right)\right) +
  \frac{1}{2} a_1 \shuffle_\Theta \Theta(a_1,a_1) - \frac{1}{2}
  a_1\shuffle_\Theta a_1 \shuffle_\Theta a_1.
\end{align}
Now, we act with $\Psi_R = \Phi_R \circ \Upsilon^{-1}$ and use
Eq.~(\ref{eq:PsiTheta1}). We thus, find the periods
\begin{align}
  \Psi_R\textbf{(}\Theta(a_1,a_1,a_1)\textbf{)} =&\, 3 \Phi_R\left(
    B_+^{\Gamma_1}\left(B_+^{\Gamma_1}\left(\Gamma_1\right)\right)\right)
  - 3 \Phi_R(\Gamma_1) \Phi_R
  \left(B_+^{\Gamma_1}\left(\Gamma_1\right)\right) +
  \Phi_R(\Gamma_1)^3, \\
  \Psi_R\textbf{(}\Theta(a_1,a_2)\textbf{)} =&\, -
  \Phi_R\left(B_+^{\Gamma_1}(\Gamma_2)\right) -
  \Phi_R\left(B_+^{\Gamma_2}(\Gamma_1)\right) + \Phi_R(\Gamma_1)
  \Phi_R( \Gamma_2), \\
  \Psi_R([a_1,a_2]) =&\, \Phi_R\left(B_+^{\Gamma_1}(\Gamma_2)\right) -
  \Phi_R\left(B_+^{\Gamma_2}(\Gamma_1)\right), \\
  \Psi_R([a_1, \Theta(a_1,a_1)]) =&\, 2\Phi_R \left( B_+^{\Gamma_1}
    \left( \Gamma_1 \cup \Gamma_1 \right) \right) - \Phi_R \left(
    B_+^{\Gamma_1}\left(B_+^{\Gamma_1}\left(\Gamma_1\right)\right)
  \right) -
  \Phi_R(\Gamma_1)\Phi_R\left(B_+^{\Gamma_1}\left(\Gamma_1\right)\right).
\end{align}
Inserting these identities together with Eq.(\ref{eq:PsiTheta1}) into
Eq.~(\ref{eq:PsiW-nnll}), we finally obtain the
next-to-next-to-leading log order Green function,
\begin{align}
  \label{eq:GR-nnll1}
  G_R(X_\mathrm{Yuk}) \big|_{n.n.l.l.} =& \, \alpha^3
  \Bigg( \begin{bmatrix} \bullet & 0 & 1
  \end{bmatrix} \Phi_R(\Gamma_3) + \alpha
  \begin{bmatrix}
    \bullet & 2
  \end{bmatrix} \Phi_R(\Gamma_2)^2 \nonumber\\ 
  &+\,
  \begin{bmatrix}
    \bullet \\ 3
  \end{bmatrix}
  \left( 3 \Phi_R\left(
      B_+^{\Gamma_1}\left(B_+^{\Gamma_1}\left(\Gamma_1\right)
      \right)\right) 
    - 3 \Phi_R(\Gamma_1) \Phi_R
    \left(B_+^{\Gamma_1}\left(\Gamma_1\right)\right) +
    \Phi_R(\Gamma_1)^3 \right) \nonumber\\
  &+\,
  \begin{bmatrix}
    \bullet & 0 \\ 1 & 1
  \end{bmatrix}
  \left( - \Phi_R\left(B_+^{\Gamma_1}(\Gamma_2)\right) -
    \Phi_R\left(B_+^{\Gamma_2}(\Gamma_1)\right) + \Phi_R(\Gamma_1)
    \Phi_R( \Gamma_2) \right)
  \nonumber\\
  &+\, \alpha
  \begin{bmatrix}
    \bullet \\ 2 \\ 2
  \end{bmatrix}
  \left( \Phi_R (\Gamma_1)^2 - 2 \Phi_R
    \left(B_+^{\Gamma_1}(\Gamma_1)\right) \right)^2 + \alpha
  \begin{bmatrix}
    \bullet & 1 \\ 2 & 0
  \end{bmatrix}
  \Phi_R(\Gamma_2) \left( \Phi_R (\Gamma_1)^2 - 2 \Phi_R
    \left(B_+^{\Gamma_1}(\Gamma_1)\right) \right) \nonumber\\
  &+\, \begin{bmatrix}\bullet\end{bmatrix}_{[a_1,a_2]} \left(
    \Phi_R\left(B_+^{\Gamma_1}(\Gamma_2)\right) -
    \Phi_R\left(B_+^{\Gamma_2}(\Gamma_1)\right)
  \right) \nonumber\\
  &+\, \begin{bmatrix}\bullet\end{bmatrix}_{[a_1,\Theta(a_1,a_1)]}
  \bigg( 2\Phi_R 
    \left( 
      B_+^{\Gamma_1} \left( \Gamma_1 \cup \Gamma_1 \right) 
    \right) -
    \Phi_R 
    \left(
      B_+^{\Gamma_1}\left(B_+^{\Gamma_1}\left(\Gamma_1\right)\right)
    \right) \nonumber\\ 
    & - \Phi_R(\Gamma_1) \Phi_R
    \left(
      B_+^{\Gamma_1}\left(\Gamma_1\right)
    \right)
  \bigg)
  \Bigg)\Bigg|_{z \to \alpha\Phi_R(\Gamma_1)}.
\end{align}
We refer to the second column of Table \ref{tab:results} for the
explicit expressions of the generating functions. {\it This shows the
  explicit dependence of the full next-to-next-to-leading log order
  Green function $G_R(X_\mathrm{Yuk}) \big|_{n.n.l.l.}$ on the Feynman
  graphs
  \begin{equation}
    \Gamma_1, \qquad \Gamma_2, \qquad
    B_+^{\Gamma_1}\left(\Gamma_1\right), \qquad
    \Gamma_3,\qquad B_+^{\Gamma_1}\left(\Gamma_2\right),\qquad
    B_+^{\Gamma_2}\left(\Gamma_1\right),\qquad B_+^{\Gamma_1} \left(
      B_+^{\Gamma_1}\left(\Gamma_1\right) \right), \qquad B_+^{\Gamma_1}
    \left( \Gamma_1 \cup \Gamma_1 \right).
  \end{equation}
  These are at most, three-loop graphs.}

\appendix

\section{Relations for the log-expansion of the QED photon
  self-energy}
\label{sec:qed-results}

We relate the next-to$^{\{j\}}$-leading log order to the first $(j+1)$
terms of perturbation theory in the QED photon self-energy Green
function $G_R(X_\mathrm{QED})$. There are two differences to the
Yukawa propagator Green function. \textbf{(}Consider the DSEs
Eqs.~(\ref{eq:DSE-solution-Yuk},\ref{eq:DSE-solution-QED}).
First, there are no insertion points in the one-loop primitive
propagator graph. The sum in Eq.~(\ref{eq:DSE-solution-QED}) starts
with $j = 2$ rather than $j = 1$ in Eq.~(\ref{eq:DSE-solution-Yuk}).
This will simplify the following calculations drastically. Secondly,
the term $
\begin{psmallmatrix}
  2j - 2 + k \\ k
\end{psmallmatrix}
$ in Eq.~(\ref{eq:DSE-solution-Yuk}) is replaced by $
\begin{psmallmatrix}
  j - 2 + k \\ k
\end{psmallmatrix}
$ in Eq.~(\ref{eq:DSE-solution-QED}). This will change the structure
$1/\sqrt{1-2z}$ in the generating functions of Yukawa theory to
$1/(1-z)$ in QED.

In the first part, we treat index-free matrices. The corresponding
shuffle products do not contain $[\cdot,\cdot]$-letters. We repeat the
same steps to derive the master differential equation as in Yukawa
theory. The two mentioned differences in the respective DSEs
Eqs.~(\ref{eq:DSE-solution-Yuk},\ref{eq:DSE-solution-QED}) change
Eq.~(\ref{eq:master}) to
\begin{align}
  \label{eq:masterQED}
  \mathcal{M}(z)' =& \sum_{m \sim \mathcal{M}} \left( -\sum_{m' \neq
      m} \frac{\mathrm{d}}{\mathrm{d}z} z^{|m| - |m'|}
    \mathcal{S}(m,m') m' z^{|m'|} +
    \delta_{|m|1}\delta_{n_\Theta(m)0} + \sum_{j\geq 2}^{\mathbf{u} m
      \mathbf{v}^T - 1} \, ( 1 - \delta_{m_{1j} \, 0} ) \, \sum_{k\geq
      1}^{\mathbf{u} m \mathbf{v}^T - j}
    \begin{pmatrix}
      j - 2 + k \\ k
    \end{pmatrix} \right. \nonumber\\
  &\left. \times\sum^{(*)} z^{|m| - 1 - \sum_i |m_i|}
    \mathcal{S}\left(m \ominus p_j, \bigoplus_i m_i \right) m_1
    z^{|m_1|} m_2 z^{|m_2|} \ldots m_k z^{|m_k|}
  \right), \\
  (*):\quad & t_i \geq 1,\quad i= 1 \ldots k, \quad \sum_{i=1}^{k} t_i =
  \mathbf{u} m \mathbf{v}^T - j,\quad \mathbf{u} m_i \mathbf{v}^T =
  t_i,\quad \sum_i \mathbf{u} m_i + \mathbf{u} p_j = \mathbf{u} m.
\end{align}
This is an ordinary equation for $\mathcal{M}(z)'$ and no differential
equation because the sum starts with $j = 2$. We therefore call
Eq.~(\ref{eq:masterQED}) {\it master equation}. We must integrate the
master equation to obtain $\mathcal{M}(z)$ such that $\mathcal{M}(0) =
0$ \textbf{(}Eq.~(\ref{eq:initial-condition})\textbf{)}.
Eqs.~(\ref{eq:def-generating-function},\ref{eq:matrices-of-generating-functions})
remain valid,
\begin{equation}
  \label{eq:def-generating-function-copy}
  \mathcal{M} = \mathcal{M}(z) = \sum_{m \sim \mathcal{M}}^{m \neq
    (0)} m \, z^{|m|}, \qquad m = \frac{1}{|m|!} \left(
    \frac{\mathrm{d}}{\mathrm{d}z}\right)^{|m|}
  \mathcal{M}(z)\Big|_{z=0}.
\end{equation}

Consider for example the case $\mathcal{M}(z)
= \begin{bsmallmatrix}\bullet\end{bsmallmatrix}$. The matrices $m\sim
\mathcal{M}$ belong to the shuffle products
$a_1^{\shuffle_\Theta\,N}$. Since there is no $B_+^{a_1}$, we already
know that $
\begin{bsmallmatrix}
  N
\end{bsmallmatrix}
= 0$ for $N>1$. In the master equation, all terms vanish except for
$\delta_{|m|1}\delta_{n_\Theta(m)0}$. Integrating the remaining
equation $\begin{bsmallmatrix}\bullet\end{bsmallmatrix}' = 1$ yields
\begin{equation}
  \label{eq:bulletQED1}
  \begin{bmatrix}\bullet\end{bmatrix} = z,
\end{equation}
which generates the matrices $
\begin{bsmallmatrix}
  N
\end{bsmallmatrix}
= \delta_{N\,1}$ as expected
\textbf{(}see Eq.~(\ref{eq:def-generating-function-copy})\textbf{)}.

In the following, we derive the generating functions up to
next-to-next-to-leading log order for the QED photon self-energy. As
in Section \ref{sec:relat-betw-next}, we consider the cases
$n_\Theta(m) = 0$ and $n_\Theta(m) \neq 0$ separately. In Section
\ref{sec:commutators-1}, we treat indexed matrices $m$.

\subsection{Generating functions for index-free matrices with
  $n_\Theta(m) = 0$}
\label{sec:index-free-n0-1}

Index-free matrices with $n_\Theta(m) = 0$ belong to shuffle products
without $\Theta(\cdot,\cdot)$- and $[\cdot,\cdot]$-letters. In full
analogy to the Yukawa propagator, only one row in $\mathcal{M}(z)$
reduces the master equation to
\begin{equation}
  \label{eq:masterQED-special}
  \mathcal{M}(z)' = \delta_{|\mathcal{M}|0} + \delta_{|\mathcal{M}|1} +
  \sum_{j \geq 2, k \geq 1}^\infty ( 1 - \delta_{\mathcal{M}_{1j} \,
    0} ) 
  \begin{pmatrix}
    j - 2 + k \\ k
  \end{pmatrix}
  \sum^{(**)}\mathcal{M}_1(z) \mathcal{M}_2(z) \ldots
  \mathcal{M}_k(z), 
\end{equation}
where
\begin{equation}
  \label{eq:starstarQED}
  (**): \quad \mathcal{M}_1 \oplus \mathcal{M}_2 \oplus \ldots \oplus
  \mathcal{M}_k \oplus \mathcal{P}_j = \mathcal{M}.
\end{equation}

\subsubsection{The generating function $
  \protect\begin{bsmallmatrix}
    \bullet & 1
  \protect\end{bsmallmatrix}
  $}

Let $\mathcal{M}(z) =
\begin{bsmallmatrix}
  \bullet & 1
\end{bsmallmatrix}
$. The sum in Eq.~(\ref{eq:masterQED-special}) is non-zero only for $j
= 2$. Then, $\mathcal{P}_j = \mathcal{P}_2 = \mathcal{M}$ and
$\mathcal{M}_i = \begin{bsmallmatrix}\bullet\end{bsmallmatrix} = z$
$\forall i \leq k$. We obtain
\begin{equation}
  \label{eq:bulletQED2}
  \begin{bmatrix}
    \bullet & 1
  \end{bmatrix}' = 1 + \sum_{k\geq
    1} \begin{bmatrix}\bullet\end{bmatrix}^k = \frac{1}{1-z} 
  \qquad \Rightarrow
  \begin{bmatrix}
    \bullet & 1
  \end{bmatrix} = \log
  \left(
    \frac{1}{1-z}
  \right).
\end{equation}

\subsubsection{The generating function $
  \protect\begin{bsmallmatrix}
    \bullet & 0 & 1
  \protect\end{bsmallmatrix}
  $}

Let $\mathcal{M}(z) =
\begin{bsmallmatrix}
  \bullet & 0 & 1
\end{bsmallmatrix}
$. The sum in Eq.~(\ref{eq:masterQED-special}) is non-zero only for $j
= 3$. Then, $\mathcal{P}_j = \mathcal{P}_3 = \mathcal{M}$ and
$\mathcal{M}_i = \begin{bsmallmatrix}\bullet\end{bsmallmatrix} = z$
$\forall i \leq k$. We obtain
\begin{equation}
  \label{eq:bulletQED3}
  \begin{bmatrix}
    \bullet & 0 & 1
  \end{bmatrix}' = 1 + \sum_{k\geq 1}
  \begin{pmatrix}
    k + 1 \\ k
  \end{pmatrix}
\begin{bmatrix}\bullet\end{bmatrix}^k = \frac{1}{(1-z)^2}
  \qquad \Rightarrow
  \begin{bmatrix}
    \bullet & 0 & 1
  \end{bmatrix} = - 1 + \frac{1}{1-z}.
\end{equation}
The term $-1$ is the integration constant such that $
\begin{bsmallmatrix}
  \bullet & 0 & 1
\end{bsmallmatrix}(0) = 0 $.

\subsubsection{The generating function $ \protect\begin{bsmallmatrix}
    \bullet & 2 \protect\end{bsmallmatrix} $}

Let $\mathcal{M}(z) =
\begin{bsmallmatrix}
  \bullet & 2 
\end{bsmallmatrix}
$. The sum in Eq.~(\ref{eq:masterQED-special}) is non-zero only for $j
= 2$. Then, $\mathcal{P}_j = \mathcal{P}_2$.
Eq.~(\ref{eq:starstarQED}) implies that for one $i \leq k$,
$\mathcal{M}_i =
\begin{bsmallmatrix}
  \bullet & 1
\end{bsmallmatrix} = \log\textbf{(}1/(1-z)\textbf{)} $ and for all other $i$, $\mathcal{M}_i
= \begin{bsmallmatrix}\bullet\end{bsmallmatrix} = z$. We obtain
\begin{equation}
  \label{eq:bulletQED4}
  \begin{bmatrix}
    \bullet & 2
  \end{bmatrix}' =  \sum_{k\geq 1}
  k
  \begin{bmatrix}\bullet\end{bmatrix}^{k-1}
  \begin{bmatrix}
    \bullet & 1
  \end{bmatrix}
  = \frac{1}{(1-z)^2} \log \left( \frac{1}{1-z} \right)
  \qquad \Rightarrow
  \begin{bmatrix}
    \bullet & 2
  \end{bmatrix}
  = 1 - \frac{1}{1-z} + \frac{1}{1-z} \log \left( \frac{1}{1-z}
  \right).
\end{equation}

\subsection{Generating functions for index-free matrices with
  $n_\Theta(m) \neq 0$}

We now treat full shuffle products that contain
$\Theta(\cdot,\cdot)$-letters but no $[\cdot,\cdot]$-letters. Here, we
have to proceed from the master equation Eq.~(\ref{eq:masterQED}). The
following generating functions simplify to zero,
\begin{equation}
  \label{eq:bulletQED5}
  \begin{bmatrix}
    \bullet \\ 2
  \end{bmatrix} =
  \begin{bmatrix}
    \bullet \\ 3
  \end{bmatrix} =
  \begin{bmatrix}
    \bullet \\ 2 \\ 2
  \end{bmatrix} = 0.
\end{equation}
The reason is that the sum in the master equation starts with $j = 2$.

\subsubsection{The generating function $ \protect\begin{bsmallmatrix}
    \bullet & 0 \protect\\ 1 & 1 \protect\end{bsmallmatrix}
  $}

There is another simplification due to the missing $(j=1)$-term. If the
first row of a matrix $m$ is of the form $
\begin{bsmallmatrix}
  N & 0 & 0 & \ldots
\end{bsmallmatrix}
$ with arbitrary $N$, then the second (and more complicated) term on
the rhs of the master equation Eq.~(\ref{eq:masterQED}) yields zero.
For example for the next-to-next-to-leading log generating function $
\begin{bsmallmatrix}
  \bullet & 0 \\ 1 & 1
\end{bsmallmatrix}
$, Eq.~(\ref{eq:masterQED}) reduces to
\begin{align}
  \begin{bmatrix}
    \bullet & 0 \\ 1 & 1
  \end{bmatrix}' =& \sum_{N = 0}^\infty \left( -
    \frac{\mathrm{d}}{\mathrm{d}z} z^{(N+1) - (N+2)}
    \mathcal{S} \left(
      \begin{bmatrix}
        N & 0 \\ 1 & 1
      \end{bmatrix},
      \begin{bmatrix}
        N + 1 & 1
      \end{bmatrix}
    \right)\,
    \begin{bmatrix}
      N + 1 & 1
    \end{bmatrix}
    z^{N+2} \right) \nonumber\\
  =& \sum_{N=0}^\infty \left( -\frac{\mathrm{d}}{\mathrm{d}z}
    \frac{1}{z} (N+1) \begin{bmatrix} N + 1 & 1
    \end{bmatrix}
    z^{N+2} \right) \nonumber\\
  =& \sum_{N=0}^\infty \left( -\frac{\mathrm{d}}{\mathrm{d}z}
    \frac{1}{z} z^2 \frac{\mathrm{d}}{\mathrm{d}z}
    \frac{1}{z} \begin{bmatrix} N + 1 & 1
    \end{bmatrix}
    z^{N+2} \right) \nonumber\\
  =\;&
  - \frac{\mathrm{d}}{\mathrm{d}z} z
  \frac{\mathrm{d}}{\mathrm{d}z} \frac{1}{z}
  \begin{bmatrix}
    \bullet & 1
  \end{bmatrix}.
\end{align}
Integration with suitable initial conditions and using
Eq.~(\ref{eq:bulletQED2}) result in
\begin{equation}
  \label{eq:bulletQED6}
  \begin{bmatrix}
    \bullet & 0 \\ 1 & 1
  \end{bmatrix} = -\frac{\mathrm{d}}{\mathrm{d}z}
  \begin{bmatrix}
    \bullet & 1
  \end{bmatrix} + \frac{1}{z} \begin{bmatrix}
    \bullet & 1
  \end{bmatrix} = - \frac{1}{1-z} + \frac{1}{z}\log \left(
    \frac{1}{1-z} \right).
\end{equation}

\subsubsection{The generating function $ \protect\begin{bsmallmatrix}
    \bullet & 1 \protect\\ 2 & 0 \protect\end{bsmallmatrix} $}

For $ \mathcal{M}(z) =
\begin{bsmallmatrix}
  \bullet & 1 \\ 2 & 0
\end{bsmallmatrix}
$, the master equation Eq.~(\ref{eq:masterQED}) reduces to 
\begin{align}
  \begin{bmatrix}
    \bullet & 1 \\ 2 & 0
  \end{bmatrix}' =& \sum_{N= 0 }^\infty \Bigg( -
  \frac{\mathrm{d}}{\mathrm{d}z} z^{(N+2) - (N+3)} \mathcal{S} \left(
    \begin{bmatrix}
      N & 1 \\ 2 & 0 
    \end{bmatrix},
    \begin{bmatrix}
      N + 2 & 1
    \end{bmatrix}
  \right) \,
  \begin{bmatrix}
    N + 2 & 1
  \end{bmatrix}
  z^{N+3} \nonumber\\
  &+ \sum_{k\geq1} \sum_{\sum_i t_i = (N + 4) - 2} z^{(N+2) - 1 -
    \sum_i t_i} \mathcal{S} \left(
    \begin{bmatrix}
      N \\ 2
    \end{bmatrix},
    \begin{bmatrix}
      N+2
    \end{bmatrix}
  \right)
  \begin{bmatrix}
    t_1
  \end{bmatrix}
  z^{t_1} \ldots
  \begin{bmatrix}
    t_k
  \end{bmatrix}
  z^{t_{k}} \Bigg)
  \nonumber\\
  =& \sum_{N=0}^\infty \Bigg( -\frac{\mathrm{d}}{\mathrm{d}z}
  \frac{1}{z}
  \begin{pmatrix}
    N+2 \\ 2
  \end{pmatrix} \begin{bmatrix} N + 2 & 1
  \end{bmatrix}
  z^{N+3} + \sum_{k\geq1} \sum_{\sum_i t_i = N + 2} \frac{1}{z}
  \begin{pmatrix}
    N+2 \\ 2
  \end{pmatrix}
  \begin{bmatrix}
    t_1
  \end{bmatrix}
  z^{t_1} \ldots
  \begin{bmatrix}
    t_k
  \end{bmatrix}
  z^{t_{k}} \Bigg) \nonumber\\
  =& \sum_{N=0}^\infty \Bigg( -\frac{\mathrm{d}}{\mathrm{d}z}
  \frac{1}{z}\frac{z^3}{2} \frac{\mathrm{d}^2}{\mathrm{d}z^2}
  \frac{1}{z} 
  \begin{bmatrix} N + 2 & 1
  \end{bmatrix}
  z^{N+3} + \sum_{k\geq1} \sum_{\sum_i t_i = N + 2} \frac{1}{z}
  \frac{z^2}{2} \frac{\mathrm{d}^2}{\mathrm{d}z^2}
  \begin{bmatrix}
    t_1
  \end{bmatrix}
  z^{t_1}
  \ldots
  \begin{bmatrix}
    t_k
  \end{bmatrix}
  z^{t_{k}} \Bigg) \nonumber\\
  =& - \frac{\mathrm{d}}{\mathrm{d}z} \frac{z^2}{2}
  \frac{\mathrm{d}^2}{\mathrm{d}z^2} \frac{1}{z}
  \begin{bmatrix}
    \bullet & 1
  \end{bmatrix}
  + \sum_{k\geq1} \frac{z}{2} \frac{\mathrm{d}^2}{\mathrm{d}z^2}
  \begin{bmatrix}\bullet\end{bmatrix}^k \nonumber\\
  =& \frac{\mathrm{d}}{\mathrm{d}z} \left(
    -\frac{z^2}{2}\frac{\mathrm{d}^2}{\mathrm{d}z^2}\frac{1}{z}\log 
    \left( \frac{1}{1-z} \right) +
    \frac{z}{2}\frac{\mathrm{d}}{\mathrm{d}z} \frac{1}{1-z} -
    \frac{1}{2} \frac{1}{1-z} \right).
\end{align}
In the last line, we used
Eqs.~(\ref{eq:bulletQED1},\ref{eq:bulletQED2}). We integrate with
suitable initial conditions and obtain
\begin{equation}
  \label{eq:bulletQED7}
  \begin{bmatrix}
    \bullet & 1 \\ 2 & 0
  \end{bmatrix} = \frac{1}{2} + \frac{1}{2(1-z)} - \frac{1}{z}\log
  \left(
    \frac{1}{1-z}
  \right).
\end{equation}

\subsection{Generating functions for indexed matrices}
\label{sec:commutators-1}

As in Yukawa theory, we only calculate the generating functions
$\begin{bsmallmatrix}\bullet\end{bsmallmatrix}_{[a_1,a_2]}$ and
$\begin{bsmallmatrix}\bullet\end{bsmallmatrix}_{[a_1,\Theta(a_1,a_1)]}$
that
belong to the shuffle products $a_1^{\shuffle_\Theta\,
  N}\shuffle_\Theta [a_1,a_2]$ and $a_1^{\shuffle_\Theta\,
  N}\shuffle_\Theta [a_1,\Theta(a_1,a_1)]$.

We find
\begin{equation}
  \label{eq:bulletQED8}
  \begin{bmatrix}\bullet\end{bmatrix}_{[a_1,\Theta(a_1,a_1)]} = 0
\end{equation}
because the sum in Eq.~(\ref{eq:DSE-solution-QED}) starts with $j=2$. 

For the generating function
$\begin{bsmallmatrix}\bullet\end{bsmallmatrix}_{[a_1,a_2]}$,
Eq.~(\ref{eq:dgl-commutator}) reduces to
\begin{equation}
  \label{eq:integration-commutator}
  \int \begin{bmatrix}\bullet\end{bmatrix}_{[a_1,a_2]} \mathrm{d}z = -
  \frac{z^2}{2}\frac{\mathrm{d}}{\mathrm{d}z}\frac{1}{z}
  \begin{bmatrix}
    \bullet & 1
  \end{bmatrix}
\end{equation}
because in Eq.~(\ref{eq:wn-contribution-1}), the first two terms are
missing. Here, the derivation of the generating function is even
simpler than in all previous cases. One only needs to differentiate
instead of integrating or solving a differential equation.

We differentiate Eq.~(\ref{eq:integration-commutator}) and use the
explicit form of $
\begin{bsmallmatrix}
  \bullet & 1
\end{bsmallmatrix}
$ in Eq.~(\ref{eq:bulletQED2}). This results in
\begin{equation}
  \label{eq:bulletQED9}
  \begin{bmatrix}\bullet\end{bmatrix}_{[a_1,a_2]} = \frac{1}{2(1-z)} -
  \frac{1}{2(1-z)^2}. 
\end{equation}

\subsection{Results}
\label{sec:results-1}

We repeat the steps in Section \ref{sec:results} to write the
next-to$^{\{j\}}$-leading log order as a function of terms up to
$\mathcal{O}(j+1)$ in the log-expansion
\textbf{(}Eq.~(\ref{eq:log-expansion})\textbf{)}. We show this up to
$j \leq 2$ and use
Eqs.~(\ref{eq:GR-ll1},\ref{eq:GR-nll1},\ref{eq:GR-nnll1}), which are universally valid
(in the QED case below several terms vanish):
\begin{align}
  G_R(X_\mathrm{Yuk}) \big|_{l.l.}
  =&\, \begin{bmatrix}\bullet\end{bmatrix}|_{z \to
    \alpha\Phi_R(\Gamma_1)}, \\
  G_R(X_\mathrm{Yuk}) \big|_{n.l.l.} =&\, \begin{bmatrix} \bullet & 1 
  \end{bmatrix}\big|_{z \to \alpha\Phi_R(\Gamma_1)}
  \alpha^2\Phi_R(\Gamma_2) +
    \begin{bmatrix}
      \bullet \\ 2
    \end{bmatrix}\bigg|_{z \to \alpha\Phi_R(\Gamma_1)}
    \alpha^2 \left( \Phi_R (\Gamma_1)^2 - 2 \Phi_R
      \left(B_+^{\Gamma_1}(\Gamma_1)\right)\right),\\
    G_R(X_\mathrm{Yuk}) \big|_{n.n.l.l.} =& \, \alpha^3
    \Bigg( \begin{bmatrix} \bullet & 0 & 1
  \end{bmatrix} \Phi_R(\Gamma_3) + \alpha
  \begin{bmatrix}
    \bullet & 2
  \end{bmatrix} \Phi_R(\Gamma_2)^2 \nonumber\\ 
  &+\,
  \begin{bmatrix}
    \bullet \\ 3
  \end{bmatrix}
  \left( 3 \Phi_R\left(
      B_+^{\Gamma_1}\left(B_+^{\Gamma_1}\left(\Gamma_1\right)\right)
    \right)
    - 3 \Phi_R(\Gamma_1) \Phi_R
    \left(B_+^{\Gamma_1}\left(\Gamma_1\right)\right) +
    \Phi_R(\Gamma_1)^3 \right) \nonumber\\
  &+\,
  \begin{bmatrix}
    \bullet & 0 \\ 1 & 1
  \end{bmatrix}
  \left( - \Phi_R\left((B_+^{\Gamma_1}(\Gamma_2)\right) -
    \Phi_R\left(B_+^{\Gamma_2}(\Gamma_1)\right) + \Phi_R(\Gamma_1)
    \Phi_R( \Gamma_2)
  \right)
  \nonumber\\ 
  &+\, \alpha
  \begin{bmatrix}
    \bullet \\ 2 \\ 2
  \end{bmatrix}
  \left( \Phi_R (\Gamma_1)^2 - 2 \Phi_R
    \left(B_+^{\Gamma_1}(\Gamma_1)\right) \right)^2 + \alpha
  \begin{bmatrix} \bullet & 1 \\ 2 & 0
  \end{bmatrix} \Phi_R(\Gamma_2) \left( \Phi_R (\Gamma_1)^2 - 2 \Phi_R
    \left(B_+^{\Gamma_1}(\Gamma_1)\right) \right) \nonumber\\
  &+\, \begin{bmatrix}\bullet\end{bmatrix}_{[a_1,a_2]} \left(
    \Phi_R\left(B_+^{\Gamma_1}(\Gamma_2)\right) -
    \Phi_R\left(B_+^{\Gamma_2}(\Gamma_1)\right)
  \right) \nonumber\\
  &+\, \begin{bmatrix}\bullet\end{bmatrix}_{[a_1,\Theta(a_1,a_1)]}
  \bigg( 2\Phi_R \left( B_+^{\Gamma_1} \left( \Gamma_1 \cup \Gamma_1
    \right) \right) - \Phi_R \left(
    B_+^{\Gamma_1}\left(B_+^{\Gamma_1}\left(\Gamma_1\right)\right)
  \right) \nonumber\\
  &-
  \Phi_R(\Gamma_1)\Phi_R\left(B_+^{\Gamma_1}\left(\Gamma_1\right)\right)
  \bigg) \Bigg)\Bigg|_{z \to \alpha\Phi_R(\Gamma_1)}.
\end{align}
For the leading log and next-to-leading log expansions, we use the
explicit generating functions in the third column of Table
\ref{tab:results}. For the next-to-next-to-leading log order, we only
discard the zero functions. Thus, we finally obtain
\begin{align}
  G_R(X_\mathrm{QED}) \big|_{l.l.} =&\, \alpha\Phi_R(\Gamma_1), \\
  G_R(X_\mathrm{QED}) \big|_{n.l.l.} =&\, \alpha^2 \Phi_R(\Gamma_2)
  \log \left( \frac{1}{1-\alpha\Phi_R(\Gamma_1)}
  \right),\\
  G_R(X_\mathrm{QED}) \Big|_{n.n.l.l.} =&\, \alpha^3
  \Bigg( \begin{bmatrix} \bullet & 0 & 1
  \end{bmatrix} \Phi_R(\Gamma_3) + \alpha
  \begin{bmatrix}
    \bullet & 2
  \end{bmatrix} \Phi_R(\Gamma_2)^2 \nonumber\\
  &+ \begin{bmatrix}
    \bullet & 0 \\ 1 & 1
  \end{bmatrix}
  \left( - \Phi_R\left((B_+^{\Gamma_1}(\Gamma_2)\right) -
    \Phi_R\left(B_+^{\Gamma_2}(\Gamma_1)\right) + \Phi_R(\Gamma_1)
    \Phi_R( \Gamma_2) \right) \nonumber\\
  &+ \alpha
  \begin{bmatrix} \bullet & 1 \\ 2 & 0
  \end{bmatrix} \Phi_R(\Gamma_2) \left( \Phi_R (\Gamma_1)^2 - 2 \Phi_R
    \left(B_+^{\Gamma_1}(\Gamma_1)\right) \right) \nonumber\\
  &+ \begin{bmatrix}\bullet\end{bmatrix}_{[a_1,a_2]} \left(
    \Phi_R\left(B_+^{\Gamma_1}(\Gamma_2)\right) -
    \Phi_R\left(B_+^{\Gamma_2}(\Gamma_1)\right) \right)
  \Bigg)\Bigg|_{z \to \alpha\Phi_R(\Gamma_1)}.
\end{align}
See the third column of Table \ref{tab:results} for the remaining
generating functions.

\section{Some multiplicities of shuffles in filtered words}
\label{sec:summary}

We list some multiplicities $m$ that are generated by the generating
functions obtained so far, see Table \ref{tab:results}. We treat the
Yukawa fermion propagator and the QED photon self-energy separately.
If sequences are known, we say so explicitly and refer to \cite{OEIS}.

\subsection{Yukawa fermion propagator}
\label{sec:rationals-yukawa}

The generating functions that are necessary to simplify the
log-expansion of the Yukawa fermion propagator up to
next-to-next-to-leading log order are collected in the second column
of Table \ref{tab:results}. Some of the corresponding multiplicities
are collected in Table \ref{tab:yukawa}.

$\begin{bsmallmatrix}\bullet\end{bsmallmatrix}$ generates the
exponential sequence of double factorial odd numbers A001147,
\begin{equation}
  \begin{bmatrix}
    N
  \end{bmatrix}
  = \frac{(2N-3)(2N-5)(2N-7) \ldots 1}{N!} =: \frac{(2N -
    3)!!}{N!}.
\end{equation}

Furthermore, $
\begin{bsmallmatrix}
  \bullet & 1
\end{bsmallmatrix}
$ generates the exponential series for the scaled sums of odd
reciprocals A004041. The formula is
\begin{equation}
  \begin{bmatrix}
    N & 1
  \end{bmatrix} = \frac{(2N + 1)!!}{(N+1)!} \sum_{k = 0}^{N}
  \frac{1}{2k + 1}.
\end{equation}
We finally find
\begin{equation}
  \begin{bmatrix}
    N & 0 & 1
  \end{bmatrix} = \frac{(2N + 2)!}{N!(N+1)! 2^{N+1}}
\end{equation}
(exponential sequence A001879). As far as we are concerned, there are
no known sequences for the other rationals in Table \ref{tab:yukawa}.

\begin{table}[!]
  \centering
  \begin{tabular}{c|cccccccc}
    \hline\hline
    \backslashbox[2cm]{~~~~$m$}{$N$~~~~} & 0 & 1 & 2 & 3 & 4 & 5\\
    \hline\\[-5pt]
    $
    \begin{bmatrix}
      N
    \end{bmatrix}
    $ & 0 & 1 & $\frac{1}{2}$ & $\frac{1}{2}$ & $\frac{5}{8}$ &
    $\frac{7}{8}$ \\[10pt]
    $\begin{bmatrix}
      N & 1
    \end{bmatrix}$ & 1 & 2 & $\frac{23}{6}$ & $\frac{22}{3}$ &
    $\frac{563}{40}$ & $\frac{1627}{60}$ \\[10pt]
    $\begin{bmatrix} N \\ 2
    \end{bmatrix}$ & $-\frac{1}{2}$ & $-1$ & $-\frac{23}{12}$ &
    $-\frac{11}{3}$ & $-\frac{563}{80}$ & $-\frac{1627}{120}$ \\[10pt]
    $\begin{bmatrix} N & 0 & 1
    \end{bmatrix}$ & 1 & 3 & $\frac{15}{2}$ & $\frac{35}{2}$ &
    $\frac{315}{8}$ & $\frac{693}{8}$\\[10pt]
    $\begin{bmatrix} N & 2
    \end{bmatrix}$ & $\frac{3}{2}$ & $\frac{41}{6}$ & $\frac{265}{12}$
    &
    $\frac{3707}{60}$ & $\frac{114961}{720}$ & $\frac{219803}{560}$ \\[10pt]
    $\begin{bmatrix} N \\ 3
    \end{bmatrix}$ & $\frac{1}{2}$ & $\frac{11}{6}$ & $\frac{61}{12}$
    &
    $\frac{253}{20}$ & $\frac{7141}{240}$ & $\frac{113623}{1680}$ \\[10pt]
    $\begin{bmatrix} N & 0 \\ 1 & 1
    \end{bmatrix}$ & $-2$ & $-\frac{20}{3}$ & $-\frac{53}{3}$ &
    $-\frac{214}{5}$ & $-\frac{5933}{60}$ & $-\frac{46597}{210}$ \\[10pt]
    $\begin{bmatrix} N \\ 2 \\ 2
    \end{bmatrix}$ & $\frac{1}{24}$ & $\frac{3}{8}$ &
    $\frac{389}{240}$ & $\frac{1291}{240}$ & $\frac{314431}{20160}$ &
    $\frac{93403}{2240}$
    \\[15pt]
    $\begin{bmatrix} N & 1 \\ 2 & 0
    \end{bmatrix}$ & $-\frac{5}{6}$ & $-\frac{25}{6}$ &
    $-\frac{857}{60}$ & $-\frac{833}{20}$ & $-\frac{559579}{5040}$ &
    $-\frac{156603}{560}$
    \\[10pt]
    $
    \begin{bmatrix}
      N
    \end{bmatrix}
    _{[a_1,a_2]}$ & $-1$ & $-6$ & $-\frac{71}{3}$ &
    $-\frac{155}{2}$ & $-\frac{9129}{40}$ & $-\frac{18823}{30}$
    \\[10pt]
    $
    \begin{bmatrix}
      N
    \end{bmatrix}
    _{[a_1,\Theta(a_1,a_1)]}$ & $\frac{1}{2}$ & $3$ &
    $\frac{71}{6}$ &
    $\frac{155}{4}$ & $\frac{9129}{80}$ & $\frac{18823}{60}$\\[5pt]
    \hline\hline
  \end{tabular}
  \caption{List of some multiplicities $m$ occurring in the filtration of
    the Yukawa fermion propagator graphs. These are derived from the generating
    functions up to next-to-next-to-leading log order in the second
    column in Table \ref{tab:results}.}  
  \label{tab:yukawa}
\end{table}

\subsection{QED photon self-energy}
\label{sec:rationals-qed}

The generating functions that are necessary to simplify the
log-expansion of the photon self-energy Green function up to
next-to-next-to-leading log order are given in the third column of
Table \ref{tab:results}. Some of the corresponding rationals are
listed in Table \ref{tab:photon}. The reader immediately checks that
these numbers look much simpler than in the Yukawa case. Indeed, four
rows only contain zero numbers (the first line is almost
zero, $
\begin{bsmallmatrix}
  N
\end{bsmallmatrix}
= \delta_{N\,1} $).

We find the trivial sequences
\begin{equation}
  \begin{bmatrix}
    N & 0 & 1
  \end{bmatrix} = 1, \qquad
  \begin{bmatrix}
    N & 0 \\ 1 & 1
  \end{bmatrix} = -\frac{N+1}{N+2}, \qquad
  \begin{bmatrix}
    N & 1 \\ 2 & 0
  \end{bmatrix} = \frac{N+1}{2(N+3)}, \qquad 
  \begin{bmatrix}
    N
  \end{bmatrix}
  _{[a_1,a_2]} = - \frac{N+1}{2}.
\end{equation}
One also finds that $
\begin{bsmallmatrix}
  N & 2
\end{bsmallmatrix}
$ is the exponential series of the generalized Stirling numbers
A001705,
\begin{equation}
  \begin{bmatrix}
    N & 2
  \end{bmatrix} = \frac{1}{N+2}\sum_{k=0}^{N}\frac{k+1}{N+1-k}.
\end{equation}

\begin{table}[!]
  \centering
  \begin{tabular}{c|cccccccc}
    \hline\hline
    \backslashbox[2cm]{~~~~$m$}{$N$~~~~} & 0 & 1 & 2 & 3 & 4 & 5\\
    \hline\\[-5pt]
    $
    \begin{bmatrix}
      N
    \end{bmatrix}
    $ & 1 & 0 & 0 & 0 & 0 & 0 \\[10pt]
    $\begin{bmatrix}
      N & 1
    \end{bmatrix}$ & 1 & $\frac{1}{2}$ & $\frac{1}{3}$ & $\frac{1}{4}$
    &
    $\frac{1}{5}$ & $\frac{1}{6}$ \\[10pt]
    $\begin{bmatrix} N \\ 2
    \end{bmatrix}$ & 0 & 0 & 0 & 0 & 0 & 0 \\[10pt]
    $\begin{bmatrix} N & 0 & 1
    \end{bmatrix}$ & 1 & 1 & 1 & 1 & 1 & 1\\[10pt]
    $\begin{bmatrix} N & 2
    \end{bmatrix}$ & $\frac{1}{2}$ & $\frac{5}{6}$ & $\frac{13}{12}$ &
    $\frac{77}{60}$ & $\frac{29}{20}$ & $\frac{223}{140}$ \\[10pt]
    $\begin{bmatrix} N \\ 3
    \end{bmatrix}$ & 0 & 0 & 0 & 0 & 0 & 0 \\[10pt]
    $\begin{bmatrix} N & 0 \\ 1 & 1
    \end{bmatrix}$ & $-\frac{1}{2}$ & $-\frac{2}{3}$ & $-\frac{3}{4}$
    & $-\frac{4}{5}$ & $-\frac{5}{6}$ & $-\frac{6}{7}$ \\[10pt]
    $\begin{bmatrix} N \\ 2 \\ 2
    \end{bmatrix}$ & 0 & 0 & 0 & 0 & 0 & 0 \\[15pt]
    $\begin{bmatrix} N & 1 \\ 2 & 0
    \end{bmatrix}$ & $\frac{1}{6}$ & $\frac{1}{4}$ & $\frac{3}{10}$ &
    $\frac{1}{3}$ & $\frac{5}{14}$ & $\frac{3}{8}$ \\[10pt]
    $
    \begin{bmatrix}
      N
    \end{bmatrix}
    _{[a_1,a_2]}$ & $-\frac{1}{2}$ & $-1$ & $-\frac{3}{2}$ & $-2$
    & $-\frac{5}{2}$ & $-3$
    \\[10pt]
    $
    \begin{bmatrix}
      N
    \end{bmatrix}
    _{[a_1,\Theta(a_1,a_1)]}$ & 0 & 0 & 0 & 0 & 0 & 0\\[5pt]
    \hline\hline
  \end{tabular}
  \caption{List of some multiplicities $m$ occurring in the filtration of
    the QED photon self-energy graphs. These are derived from the
    generating functions up to next-to-next-to-leading log order in
    the third column in Table \ref{tab:results}} 
  \label{tab:photon}
\end{table}
\clearpage
\section*{References}
\bibliographystyle{apsrev}

\end{document}